\documentclass{esub2acm}

\usepackage{theorem}
\usepackage{latexsym}
\usepackage{amsfonts}
\usepackage{amssymb}
\usepackage{amsmath}
\usepackage{float}

\newtheorem{definition}{Definition}

\newtheorem{lemma}{Lemma}

\newtheorem{problem}{Problem}
\newtheorem{solution}{Solution}
\newenvironment{example}{\medskip \noindent{\bf\em Example.}\em}{\medskip}

\floatstyle{plain}
\floatname{algorithm}{\footnotesize Alg.}
\newfloat{algorithm}{htbp}{1op}

\newcommand{\polylog}{\textrm{polylog}}

\newcommand{\tf}{\mathsf{tf}}
\newcommand{\df}{\mathsf{df}}
\newcommand{\dmin}{\mathsf{dmin}}
\newcommand{\idf}{\mathsf{idf}}
\newcommand{\occ}{\mathsf{occ}}
\newcommand{\docc}{\mathsf{docc}}
\newcommand{\List}{\mathsf{list}}
\newcommand{\Top}{\mathsf{top}}
\newcommand{\T}{\mathcal{T}}
\newcommand{\Ss}{\mathcal{S}}
\newcommand{\D}{\mathcal{D}}
\newcommand{\tsearch}{t_\mathsf{search}}
\newcommand{\taccess}{t_\mathsf{SA}}
\newcommand{\CSA}{\mathsf{CSA}}
\newcommand{\rmq}{\textsc{rmq}}
\newcommand{\rank}{\mathsf{rank}}
\newcommand{\select}{\mathsf{select}}

\renewcommand{\citeN}[1]{\shortciteN{#1}}
\renewcommand{\citeANP}[1]{\shortciteANP{#1}}

\begin{document}

\title{Spaces, Trees and Colors: The Algorithmic Landscape of 
	Document Retrieval on Sequences}

\author{%
Gonzalo Navarro \\
Department of Computer Science, University of Chile \\
}


\begin{abstract}
Document retrieval is one of the best established information retrieval
activities since the sixties, pervading all search engines. Its aim is to
obtain, from a collection of text documents, those most relevant to a pattern
query. Current technology is mostly oriented to ``natural language'' text
collections, where inverted indexes are the preferred solution. As 
successful as this paradigm has been, it fails to properly handle various
East Asian languages and other scenarios where the ``natural language''
assumptions do not hold.
In this survey we cover the recent research in extending the document retrieval
techniques to a broader class of sequence collections, which has 
applications in bioinformatics, data and Web mining, chemoinformatics, 
software engineering, multimedia information retrieval, and many other fields.
We focus on the algorithmic aspects of the techniques, uncovering a rich world 
of relations between document retrieval challenges and fundamental problems on 
trees, strings, range queries, discrete geometry, and other areas.
\end{abstract}

\category{E.1}{Data structures}{}
\category{E.2}{Data storage representations}{}
\category{E.4}{Coding and information theory}{Data compaction and compression}
\category{F.2.2}{Analysis of algorithms and problem complexity}
        {Nonnumerical algorithms and problems}
        [Pattern matching, Computations on discrete structures, Sorting and
searching]
\category{H.2.1}{Database management}
                {Physical design}
                [Access methods]
\category{H.3.2}{Information storage and retrieval}
                {Information storage}
                [File organization]
\category{H.3.3}{Information storage and retrieval}
        {Information search and retrieval}
        [Search process]

\terms{Algorithms}
\keywords{Text indexing, information retrieval, string searching, colored
range queries, compact data structures, orthogonal range searches.}

\begin{bottomstuff}

Funded by Millennium Nucleus Information and Coordination in Networks ICM/FIC P10-024F.

\begin{authinfo}
\address{%
	Gonzalo Navarro, Blanco Encalada 2120, Santiago, Chile.
	E-mail: {\tt gnavarro@dcc.uchile.cl}.
	Web: {\tt http://www.dcc.uchile.cl/gnavarro}.}

\end{authinfo}
\permission

\end{bottomstuff}

\markboth{G. Navarro}{Document Retrieval on General Sequence Collections}

\maketitle

%
%

\section{Introduction}

Retrieving useful information from huge masses of data is undoubtedly one of 
the most important activities enabled by computers in the Information Age, and
text is the preferred format to represent and retrieve most of this data.
Albeit images and videos have gained much importance on the Internet, 
even on those supports the searches are mostly carried out on 
text (e.g., Google finds images based on the text
content around them in Web pages). The problem of finding the relevant 
information in large masses of text was already pressing in the sixties, 
where the basis of modern Information Retrieval techniques were laid out 
\cite{Sal68}. Nowadays, this has become one of the most important
research topics in Computer Science \cite{CMS09,BCC10,BYRN11}.

Interestingly, besides a large degree of added complexity to permanently
improve the search ``quality'' (i.e., how the returned information matches
the need expressed by the query), the core of the approach has not changed
much since \citeANP{Sal68}'s time. One assumes that there is a {\em collection 
of documents}, each of which is a {\em sequence of words}. This collection is 
{\em indexed}, that is, preprocessed in some form. This index is able to
answer {\em queries}, which are words, or sets of words, or sets of
{\em phrases} (word sequences). A {\em relevance formula} is used to establish
how relevant is each of the documents for the query. The task of the index
is to return a set of documents most relevant to the query, according to the
formula.

In the original vector space model \cite{Sal68}, a set of distinguished words
(called {\em terms}) was extracted from the documents. A {\em weight}
$w(t,d)$ for term $t$ in document $d$ was defined using the assumption that
a term appearing many times in a document was important in it. Thus a component
of the weight was the {\em term frequency}, $\tf(t,d)$, which is the number of
times $t$ appears in $d$. A second component was aimed to downplay the role
of terms that appeared in many documents (such as articles and prepositions),
as those do not really distinguish a document from others. The so-called
{\em inverse document frequency} was defined as $\idf(t) = \lg (D/\df(t))$,
where $D$ is the total number of documents and $\df(t)$ is the number of
documents where $t$ appears.\footnote{We use $\lg$ to specify logarithm
in base 2 (when it matters).} Then $w(t,d) = \tf(t,d) \times \idf(t)$
was the formula used in the famous ``tf-idf'' model. The query was a set
of terms, $Q = \{ q_1, q_2, \ldots, q_m\}$, and the relevance of a document
$d$ for query $Q$ was $w(Q,d) = \sum_{i=1}^m w(q_i,d)$. Then the system
returned the {\em top-$k$ documents} for $Q$, that is, $k$ documents $d$ with 
the highest $w(Q,d)$ value. When computers became more powerful, the so-called {\em
full-text model} took every word in the documents as a querieable term.

As said, this simple model has been sophisticated in recent years up to an 
amazing degree, including some features that are possible due to the social
nature of the Internet: the intrinsic value of the documents, the links
between documents, the fields where the words appear, the feedback and
profile of the user, the behavior of other users that made similar queries, 
and so on. Yet, the core of the idea is still to find documents where
the query terms appear many times.

The {\em inverted index} has always been the favorite structure to support
these searches. The essence of this structure could not be simpler: given the
{\em vocabulary} of all the querieable terms, the index stores a list of the
documents $d$ where each such term $t$ appears, plus information to compute its
weight $w(t,d)$ in each. Much research has been carried out to efficiently
store and access inverted indexes \cite{WMB99,BCC10,BYRN11}, without changing
its essential organization. All modern search engines use
variants of inverted indexes.

Despite the immense success of this information retrieval model and 
implementation, it has a clear limitation: it strongly relies on the fact 
that the vocabulary of all the querieable terms has a manageable
size. The empirical law proposed by \citeN{Hea78} establishes
that the vocabulary of a collection of size $n$ grows like $\Theta(n^\beta)$ for
some constant $0<\beta<1$, and it holds very accurately in many Western
languages. The model, therefore, restricts the queries to be whole words,
not parts of words. It is not even obvious how to deal with phrases. One could 
extend the concepts of $\tf$ and $\idf$ to phrases and parts of words, but 
this would be quite difficult to implement with an inverted index: in principle,
it would have to store the list of documents where every conceivable text 
substring appears!

Such a limitation causes problems in highly synthetic languages such as
Finnish, Hungarian, Japanese, German, and many others where long words are
built from particles. But it is more striking in languages where word
separators are absent from written text and can only be inferred by
understanding its meaning: Chinese, Korean, Thai, Japanese (Kanji), Lao,
Vietnamese, and many others. Indeed, ``segmenting'' those texts into words
is considered a research problem belonging to the area of Natural Language
Processing (NLP); see, e.g., \citeN{RX12}. 

Out of resorting to expensive and heuristic NLP techniques, a solution
for those cases is to treat the text as an uninterpreted sequence of symbols 
and allow queries to find {\em any substring} in those sequences. The problem,
as said, is that inverted indexes cannot handle those queries. Instead,
{\em suffix trees}
\cite{Wei73,McC76,Apo85} and {\em suffix arrays} \cite{GBYS92,MM93}, and their
recent space-efficient versions \cite{NM07} are data structures that efficiently
solve the {\em pattern matching} problem, that is, they list all the
positions in the sequences where a pattern appears as a substring. 
However, they are not
easily modified to handle {\em document retrieval} problems, such as listing
the documents, or just the most relevant documents, where the pattern appears.

Extending the document retrieval technology to efficiently handle collections
of general sequences is not only interesting to enable classical Information
Retrieval on those languages where the basic assumptions of inverted indexes
do not hold. It also opens the door to using document retrieval techniques in a
number of areas where similar queries are of interest:

\paragraph*{Bioinformatics} Searching and mining collections of 
DNA, gene, and amino acid sequences is at the core of most Bioinformatic
tasks. Genes can be regarded as documents formed by sequences of base pairs
({\tt A}, {\tt C}, {\tt G}, {\tt T}),
proteins can be seen as documents formed by amino acid sequences (an alphabet
of size 20), and
even genomes can be modeled as documents formed by gene sequences (here each
gene is identified with an integer number).
Many searching and mining problems are solved with 
suffix trees \cite{Gus97}, and some are best recast into document retrieval 
problems. Some examples are listing the proteins where a certain amino acid 
sequence appears, or the genes where a certain DNA marker appears often, 
or the 
genomes where a certain set of genes appear, and so on. Further, bioinformatic
databases integrate not only sequence data but also data on structure, function,
metabolics, location, and other items that are not always natural
language. See, for example, \citeN{Bar11}.

\paragraph*{Software repositories} Handling a large software repository requires
managing a number of versions, packages, modules, routines, etc., which can
be regarded as documents formed by sequences in some formal language (such as
a programming or a specification language). In maintaining such repositories
it is natural to look for modules implementing some function, functions that
use some expression in their code, packages where some function is frequently
used, and also higher-level information mined from the raw data. Those are, 
again, typical document retrieval queries. See, for example,
\citeN{LBNRLB09}.

\paragraph*{Chemoinformatics} Databases storing sets of complex molecules where
certain compounds are sought are of much interest for pharmaceutical companies,
to aid in the process of drug design, for example. The typical technique is
to describe compounds by means of short strings that can then be searched.
Here the documents can be long molecules formed by many compounds, or sets
of related molecules.
This is a recent area of research that has grown very fast in relatively few
years, see for example \citeN{Bro05}.

\paragraph*{Symbolic music sequences} As an example of multimedia sequences, 
consider collections of
symbolic music (e.g., in MIDI format). One may wish to look for
pieces containing some sequence, pieces where some sequence appears often,
and so on. This is useful for many tasks, including music retrieval and
analysis, determining authorship, detecting plagiarism, and so on. See,
for example, \citeN{TWV05}.


\medskip

These applications display a wide range of document sizes,
alphabets, and types of queries (list documents where a pattern appears, or
appears often enough, or most often, or find the patterns occurring most often,
etc.). Moreover, while exact matching is adequate for software repositories,
approximate searching should be permitted on DNA, some octave invariance
should be allowed on MIDI, and so on.

In this survey we focus on a basic scenario that has been challenging enough to
attract most of the research in this area, and that is general enough to be
useful in a wide number of cases. We consider document listing and top-$k$
document retrieval, and occasionally some extension, of single-string patterns 
that are matched exactly against sequence collections on arbitrary integer 
alphabets. In many cases we use the term frequency as the relevance measure, 
whereas in other cases we cover more general measures. Before the Conclusions 
we discuss more complex scenarios.

Soon in the survey, the relation between the document retrieval problems we
consider and analogous problems on sequences of {\em colors} (or 
categories) becomes apparent. Thus problems such as listing the different
colors, or count the different colors, or find the $k$ most frequent colors,
in a range of a sequence arise. Those so-called {\em color range queries}
are not only algorithmically interesting by themselves, but have immediate 
applications in some further areas related to data mining:

\paragraph*{Web mining} Web sites collect information on how users access them, 
in some cases to charge for the access, but in all cases
those access logs are invaluable tools to learn about user access patterns, 
favorite contents, and so on. Color range queries allow one, for example, to
determine the number of unique users that have accessed a site, the most 
frequently visited pages in the site, the frequencies of different types of
queries in a search engine, and so on. See, for example, \citeN{Liu07}.

\paragraph*{Database tuning} Monitoring the usage of high-performance 
database servers is important to optimize their behavior and predict potential
problems. Color range queries are useful, for example, to analyze the number
of open sessions in a time period, the most frequent queries or most 
frequently accessed tables, and so on. \citeN{SB03} give a comprehensive 
overview.


\paragraph*{Social behavior} The analysis of words used on tweets, sites 
visited, topics queried, ``likes'', and many other aspects of social behavior is
instrumental to understand social phenomena and exploit social networks. 
Queries like
finding the most frequent words used in a time period, the number of distinct
posters in a blog, the most visited pages in a time period, and so on, are 
natural color range queries. See, for example, \citeN{Sil10}.

\paragraph*{Bioinformatics again} Pattern discovery, such as finding frequent 
$q$-mers (strings of length $q$) in areas of interest in genomes, plays an important 
role in bioinformatics. For fixed $q$ (which is the usual practice) one can
see the genome as a sequence of overlapping $q$-mers, and thus pattern
discovery becomes a problem of detecting frequent colors ($q$-mers) in a 
range of a color sequence. See, again, \citeN{Gus97}.

\medskip

Unlike inverted indexes, which are algorithmically simple, the solutions
for general document retrieval (and color queries) 
have a rich algorithmic structure, with many
connections to fundamental problems on trees, strings, range queries,
discrete geometry, and other fields. Our main goal is to emphasize 
the fascinating algorithmic and data structuring aspects of the current 
document retrieval solutions. Thus, although we show the best existing 
results, we focus on the important algorithmic ideas, leaving the more 
technical details for further reading. In the way, we also fix some 
inaccuracies found in the literature, and propose some new solutions.

Before starting, we bring the readers' attention to recent surveys 
that cover topics with
some relation to ours and that, although the intersection is small, may provide 
interesting additional reading \cite{HSV10,Nav12,Lew13}.

\section{Notation and Basic Concepts}
\label{sec:def}

\subsection{Notation on Strings}

A {\em string} $S=S[1,n]$ is a sequence of {\em characters}, each of which is
an element of a set $\Sigma$ called an {\em alphabet}. We will assume $\Sigma
= [1..\sigma] = \{1,2,\ldots,\sigma\}$.
The {\em length} (number of characters) of $S[1,n]$ is denoted $|S|=n$.
We denote by $S[i]$ the $i$-th character of $S$, and $S[i,j] = S[i]
\ldots S[j]$ a {\em substring} of $S$. When $i > j$ it holds
$S[i,j]=\varepsilon$, the empty string of length $|\varepsilon|=0$.
A {\em prefix} of $S$ is a substring of the form $S[1,j]$ and a {\em suffix} 
is of the form $S[i,n]$.
By $SS'$ we denote the {\em concatenation} of strings $S$ and $S'$, where the
characters of $S'$ are appended at the end of $S$. A single character can 
stand for a string of length 1, thus $cS$ and $Sc$, for $c\in\Sigma$, also
denote concatenations.

The {\em lexicographical order} ``$\prec$'' among strings is defined as follows.
Let $a,b \in \Sigma$ and let $S$ and $S'$ be strings. Then $aS \prec bS'$ if 
$a<b$, or if $a=b$ and $S \prec S'$. Furthermore, $\varepsilon \prec S$ for 
any $S \not= \varepsilon$.

\subsection{Model and Formal Problem}

We model the document retrieval problems to be considered in the following way.

\newpage

\begin{itemize}
\item There is a collection $\D$ of $D$ documents $\D=\{T_1,\ldots,T_D$\}.
\item Each document $T_d$ is a nonempty string over alphabet 
$\Sigma = [1..\sigma]$.
\item We define $\T = T_1\$T_2\$\ldots T_D\$$ as a string over
$\Sigma \cup \{\$\}$, $\$=0<c$ for any $c\in\Sigma$, which concatenates all
the texts in $\D$ using a separator symbol.
\item The length of $\T$ is $|\T|=n$ and the length of each $T_d$
is $|T_d|=n_d$.
\item Queries consist of a single pattern string $P[1,m]$ over $\Sigma$.
\end{itemize}

We define now the problems we consider. First we define the set of occurrence
positions of pattern $P$ in a document $T_d$.

\begin{definition}[Occurrence Positions]
Given a document string $T_d$ and a pattern string $P$, 
the {\em occurrence positions} (or just {\em occurrences}) of $P$ in 
$T_d$ are the set $\occ(P,T_d) = \{ 1+|X|,~\exists Y,~T_d = XPY \}$. 
\label{def:occ}
\end{definition}

Now we define the document retrieval problems we consider. We start with the
simplest one.

\begin{problem}[Document Listing] \label{prob:listing}
Preprocess a document collection $\D$ so that, given a pattern string 
$P$, one can compute $\List(\D,P) = \{ d,~\occ(P,T_d) \not= \emptyset \}$, 
that is, the documents where $P$ appears. We call $\docc=|\List(\D,P)|$ the 
size of the output.
\end{problem}

Variants of the document listing problem, which we will occasionally consider,
include computing the term frequency for each reported document, and 
computing the document frequency of $P$. Those functions
are typically used in relevance formulas (recall measures $\tf$ and $\idf$
in the Introduction). 

\begin{definition}[Document and Term Frequency] \label{def:tfdf}
The {\em document frequency} of $P$ in a document collection $\D$ is defined
as $\df(P) = \docc = |\List(\D,P)|$, that is, the number of documents where
$P$ appears. The {\em term frequency} of $P$ in document $d$ is defined as
$\tf(P,d) = |\occ(P,T_d)|$, that is, the number of times $P$ appears in $T_d$.
\end{definition}

Our second problem relates to ranked retrieval, that is, reporting only some
important documents instead of all those where $P$ appears.

\begin{problem}[Top-$k$ {[Most Frequent]} Documents] \label{prob:topk}
Preprocess a document collection $\D$ so that, given a pattern 
string $P$ and a threshold $k$, one can compute $\Top(\D,P,k)
\subseteq \List(\D,P)$ such that $|\Top(\D,P,k)|=\min(k,\df(P))$ and, for any
$d \in \Top(\D,P,k)$ and $d' \in \List(\D,P)\setminus
\Top(\D,P,k)$, it holds $|\occ(P,T_d)| \ge |\occ(P,T_{d'})|$. That is, find
$k$ documents where $P$ appears the most times. This latter condition can be
generalized to any other function of $\occ(P,T_d)$ and $\occ(P,T_{d'})$.
\end{problem}

A simpler variant of this problem arises when the importance of the documents
is fixed and independent of the search pattern (as in Google's PageRank).

\begin{problem}[Top-$k$ Most Important Documents] \label{prob:topkimport}
Preprocess a document collection $\D$ so that, given a 
pattern string $P$ and a threshold $k$, one can compute $\Top(\D,P,k)
\subseteq \List(\D,P)$ such that $|\Top(\D,P,k)|=\min(k,\df(P))$ and, for any
$d \in \Top(\D,P,k)$ and $d' \in \List(\D,P)\setminus
\Top(\D,P,k)$, it holds $W(d) \ge W(d')$, where $W$ is a fixed weight
function assigned to the documents.
\end{problem}

\subsection{Some Fundamental Problems and Data Structures}

Before entering into the main part of the survey, we cover here a few 
fundamental problems and existing solutions to them. Understanding the
problem definitions and the complexities of the solutions is sufficient
to follow the survey. Still, we give pointers to further reading for the 
interested readers. Rather than giving early isolated illustrations of these
data structures, we will exemplify them later, when they become used in 
the document retrieval structures.

\subsubsection{Some Compact Data Structures} \label{sec:cds}

Many document retrieval solutions require too much space in their simplest
form, and thus compressed representations are used to reduce their space up to
a manageable level. We enumerate some basic problems that arise 
and the compact data structures to handle them. Most of these are covered in
detail in a previous survey \cite{NM07}, so we only list the results here.
All the compact data structures we will use, and the document retrieval 
solutions we build on them, assume the RAM model of computation, where the
computer manages in constant time words of size $\Theta(\lg n)$, as it must
be possible to address an array of $n$ elements. The typical arithmetic and
bit manipulation operations can be carried out on words in constant time.

A basic problem is to store a sequence over an integer alphabet 
so that any sequence position can be accessed and also two complementary
operations called $\rank$ and $\select$ can be carried out on it.

\begin{problem}[Rank/Select/Access on Sequences] \label{prob:ranksel}
Represent a sequence $C[1,n]$
over alphabet $[1..D]$ so that one can answer three queries on it: (1) accessing
any $C[i]$; (2) computing $\rank_c(C,i)$, the number of times symbol 
$c \in [1..D]$ occurs in $C[1,i]$; (3) computing $\select_c(C,j)$, the position 
of the $j$th occurrence of symbol $c$ in $C$. It is assumed that 
$\rank_c(C,0)=\select_c(C,0) =0$.
\end{problem}

A basic case arises when the sequence is a bitmap $B$ over alphabet $\{0,1\}$. 
Then the problem can be solved in constant time and using sublinear extra 
space.

\begin{solution}[Rank/Select/Access on Bitmaps] \label{thm:bitmaps} 
{\rm \cite{Mun96,Cla96}}
By storing $o(n)$ bits on top of $B[1,n]$ one can solve the three
queries in constant time.
\end{solution}

There exist also solutions suitable for the case where $B$ contains few 1s or
few 0s. From the various solutions, the following one is suitable for this
survey.
Note that access queries can be solved using $B[i]=\rank_1(B,i)-\rank_1(B,i-1)$.

\begin{solution}[Rank/Select/Access on Bitmaps] \label{thm:comprbitmaps} 
{\rm \cite{RRR07}}
A bitmap $B[1,n]$ with $m$ 1s (or $m$ 0s) can be stored in $m\lg\frac{n}{m} +
O(m) + o(n)$ bits so that the three queries are solved in constant time.
\end{solution}

A weaker representation can only compute $\rank_1(B,i)$, only in
those positions where $B[i]=1$, and it cannot determine whether this is the
case. This is called a {\em monotone minimum perfect hash function (mmphf)}
and can be stored in less than the space required for compressed bitmaps.
We will use it in Section~\ref{sec:tf}.

\begin{solution}[Mmphfs on Bitmaps] \label{thm:mmphf} {\rm \cite{BBPV09}}
A bitmap $B[1,n]$ with $m$ 1s can be stored in 
$O(m\lg\lg\frac{n}{m})$ bits so that $\rank_1(B,i)$, if
$B[i]=1$, is computed in $O(1)$ time. If $B[i]=0$ the query returns 
an arbitrary value. Alternatively, the bitmap can be stored in 
$O(m\lg\lg\lg\frac{n}{m})$ bits and the query time is $O(\lg\lg\frac{n}{m})$.
\end{solution}

There are also various efficient solutions for general sequences. 
One uses wavelet trees \cite{GGV03,Nav12}, which we describe soon for
other applications. When we only need to solve
Problem~\ref{prob:ranksel} and the sequence does not offer relevant compression 
opportunities, as will be the case in this survey, the following result is 
sufficient (although the results can be slightly improved \cite{BN12}).
We will use it mostly in Section~\ref{sec:tf} as well.

\begin{solution}[Rank/Select/Access on Sequences] \label{thm:seqs} 
{\rm \cite{GMR06,GOR10}}
A sequence $C[1,n]$ over alphabet $[1..D]$ can be stored in $n\lg D + o(n\lg D)$
bits so that query $\rank$ can be solved in time $O(\lg\lg D)$ and, either
$C[i]$ can be accessed in $O(1)$ time and query $\select$ can be solved in
$O(\lg\lg D)$ time, or vice versa.
\end{solution}

%

\subsubsection{Range Minimum Queries (RMQs) and Lowest Common Ancestors (LCAs)} \label{sec:rmq}

Many document retrieval solutions make heavy use of the following problem on
arrays of integers.

\begin{problem}[Range Minimum Query, RMQ] \label{prob:RMQ}
Preprocess an array 
$L[1,n]$ of integers so that, given a range $[sp,ep]$, we can output 
the position of a minimum value in $L[sp,ep]$, $\rmq_L(sp,ep) =
\mathrm{argmin}_{i\le p\le j} L[p]$.
\end{problem}

The RMQ problem has a rich history, which we partially cover in 
Appendix~\ref{app:rmq}. An interesting data structure related to it is the
{\em Cartesian tree} \cite{Vui80}.

\begin{definition}[Cartesian Tree] \label{def:cartesian}
The {\em Cartesian tree} of an array $L[1,n]$ is a binary tree whose root 
corresponds to the position $p$ of the minimum in $L[1,n]$, and the left and 
right children are, recursively, Cartesian trees of $L[1,p-1]$ and $L[p+1,n]$, 
respectively. The Cartesian tree of an empty array interval is a null pointer.
\end{definition}

Cartesian trees are instrumental in relating RMQs with the following problem.

\begin{problem}[Lowest Common Ancestor, LCA] \label{prob:LCA}
Preprocess a tree
so that, given two nodes $u$ and $v$, we can output the deepest tree
node that is an ancestor of both $u$ and $v$.
\end{problem}

The main results we need on RMQs are summarized in the following two
solutions. The first is a classical result stating that the problem can be 
solved in linear space and optimal time.

\begin{solution}[RMQ] \label{thm:rmqlinear} {\rm \cite{HT84,SV88,BV93,BF00}}
The problem can be solved using linear space and preprocessing
time, and constant query time.
\end{solution}

The second result shows that, by storing just $O(n)$ bits from the original
array $L$, we can solve RMQs {\em without accessing $L$ at query time}. This
is relevant for the compressed solutions.

\begin{solution}[RMQ] \label{thm:rmq} {\rm \cite{FH11}}
The problem can be solved using $2n+o(n)$ bits, linear
preprocessing time, and constant query time, without accessing the original
array at query time. This space is asymptotically optimal.
\end{solution}

Similarly, the related LCA problem can be solved in constant time using linear
space, and even on a tree representation that uses $2n+o(n)$ bits for a tree
of $n$ nodes (e.g., that of \citeN{SN10}).

\subsubsection{Wavelet Tres}

Finally, we introduce the wavelet tree data structure \cite{GGV03}, which is 
used for many document retrieval solutions, and whose structure is necessary
to understand in this survey.

A {\em wavelet tree} over a sequence $C[1,n]$ is a perfectly balanced binary 
tree where each node handles a range of the alphabet. The root handles the 
whole alphabet and the leaves handle individual symbols. At each node, the
alphabet is divided by half and the left child handles the smaller half of
the symbols and the right child handles the larger half. Each node $v$
represents (but does not store) a subsequence $C_v$ of $C$ containing the 
symbols of $C$ that the node handles. What each internal node $v$ stores is
just a bitmap $B_v$, where $B_v[i]=0$ iff $C_v[i]$ belongs to the range of 
symbols handled by the left child of $v$, otherwise $B_v[i]=1$.

Over an alphabet $[1..D]$, the wavelet tree has height $\lceil \lg D\rceil$ and 
stores $n$ bits per level, thus its total space is $n\lceil\lg D\rceil$ bits,
that is, the same of a plain representation of $C$. For it to be functional,
we need that the bitmaps $B_v$ can answer $\rank$ and $\select$ queries, thus
the total space becomes $n\lg D + o(n\lg D)$ bits. Within this space, the
wavelet tree actually represents $C$: to recover $C[i]$, we start at the root
node $v$. If $B_v[i]=0$ then $C[i]$ belongs to the first half of the alphabet,
so we continue the search on the left child of the root, with the new position
$i \leftarrow \rank_0(B_v,i)$. Else we continue on the right child, with
$i \leftarrow \rank_1(B_v,i)$. When we arrive at a leaf handling symbol $d$,
it holds $C[i]=d$. The process takes $O(\lg D)$ time. Within this time the
wavelet tree can also answer $\rank$ and $\select$ queries on $C$, as well as many
other operations. See \citeN{Mak12} and \citeN{Nav12} for full surveys.

An example of wavelet trees is given in Appendix~\ref{app:wtree}, in a
proper context. A formal definition of wavelet trees follows.

\begin{definition}[Wavelet Tree] \label{def:wt}
A {\em wavelet tree} over a sequence $C[1,n]$ on alphabet $[1..D]$ is a 
perfectly balanced binary tree where the $i$th node of height $h$ (being the
leaves of height $0$) is associated to the symbols $d$ such that 
$\lceil d/2^h \rceil = i$. The node $v$ handling symbols $[a..b] \subseteq [1..D]$
represents the subsequence $C_v$ of $C$ consisting of the symbols in $[a..b]$.
For each node $v$ we store a bitmap $B_v[1,|C_v|]$ where $B_v[i]=0$ iff 
the left child of $v$ is associated with symbol $C_v[i]$.
\end{definition}

\section{Occurrence Retrieval Indexes}

In this section we cover indexes that handle collections of general sequences,
but that address the more traditional problem of finding or counting all the
occurrences of a pattern $P$ in a text $\T$ (i.e., computing $\occ(P,\T)$ or
$|\occ(P,\T)|$). We focus on those upon which document retrieval indexes are
built: suffix trees, suffix arrays, and compressed suffix arrays.

\subsection{Generalized Suffix Trees}
\label{sec:stree}

Consider a text $\T[1,n] = T_1\$T_2\$\ldots T_D\$$ over alphabet 
$\Sigma \cup \{\$\}$. Now consider the $n$ suffixes of the form
$\Ss = \{ T_d[i,n_d]\$, 1 \le d \le D, 1 \le i \le n_d+1 \}$.
The {\em Generalized Suffix Tree (GST)} of $\T$ is a data structure
storing those $n$ strings in $\Ss$.\footnote{For technical reasons each 
``$\$$'' symbol should be different, but this is not done in practice. We
prefer to ignore this issue for simplicity.}

To describe the GST, we start with a tree where the edges are labeled with
symbols in $\Sigma \cup \{\$\}$, and where each string in $\Ss$ can be read
by concatenating the labels from the root to a leaf. 
No two edges leaving a node have the same
label, and they are ordered left to right according to 
those labels. The {\em string label} of a node $v$, $str(v)$, is the
concatenation of the characters labeling the edges from the root to the node.
Thus, each string label in the tree is a unique prefix in $\Ss$,
and there is exactly one tree leaf per string in $\Ss$.

To obtain a GST from this tree we carry out three steps: (1) remove all nodes
$v$ with just one child $u$, prepending the label of the edge that connects 
$v$ to its parent to the label that connects $u$ and $v$ (now edges
will be labeled with strings); (2) attach to leaves the starting position
of their suffix in $\T$; (3) retain only the first character and the length
of the strings labeling the edges, that is, labels will be of the form
$(c,\ell)$, with $c \in \Sigma\cup\{\$\}$ and $\ell > 0$. 

\begin{figure}[t]
\centerline{\includegraphics[width=0.9\textwidth]{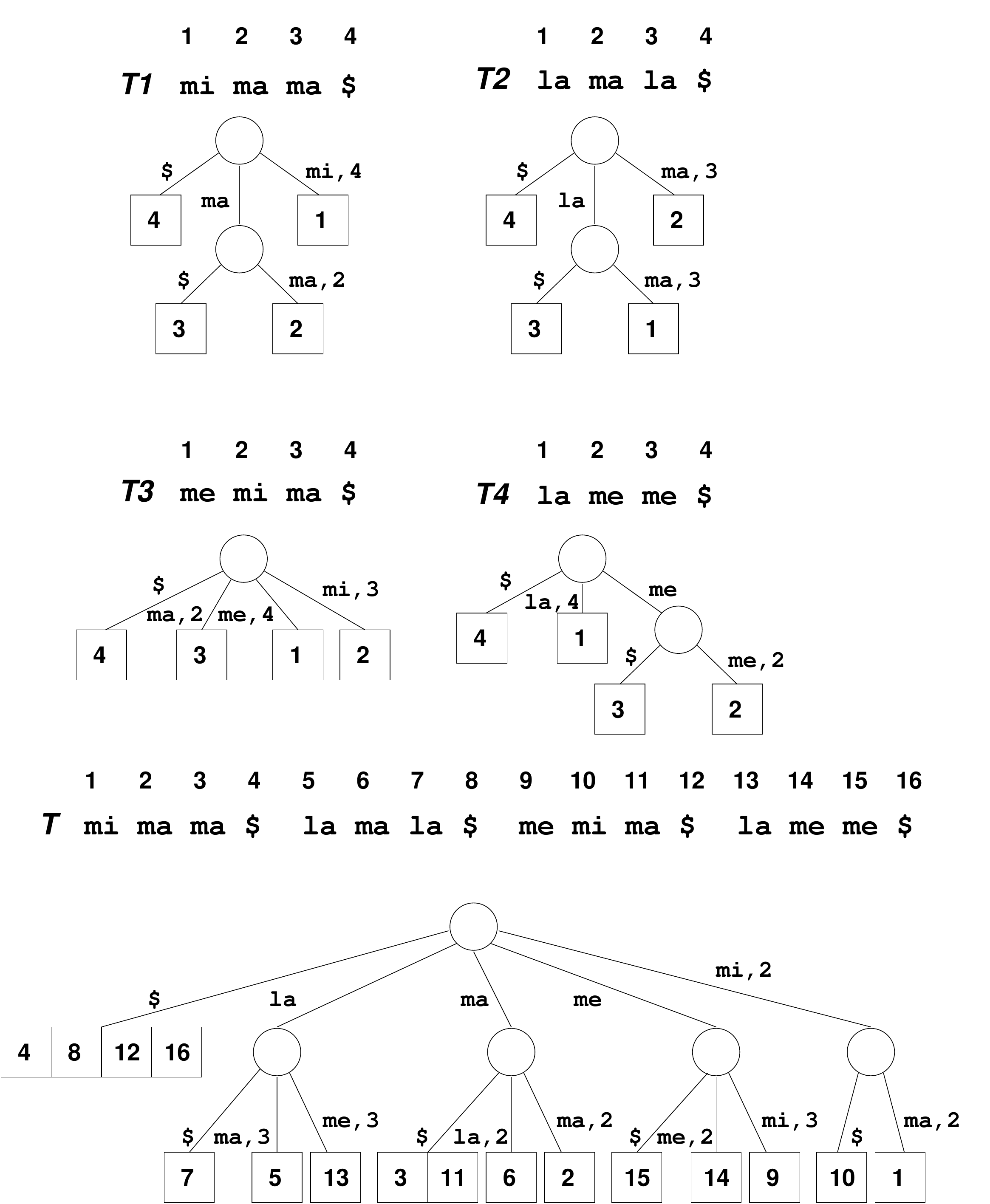}}
\caption{The individual suffix tree of each document and the GST of the 
concatenated text $\T$, in our running example text collection. For legibility
we omit the edge length when it is 1.}
\label{fig:docs}
\end{figure}

\begin{example}
We introduce our running example text collection. To combine readability and
manageability, we consider an alphabet of {\em syllables} on a hypothetic
language.\footnote{Suspiciously close to Spanish.} The alphabet of our texts
will be $\Sigma = \{ \mathsf{la}, \mathsf{ma}, \mathsf{me}, \mathsf{mi} \}$.
Our document collection $\D = \{ T_1, T_2, T_3, T_4 \}$ will have $D=4$ texts,
$T_1 = \textsf{"mi ma ma"}$, $T_2 = \textsf{"la ma la"}$, 
$T_3 = \textsf{"me mi ma"}$, $T_4 = \textsf{"la me me"}$.
Their lengths are $n_1 = n_2 = n_3 = n_4 = 3$. They are concatenated into 
a single text
$$\T = \textsf{"mi ma ma \$ la ma la \$ me mi ma \$ la me me \$"}$$
of length $n=16$.
Fig.~\ref{fig:docs} shows the individual suffix trees of the texts, plus 
the GST of $\T$ (which we also call the GST of $\D$).

Note that, because we do not use a distinct ``\$'' terminator per document,
some anomalies arise in our example, with leaves corresponding to several
symbols. As explained, those do not cause any problem in practice.
\end{example}

As mentioned, suffix trees (which are GSTs of only one text $T_1\$$) and 
GSTs are used for many complex tasks \cite{Apo85,Gus97,MR02}, yet in this
survey we will only describe the simplest one as an occurrence retrieval
index. In this case, given a pattern $P[1,m]$, we find the so-called
{\em locus} of $P$, that is, the highest suffix tree node $v$ 
such that $P$ is a prefix of $str(v)$. The locus can be found by a well-known
traversal procedure, in $O(m)$ time on integer alphabets%
\footnote{If $\sigma$ is not taken as a constant we require perfect hashing
to obtain $O(m)$ time and linear space for the structure; otherwise
$O(m\lg\sigma)$ time is achieved with binary search on the children.}
(see the given references for details).
Once we find the locus $v$, $\occ(P,\T)=\occ(str(v),\T)$ is the set of 
positions attached to the leaves descending from $v$. Thus $\occ(P,\T)$ can 
be listed in time $O(m+|\occ(P,\T)|)$, or we can record $|\occ(str(v),\T)|$ 
in each node $v$ so that we can count the number of times $P$ occurs in $\T$ 
in time $O(m)$.


\begin{example}
A search for $P=\textsf{"mi"}$ or for $P=\textsf{"mi ma"}$ 
in the GST will end up in the rightmost child $v$ of the root.
Hence $\occ(P,\T) = \{ 10, 1 \}$.
\end{example}


A formal succinct definition of GSTs, plus a couple of key concepts, follows.

\begin{definition}[Generalized Suffix Tree, GST] \label{def:stree}
The {\em generalized suffix tree} of a text collection 
$\D = \{ T_1, T_2, \ldots, T_D\}$ is
a path-compressed trie storing all the suffixes of $\T = T_1\$T_2\$
\ldots T_D\$$, where the $\$$ is a special character.
The {\em string label} $str(v)$ of a node $v$ is the concatenation of the
string labels of the edges from the root to $v$. The {\em locus} of a pattern
$P$ is the highest node $v$ such that $P$ is a prefix of $str(v)$.
\end{definition}

Note that the search times are independent of the length of the text $\T$, 
which is a remarkable property of suffix trees. Other good
properties are that it takes linear space (i.e., $O(n)$ words) since 
it has $n$ leaves and no unary nodes, and that it can be built in linear
time for constant alphabets \cite{Wei73,McC76,Ukk95}, and also on integer
alphabets \cite{Far97,KSB06}.

The GST is a useful tool to group the $O(n^2)$ possible substrings
of $\T$ (and hence possible search patterns) into $O(n)$ nodes, where each
node represents a group of substrings that share the same occurrence positions
in $\T$. This allows one to store, in linear space, information that is useful
for document retrieval. For example, one can associate to each GST $v$ node the
number of distinct documents where $str(v)$ appears, $\df(str(v))$, which 
allows us to solve in
$O(m)$ time and linear space the problem of computing $\docc$. This
ability has been exploited several times for document retrieval.
Note, on the other hand, that the linear space of GSTs is actually $O(n\lg n)$ 
bits, as the tree pointers must at least distinguish between the $O(n)$ 
different nodes (thus $\lg n + O(1)$ bits are needed for each pointer), and
some further data is stored. Thus their space usage is a concern in 
practice.

\subsection{Suffix Arrays}
\label{sec:sarrays}

The {\em suffix array} \cite{MM93,GBYS92} of a text $\T$ is a permutation of 
the (starting positions of) suffixes of $\T$, so that the suffixes are 
lexicographically sorted. Alternatively, the suffix array of $\T$ is the
sequence of positions attached to the leaves of the suffix tree, read left
to right.

\begin{definition}[Suffix Array] \label{def:sarray}
The {\em suffix array} of a collection $\D = \{ T_1, T_2, \ldots, T_D\}$ is
an array $A[1,n]$ containing a permutation of $[1..n]$, such that 
$\T[A[i],n] \prec \T[A[i+1],n]$ for all $1 \le i < n$, where 
$\T[1,n] = T_1\$T_2\$ \ldots T_D\$$.
\end{definition}

Suffix arrays also take linear space and can be built in $O(n)$ time, without
the need of building the suffix tree first \cite{KSPP05,KA05,KSB06}. They use 
less space than suffix trees, but still their space usage is high.

\begin{figure}[t]
\centerline{\includegraphics[width=0.8\textwidth]{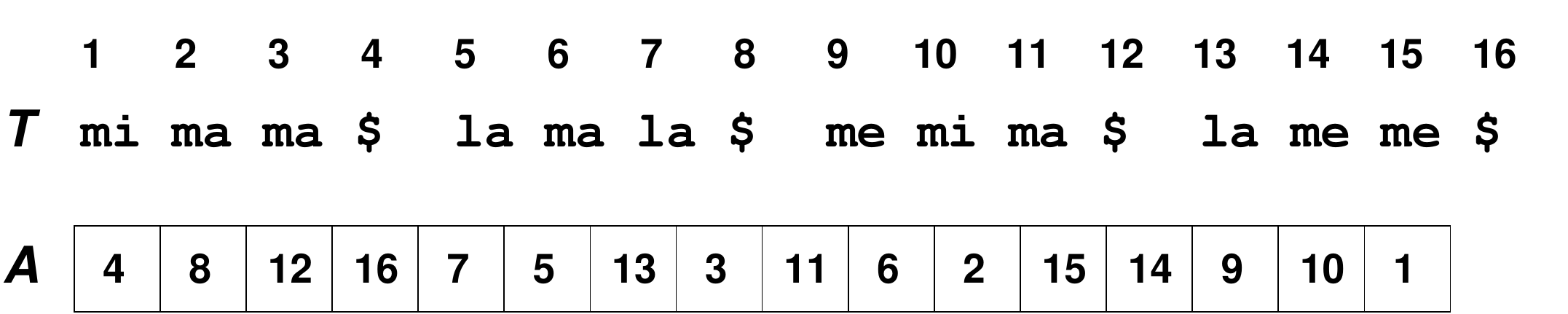}}
\caption{The suffix array of text $\T$ on our running example.}
\label{fig:docsa}
\end{figure}

\begin{example}
Fig.~\ref{fig:docsa} illustrates the suffix array for our example. 
\end{example}

An
important property of suffix arrays is that each subtree of the suffix tree 
corresponds to an interval of the suffix array, namely the one containing its
leaves. In particular, having $A$ and $\T$, one can just binary search the
suffix array interval corresponding to the occurrences of a pattern $P[1,m]$,
in $O(m\lg n)$ time (that is, $O(\lg n)$ comparisons of $m$ symbols).%
\footnote{By storing
more data, this can be reduced to $O(m+\lg n)$ time \cite{MM93}.}
Another way to see this is that, since suffixes are sorted in $A$, all those
starting with $P$ form a contiguous range. Once we determine that all the
occurrences of $P$ are listed in $A[sp,ep]$, we have $|\occ(P,\T)| = ep-sp+1$
and $\occ(P,\T) = \{ A[sp], A[sp+1], \ldots, A[ep] \}$.

\subsection{Compressed Suffix Arrays} \label{sec:csa}

A {\em compressed suffix array (CSA)} over a text collection $\D$ is a data 
structure that emulates a suffix array on $\T$ within less space, usually 
providing even richer functionality. At most, a CSA must use $O(n\lg \sigma)$ 
bits of space, that is, proportional to the size of the text stored in plain
form (as opposed to the $O(n\lg n)$ bits of classical suffix arrays). There are,
however, several CSAs using as little as $nH_k(\T) + o(n\lg\sigma)$ bits, 
where $H_k(\T) \le \lg\sigma$ is the $k$-th order empirical entropy of $\T$. 
This is a lower bound to the bits-per-symbol 
achievable on $\T$ by any statistical compressor that encodes 
each symbol according to the $k$ symbols that precede it in the text 
\cite{Man01}. That is, $nH_k(\T)$ is the least space a statistical 
encoder can achieve on $\T$.

CSAs are well covered in a relatively recent survey \cite{NM07}, so we only
summarize the operations they support. First, given a pattern $P$, they find
the interval $A[sp,ep]$ of the suffixes that start with $P$, in time
$\tsearch(m)$. Second, given a cell $i$, they return $A[i]$, in time
$\taccess$. Third, they are generally able to emulate the inverse permutation
of the suffix array, $A^{-1}[j]$, also in time $\taccess$. This corresponds to
asking which cell of $A$ points to the suffix $\T[j,n]$. Finally, many CSAs
are {\em self-indexes}, meaning that they are able to extract any substring
$\T[i,j]$ without accessing $\T$, so the text itself can be discarded. Those
CSAs replace $\T$ by a (usually) compressed version that can in addition
be queried. The following definition captures the minimum functionality we
need from a CSA in this article.

\begin{definition}[Compressed Suffix Array, CSA] \label{def:csa}
A {\em compressed suffix array} is a data structure that simulates
a suffix array on text $\T[1,n]$ over alphabet $[1..\sigma]$ using at most 
$O(n\lg\sigma)$ bits. It finds the interval
$A[sp,ep]$ of a pattern $P[1,m]$ in time $\tsearch(m)$, and computes any
$A[i]$ or $A^{-1}[j]$ in time $\taccess$.
\end{definition}

For example, a recent CSA \cite{BN11} requires 
$nH_k(\T)(1+o(1))+O(n)$ bits of space and offers time complexities
$\tsearch=O(m)$ and $\taccess=O(\lg n)$. Another recent one
\cite{BGNN10} uses $nH_k(\T)(1+o(1))+o(n)$ bits and offers
$\tsearch=O(m\lg\lg\sigma)$ and $\taccess=O(\lg n (\lg\lg\sigma)^2)$.
Both are self-indexes. Many older ones can be found in the 
survey \cite{NM07}.

\section{Document Listing}
\label{sec:doclist}

\citeN{Mut02} gave an optimal solution to the document listing problem
(Problem~\ref{prob:listing}), within linear space, that is, 
$O(n\lg n)$ bits (see \citeN{JL93} and \citeN{MMSZ98} for previous work). 
\citeANP{Mut02} introduced the 
so-called {\em document array}, which has been used frequently since then.

\begin{definition}[Document Array] \label{def:darray}
Given a document collection $\D$, its text $\T$, and the suffix array $A[1,n]$
of $\T$, the {\em document array} $C[1,n]$ contains in each $C[i]$ the number
of the document suffix $A[i]$ belongs to.
\end{definition}

It is not hard to see that all we need for document listing is to determine
the interval $A[sp,ep]$ corresponding to the pattern and then output the
set of distinct values in $C[sp,ep]$. This gives rise to the following 
algorithmic problem.

\begin{problem}[Color Listing] \label{prob:CL}
Preprocess an array $C[1,n]$ of
{\em colors} in $[1..D]$ so that, given a range $[sp,ep]$, we can
output the different colors in $C[sp,ep]$.
\end{problem}

To solve this problem, \citeANP{Mut02} defines a second array, which is also 
fundamental for many related problems.

\begin{definition}[Predecessors Array] \label{def:carray}
Given an array $C[1,n]$, the {\em predecessors array} of $C$ is
$L[1,n]$ such that $L[i] = \max \{ 1 \le j < i,~C[j] = C[i]\} \cup \{0\}$.
\end{definition}

That is, array $L$ links each position in $C$ to the previous occurrence
of the same color, or to position 0 if this is the first 
occurrence of that color in $C$. 

\begin{figure}[t]
\centerline{\includegraphics[width=0.8\textwidth]{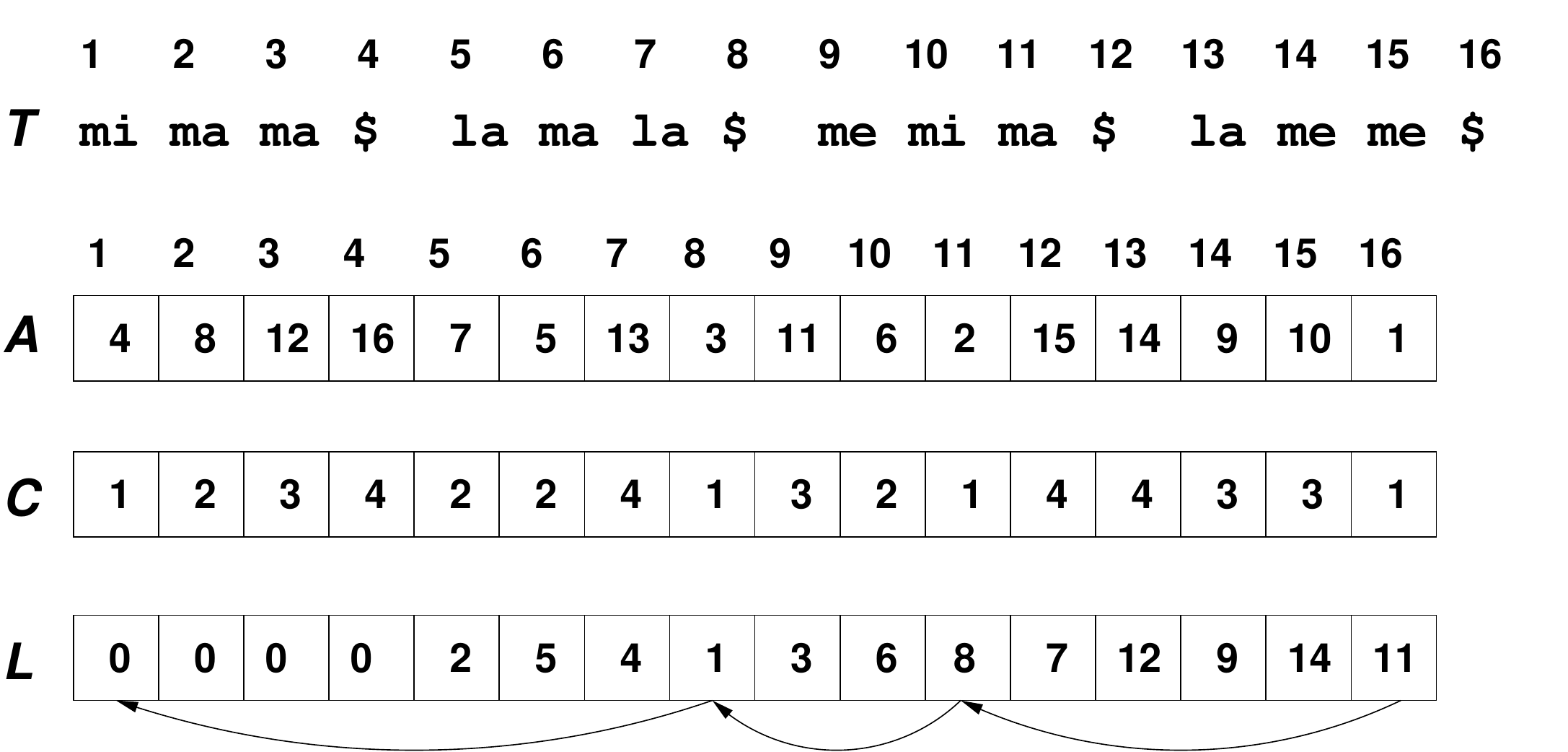}}
\caption{The $C$ and $L$ arrays for our running example.}
\label{fig:muthu}
\end{figure}

\begin{example}
Fig.~\ref{fig:muthu} illustrates arrays $C$ and $L$ on our running example.
We show how $L$ acts as a linked list of the occurrences of color 1.
\end{example}

\citeANP{Mut02}'s solution to color listing is then based on the
following lemma, which is immediate to see.

\begin{lemma} {\rm \cite{Mut02}} \label{lem:L}
If a color $d$ occurs in $C[sp,ep]$, then its leftmost occurrence $p
\in [sp,ep]$ is the only one where it holds $L[p] < sp$.
\end{lemma}

From the lemma, it follows that all we have to do for color listing is to
find all the values smaller than $sp$ in $L[sp,ep]$. To do this in optimal
time, \citeANP{Mut02} makes use of RMQs (more precisely,
Solution~\ref{thm:rmqlinear}).
The algorithm proceeds recursively. It starts with the interval $[i,j] =
[sp,ep]$. It first finds $p = \rmq_L(i,j)$. If $L[p] < sp$, then $C[p]$ is
a new distinct color in $C[sp,ep]$ and can be reported immediately. Then we
continue recursively with the intervals $[i,p-1]$ and $[p+1,j]$. If,
instead, $L[p] \ge sp$, then position $p$ is not the first occurrence 
of color $C[p]$ in $C[sp,ep]$, and moreover no position in $C[i,j]$ is the first
of its color. Thus we terminate the recursion for the current interval $[i,j]$.
Note that we always compare $L[p]$ with the original $sp$ limit, even 
inside a recursive call with a smaller $[i,j]$ interval.

The recursive calls define a binary tree: at each internal node (where
$L[p] < sp$) one 
distinct color appearing in $C[sp,ep]$ is reported, and two further calls are 
made. Leaves of the recursion tree (where $L[p] \ge sp$)
report no colors. Hence the recursion tree 
has twice as many nodes as colors reported, and thus the algorithm is optimal
time. Indeed, it is interesting to realize that what this algorithm is doing 
is to incrementally build the top part of the Cartesian tree of $L[sp,ep]$, 
recall Definition~\ref{def:cartesian}.

\begin{figure}[t]
\centerline{\includegraphics[width=0.8\textwidth]{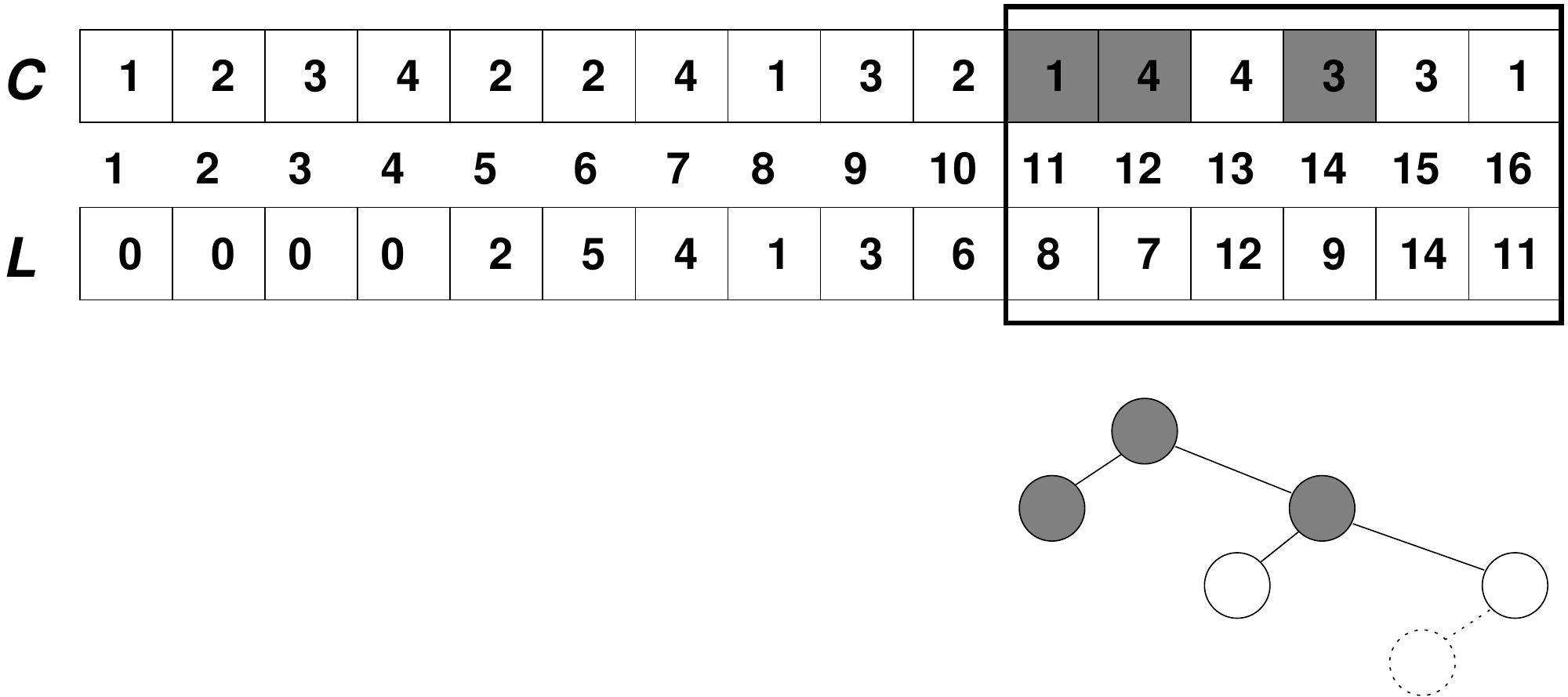}}
\caption{Color listing on $C[11,16]$ in our running example.}
\label{fig:muthu2}
\end{figure}

\begin{example}
For a relevant example, consider color listing over
$C[sp,ep]=C[11,16]$ in the array of Fig.~\ref{fig:muthu2} (this corresponds to 
document listing of a lexicographic pattern range 
$[\textsf{"ma ma"},\textsf{"mi ma"}]$, which is perfectly possible on 
suffix arrays).

We start with $p=\rmq_L(i,j)=\rmq_L(sp,ep)=\rmq_L(11,16)=12$. 
Since $L[p]=L[12]=7<11=sp$, we report color $C[p]=C[12]=4$. 
Now we continue on the left subinterval, $L[i,p-1]=L[11,11]$. 
Here obviously we have $p=11$, and since $L[11]=8<11$ we report
$C[11]=1$.
Now we go to the right of the initial recursive call, for $L[13,16]$.
We compute $p=\rmq_L(13,16)=14$. Since $L[14]=9<11$, we report
$C[14]=3$, and recurse on both sides of $p$.
The left side is $L[13,13]$. Since $L[13]=12\ge 11$, we do not report
this position and terminate the recursion.
The right side is $L[15,16]$.
Once again, we compute $p=\rmq_L(15,16)=16$, and since $L[16]=11\ge 11$, we
also terminate the recursion here. Note we have not needed to examine
$L[15]$ to know it does not contain new colors.
We have correctly reported the colors 4, 1, and 3.

In the bottom part of the figure we show the part of the Cartesian tree
of $L[11,16]$ we have uncovered, or what is the same, the tree of the
recursive calls. Shaded nodes represent colors reported (also marked in $C$),
empty nodes represent cells where the recursion ended, and dotted nodes are
the part we have not visited of the Cartesian tree (usually much more than just
one node).
\end{example}

The algorithm is not only optimal time but also {\em real time}: As it reports
the color before making the recursive calls, the top part of the Cartesian tree
can be thought of as generated in preorder. Thus it takes $O(1)$ time to list
each successive result.

\begin{solution}[Color Listing] {\rm \cite{Mut02}} \label{thm:clmuthu}
The problem can be solved in linear space and real time.
\end{solution}

In addition to this machinery, \citeANP{Mut02} uses a suffix tree on $\T$ to 
compute the interval $[sp,ep]$. This immediately gives an optimal solution to 
the document listing problem. 

\begin{solution}[Document Listing] {\rm \cite{Mut02}} \label{thm:dlmut}
The problem can be solved in $O(m+\docc)$ time and
$O(n\lg n)$ bits of space.
\end{solution}

\citeN{Mut02} considered other more complex variants of the problem, such as
listing the documents that contain $t$ or more occurrences of the pattern,
or that contain two occurrences of the pattern within distance $t$. Those
also lead to interesting, albeit more complex, algorithmic problems.


In practical terms, \citeANP{Mut02}'s solution may use too much space, even if
linear. It has been shown, however, that his very same idea can be implemented
within much less space: just $O(n)$ bits on top of a CSA 
(Definition~\ref{def:csa}). We cover those developments in the next
section.

\section{Document Listing in Compressed Space}
\label{sec:dlcompr}

\citeN{Sad07} addressed the problem of reducing the space of \citeANP{Mut02}'s
solution. He replaced the suffix tree by a CSA,
and proposed the first RMQ solution that did not need to access $L$ (this one 
used $4n+o(n)$ bits; the one we have referenced in Solution~\ref{thm:rmq} 
uses the optimal $2n+o(n)$). Thus array $L$ was not
necessary for computing RMQs on it. \citeANP{Mut02}'s algorithm, however,
needs also to ask if $L[p] < sp$ in order to determine whether this is the
first occurrence of color $C[p]$. \citeANP{Sad07} uses instead
a bitmap $V[1,D]$ (set initially to all 0s), 
so that if $V[C[p]] = 0$ then the document has
not yet been reported, so we report it and set $V[C[p]] \leftarrow 1$. 

Just as before, \citeANP{Sad07} ends the recursion at an interval $[i,j]$
when its minimum position $p$ satisfies $V[C[p]] = 1$. There is a delicate
point about the correctness of this algorithm, which is not stressed in that
article. Replacing the check $L[p]<sp$ by $V[C[p]]=0$ only works if we
first process recursively the left interval, $[i,p-1]$, and then the right 
interval, $[p+1,j]$. In this case one can see that the leftmost
occurrence of each color is found and the algorithm visits the same cells
of \citeANP{Mut02}'s (we prove this formally in Appendix~\ref{app:sadaproof},
and also give an example where an error occurs otherwise). 

\citeANP{Sad07}'s technique yields the following solution, which uses $O(n)$
bits on top of the original array, as opposed to \citeANP{Mut02}'s
$O(n\lg n)$ bits used for $L$.

\begin{solution}[Color Listing] {\rm \cite{Sad07}} \label{thm:clsada}
The problem can be solved using $O(n)$ bits of space on top of array $C$,
and in real time.
\end{solution} 

The reader may have noticed that we should reinitialize $V$ to all zeros
before proceeding to the next query. A simple solution is to remember the
documents output by the algorithm, so as to reset those entries of $V$ after
finishing. This requires $\docc \lg D \le D \lg D$ bits, which may be 
acceptable. Otherwise, array $V$ could be restored by rerunning 
the algorithm and using the bits with the reverse meaning. In practice, however,
this doubles the running time. A more practical alternative is to use 
a classical solution to initialize arrays in constant time \cite{Meh84}.
Although this solution requires $O(D\lg D)$ extra bits, we show in 
Appendix~\ref{app:constinit} how to reduce the space to $O(D)=O(n)$ bits,
and even $D+o(D)$ bits.

\medskip

By combining Solution~\ref{thm:clsada} with any CSA, we also obtain a slightly 
improved version of Solution~\ref{thm:dlmut}.

\begin{solution}[Document Listing] {\rm \cite{Sad07}} \label{thm:dlsadalarge}
The problem can be solved in time 
$O(\tsearch(m) + \docc)$ and $|\CSA|+n\lg D + O(n)$ bits of space, 
where $\CSA$ is a CSA indexing $\D$.
\end{solution} 

Note that we have removed array $L$, but $C$ is still used to report the
actual colors. For the specific case of document listing,
\citeANP{Sad07} also replaced array $C$ by noticing that it can be easily
computed from the CSA and a bitmap $B[1,n]$ that marks with 1s the positions 
of the ``\$'' symbols in $\T$. Then, using the $\rank$ operation
(Problem~\ref{prob:ranksel}) we compute $C[i] = 1+\rank_1(B,A[i]-1)$.
While $\rank$ can be computed in constant 
time (Solution~\ref{thm:bitmaps}), the computation of $A[i]$ using the CSA
(Definition~\ref{def:csa}) requires time $\taccess$. Overall, the following
result is obtained.

\begin{solution}[Document Listing] {\rm \cite{Sad07}} \label{thm:dlsada}
The problem can be solved in time 
$O(\tsearch(m) + \docc\,\taccess)$ and $|\CSA|+O(n)$ bits of space, 
where $\CSA$ is a CSA indexing $\D$.
\end{solution}

\begin{figure}[t]
\centerline{\includegraphics[width=0.8\textwidth]{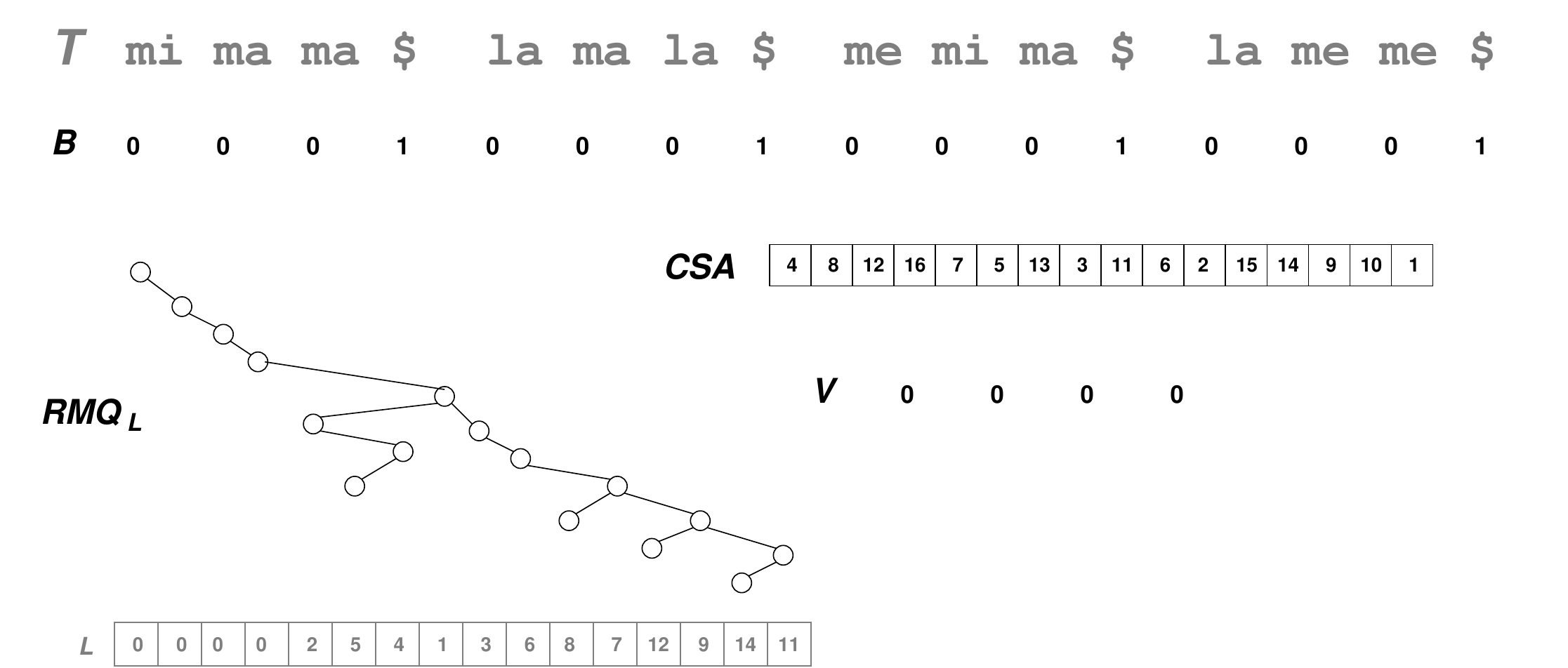}}
\caption{The structures for document listing in compressed space, on our 
running example. The grayed structures are not stored. We draw the Cartesian
tree of $L$ to represent $\textsc{rmq}_L$.}
\label{fig:dlsad}
\end{figure}

\begin{example}
Fig.~\ref{fig:dlsad} illustrates the components of \citeANP{Sad07}'s solution.
\end{example}

\citeN{HSV09} managed to reduce the $O(n)$-bit 
term in the space complexity to just $o(n)$. They group $b$ consecutive entries
in $L$. Then they create a sampled array $L'[1,n/b]$ where each entry contains 
the minimum value in the corresponding block of $L$. The RMQ data structure is
built over $L'$, and they run the algorithm over the blocks that are fully 
contained in $L[sp,ep]$. Each time a position in $L'$ is reported, they 
consider all the $b$ entries in the corresponding block of $L$, reporting all 
the documents that have not yet been reported. Only if all of them have
already been 
reported can the recursion stop. They also process by brute force the 
tails of the interval $L[sp,ep]$ that overlap blocks. Therefore they have a
multiplicative time overhead of $O(b)$ per document reported, in exchange for
reducing the $O(n)$-bit space to $O(n/b)$. 

This idea only works if we first process the left subinterval, then mark the
documents in the block where the minimum was found, and then process the
right subinterval, for the same delicate reason we have described on 
\citeANP{Sad07}'s method. Otherwise, marking as visited other documents than 
the one holding the minimum $L$ value (namely all others in the block) can 
make the recursion stop earlier than it should. We prove the correctness of 
this technique in Appendix~\ref{app:hsvproof}, where we also show that the
result can be incorrect if applied in a different order.

The other $O(n)$ bits of space come from the bitmap $B$ that marks the
positions of terminators in $\T$. Since $B$ has only $D$ bits set, it is
easily represented in compressed form using Solution~\ref{thm:comprbitmaps}.

\begin{solution}[Document Listing] {\rm \cite{HSV09}} \label{thm:dlhsv}
The problem can be solved in time
$O(\tsearch(m) + \docc\,\taccess\lg^\epsilon n)$ and
$|\CSA|+D\lg(n/D)+O(D)+o(n)$ bits of
space, where $\CSA$ is a CSA indexing $\D$, for any constant $\epsilon>0$.
\end{solution}

\section{Computing Term Frequencies}
\label{sec:tf}

As explained in the Introduction, the term frequency $\tf(P,d)$ is a
key component in many relevance formulas, and thus the problem of computing
it for the documents that are output by a document listing algorithm is
relevant. In terms of the document array, this leads to the following 
problem on colors, which as explained has its own important applications
in various data mining problems.

\begin{problem}[Color Listing with Frequencies] \label{prob:colfreq}
Preprocess an array $C[1,n]$ of colors in $[1..D]$ so that, given a 
range $[sp,ep]$, we can output the distinct colors in $C[sp,ep]$,
each with its frequency in this range.
\end{problem}

Given a color $d$, computing its frequency in $C[sp,ep]$ is easily done via
$\rank$ operations (Problem~\ref{prob:ranksel}): 
$\rank_d(C,ep)-\rank_d(C,sp-1)$. Therefore, any solution to color listing
(e.g., Solution~\ref{thm:clsada}) plus any solution to computing $\rank$
on sequences (e.g., Solution~\ref{thm:seqs}) yields a solution to color
listing with frequencies. Indeed, Solution~\ref{thm:seqs} is close to optimal
\cite{BN12}. Thus, one can solve color listing with frequencies 
using $o(n\lg D)$ additional bits and $O(\lg\lg D)$ per color output.

Somewhat surprisingly, the problem can be solved faster by noticing that we
do not need the full power of $\rank$ queries on $C$. \citeN{BNV13} replace 
the $\rank$-enabled sequence representation of $C$ by a weaker representation 
using less space and time, but sufficient for this purpose. They use one
mmphf $B_d$ (recall Solution~\ref{thm:mmphf}) for each color $d$, marking the
positions $B_d[i]=1$ where $C[i]=d$. Then, if one knew that $i$ and $i'$ are 
the first and last occurrences of color $d$ in $C[sp,ep]$, one could compute
the color frequency as $\rank_1(B_d,i')-\rank_1(B_d,i)+1$. 

Note that \citeANP{Mut02}'s algorithm naturally finds the leftmost occurrence 
$i$. With a different aim, \citeN{Sad07} had shown how to compute the rightmost
occurrence $i'$. He creates another RMQ (now meaning range maximum query) 
structure over a variant of $L$ where each element points to its successor 
rather than its predecessor. Run over this new RMQ, \citeANP{Mut02}'s 
algorithm will find $i'$ instead of $i$, for each document $d$ in $C[sp,ep]$. 

\citeN{Sad07} then uses sorting to match the $i$ and $i'$ position of each 
document $d$, whereas \citeN{BNV13} avoids the sorting but uses $O(D\lg n)$
further bits. In Appendix~\ref{app:dict} we show how both can
be avoided with a dictionary that takes just $O(\lg D)$ bits per listed
color plus $o(D)$ total space. In this dictionary we insert the 
leftmost occurrences (with document identifiers as keys) and later search for
the rightmost occurrences. Then, by using the faster variant of
Solution~\ref{thm:mmphf}, \citeN{BNV13} obtain the following result.

\begin{solution}[Color Listing with Frequencies] {\rm \cite{BNV13}} \label{thm:cltfbn}
The problem can be solved in optimal time and 
$O(n\lg\lg D)$ bits on top of array $C$.
\end{solution}

Note that, compared to Solution~\ref{thm:clsada}, that listed the colors 
without frequencies, the real time has become ``just'' optimal, and the extra
space of $O(n)$ bits has increased, yet it is still $o(n\lg D)$ as long as
$D=o(n)$. Note also that this solution cannot compute the frequency of an 
arbitrary color, but only of those output by the color listing algorithm. 

\medskip

The corresponding problem on documents is defined as follows.

\begin{problem}[Document Listing with Frequencies] \label{prob:dlfreq}
Preprocess a document collection $\D$ so that, given a pattern string 
$P$, one can compute $\{ (d,\df(P,d)),\df(P,d){>}0\}$.
\end{problem}

\citeN{VM07} were the first to propose reducing this problem
to color listing with frequencies on the document array $C$.
Their most interesting idea is that
the whole document listing problem can be reduced to
$\rank$ and $\select$ operations on $C$: Array $L$ can be simulated as
$L[i] = \select_{C[i]}(C,\rank_{C[i]}(C,i)-1)$. Thus, they simulate the original
document listing algorithm of \citeN{Mut02} over this representation of $L$, 
using the $O(n)$-bit RMQ of Solution~\ref{thm:rmq}, and using a CSA
(Definition~\ref{def:csa}) to obtain the range $C[sp,ep]$. 
Although they used wavelet trees (Definition~\ref{def:wt})
to represent $C$, using instead
Solution~\ref{thm:seqs} yields the following result.

\begin{solution}[Document Listing with Frequencies] {\rm \cite{VM07}} 
\label{thm:dltfvm}
The problem can be solved in time 
$O(\tsearch(m) + \docc\,\lg\lg D)$ and $|\CSA|+n\lg D+o(n\lg D)$ bits 
of space, where $\CSA$ is a CSA indexing $\D$. 
\end{solution}

Combining this idea with Solution~\ref{thm:cltfbn}, so that we need only
constant-time access to $C$, we immediately obtain a time-optimal solution.

\begin{solution}[Document Listing with Frequencies] {\rm \cite{BNV13}} \label{thm:dltfbn-large}
The problem can be solved in time $O(\tsearch(m)+\docc)$ and 
$|\CSA|+n\lg D+o(n\lg D)$ bits of space, where $\CSA$ is a CSA indexing $\D$. 
\end{solution}

Still, the space is much higher than the near-optimal one of 
Solutions~\ref{thm:dlsada} and \ref{thm:dlhsv}. The document array is usually
much larger than the text or its CSA. \citeN{Sad07} proposed instead a 
solution that, once the leftmost and rightmost positions $i$ and $i'$ of 
document $d$ are known, computes $\tf(P,d)$ using the local CSA of document
$d$. This doubles the total CSA space, but avoids storing the document array. 
We describe this solution in Appendix~\ref{app:dltfsada}. 

\begin{solution}[Document Listing with Frequencies] {\rm \cite{Sad07}} \label{thm:dltfsada}
The problem can be solved in time 
$O(\tsearch(m) + \docc\,\taccess)$ and $2|\CSA|+O(n)$ bits 
of space, where $\CSA$ is a CSA indexing $\D$. 
\end{solution}

If we use the mmphf-based solution of \citeN{BNV13} to compute the frequencies,
instead of doubling the CSA space, the following results are obtained by using 
variants of Solution~\ref{thm:mmphf}. 

\begin{solution}[Document Listing with Frequencies] {\rm \cite{BNV13}} \label{thm:dltfbn}
The problem can be solved in time 
$O(\tsearch(m) + \docc\,\taccess)$ and $|\CSA|+O(n\lg\lg D)$ bits 
of space, where $\CSA$ is a CSA indexing $\D$. 
By adding $O(\docc\lg\lg D)$ time, the space can be reduced to 
$|\CSA|+O(n\lg\lg\lg D)$ bits.
\end{solution}

All the solutions described build over the original algorithm of
\citeN{Mut02}. There is another line of solutions that, although not much
competitive in terms of complexities, yields good practical results and
builds on different ideas. We describe this thread in Appendix~\ref{app:wtree};
the main result obtained follows.

\begin{solution}[Document Listing with Frequencies] {\rm \cite{GNP11}} \label{thm:dltfgnp}
The problem can be solved in time 
$O(\tsearch(m) + \docc\,\lg(D/\docc))$ and $|\CSA|+n\lg D+o(n\lg D)$ bits 
of space, where $\CSA$ is a CSA indexing $\D$. 
\end{solution}

We note that, while document listing could be carried out within just $o(n)$
extra bits on top of the CSA, reporting frequencies requires significantly
more space. This is similar to what we observed for color listing. In 
Appendix~\ref{sec:newdltfcompr} (Solution~\ref{thm:dltfopt}) we will show how to
use the solutions for top-$k$ retrieval to perform document listing with
frequencies using only $o(n)$ bits on top of the CSA.

\section{Computing Document Frequencies}
\label{sec:df}

Document frequency $\df(P)$, the number of distinct documents where a pattern 
occurs, is used in many variants of the tf-idf weighting formula, as mentioned
in the Introduction. In other contexts, such as pattern mining, it is a measure
of how interesting a pattern is. 

\begin{problem}[Document Frequency] \label{prob:docfreq}
Preprocess a document collection $\D$ so that, given a 
pattern $P$, one can compute $\df(P)$, the number of documents where $P$
appears.
\end{problem}

\citeN{Sad07} showed that this problem has a good solution.
One can store $2n+o(n)$ bits associated to the GST of $\D$ so that 
$\df(P)$ can be computed in constant time once the locus node of $P$ is known. 
We describe his solution in Appendix~\ref{sec:dfsada}.

\begin{solution}[Document Frequency] {\rm \cite{Sad07}} \label{thm:docfreq}
The problem can be solved in $O(\tsearch(m))$ time using $|\CSA|+O(n)$ bits of
space.
\end{solution}

On the other hand, in terms of the document array, computing
document frequency leads to the following problem, which is harder
but of independent interest.

\begin{problem}[Color Counting] \label{prob:colcount}
Preprocess an array $C[1,n]$ of
colors in $[1..D]$ so that, given a range $[sp,ep]$,
we can compute the number of distinct colors in $C[sp,ep]$.
\end{problem}

This problem is also
called {\em categorical range counting} and it has been recently shown to
require at least $\Omega(\lg n / \lg\lg n)$ time if using space
$O(n\,\polylog(n))$ \cite{LW13}. Indeed, it is not difficult to match this
lower bound: By Lemma~\ref{lem:L}, it suffices to count the number of values
smaller than $sp$ in $L[sp,ep]$ \cite{GJS95}. This is a well-known geometric 
problem, which in simplified form follows.

\begin{problem}[Two-Dimensional Range Counting] \label{prob:rangecount}
Preprocess an
$n \times n$ grid of $n$ points so that, given a range $[r_1,r_2] \times
[c_1,c_2]$, we can count the number of points in the range.
\end{problem}

Our problem on $L$ becomes a two-dimensional range counting problem if we
consider the points $(L[i],i)$. Then our two-dimensional range is 
$[0,sp-1] \times [sp,ep]$. 

\begin{figure}[t]
\centerline{\includegraphics[width=0.5\textwidth]{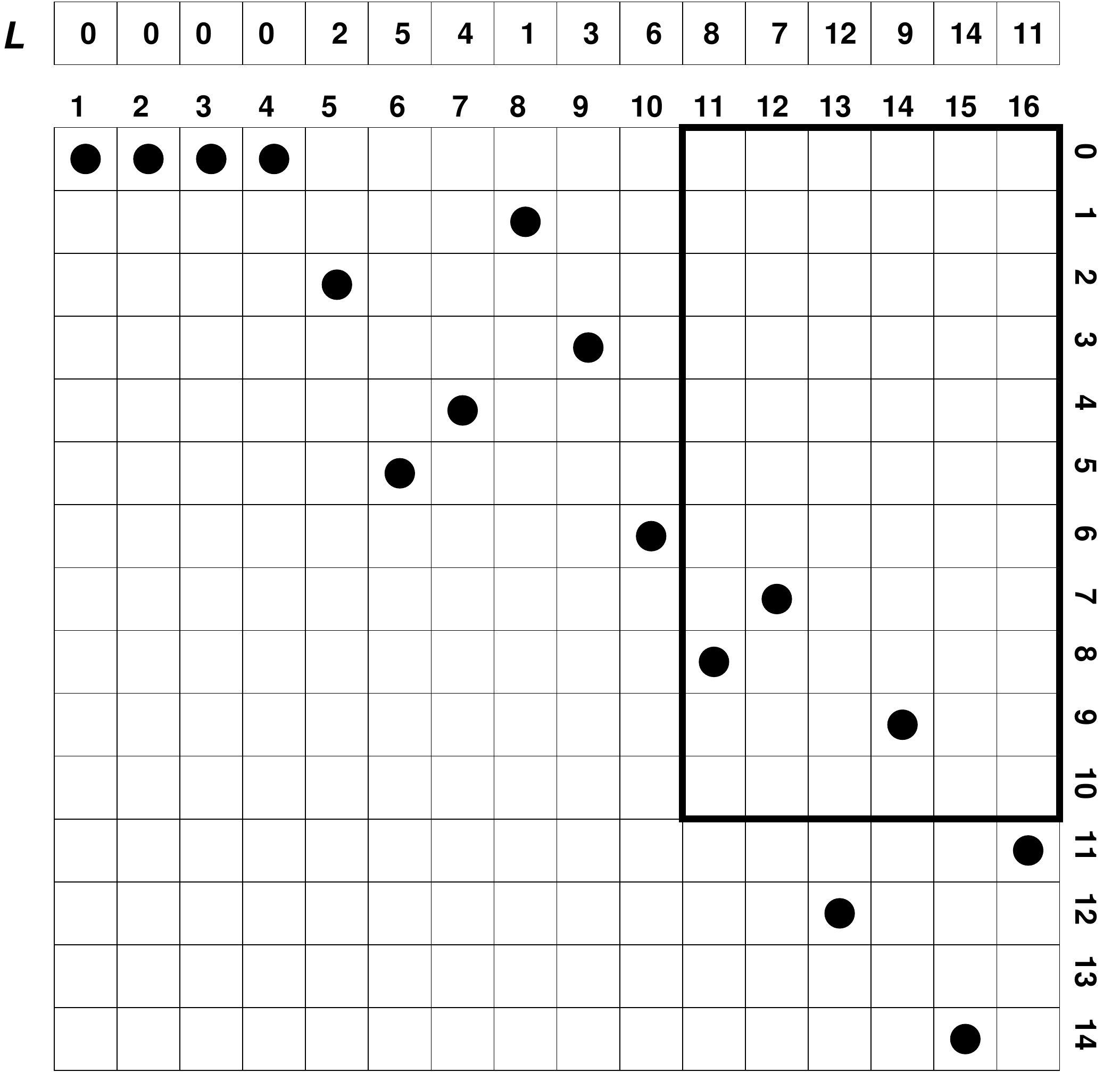}}
\caption{The grid representation of array $L$ in our running example.}
\label{fig:grid}
\end{figure}

\begin{example}
Fig.~\ref{fig:grid} shows the grid corresponding to array $L$ in our running
example, where we have highlighted the query corresponding to $C[11,16]$. As
expected, two-dimensional range counting indicates that there are $3$ points
in $[0,10] \times [11,16]$, and thus $3$ distinct colors in $C[11,16]$.
\end{example}

Two-dimensional range counting 
has been solved in $O(\lg n / \lg\lg n)$ time and $n\lg n + o(n\lg n)$ bits
of space by \citeN{BHMM09}. Unsurprisingly, this time 
is also optimal within $O(n\,\polylog(n))$ space \cite{Pat07}.

\begin{solution}[Color Counting] {\rm \cite{GJS95,BHMM09}} \label{thm:colcount}
The problem can be solved in $O(\lg n / \lg\lg n)$ time and
$n\lg n + o(n\lg n)$ bits of space.
\end{solution}

We note that the space in this solution does not account for the storage of the
color array $C[1,n]$ itself, but it is additional space (the solution does not
need to access $C$, on the other hand). \citeN{GKNP13} used this same reduction,
but resorting to binary wavelet trees instead of the faster data structure of 
\citeANP{BHMM09} Instead, they reduced the $n\lg n$ bits of this wavelet tree, 
and also made the time dependent on the query range (we ignore compressibility 
aspects of their results).

\begin{solution}[Color Counting] {\rm \cite{GKNP13}} \label{thm:colcount2}
The problem can be solved in $O(\lg (ep-sp+1))$ time and
$n\lg D + o(n\lg D) + O(n)$ bits of space.
\end{solution}

Again, this result does not consider (nor needs) the storage of the array
$C$ itself.

\section{Most Important Document Retrieval}
\label{sec:topkimport}

We move on now from the problem of listing all the documents where a pattern
occurs to that of listing only the $k$ most important ones. In the simplest
scenario, the documents have assigned a fixed {\em importance}, as defined
in Problem~\ref{prob:topkimport}. By means of the document array, this can
be recast into the following problem on colors.

\begin{problem}[Top-$k$ Heaviest Colors] \label{prob:topkcolimport}
Preprocess an 
array $C[1,n]$ of colors in $[1..D]$ with {\em weights} in $W[1,D]$ so that, 
given a range $[sp,ep]$ and a threshold $k$, we can output $k$ distinct
colors with highest weight in $C[sp,ep]$.
\end{problem}

A first observation is that, if we reorder the colors so that their weights $W$
become decreasing, $W[d] \ge W[d+1]$, then the problem becomes that of reporting
the $k$ colors with lowest identifiers in $C[sp,ep]$. This is indeed achieved 
by \citeN{GPT09} using consecutive range quantile queries on wavelet trees, 
as explained in Appendix~\ref{app:wtree} (Solution~\ref{thm:quantile}), and 
it is easy to adapt the subsequent improvement \cite{GNP11} to stop after the 
first $k$ documents are reported. It is not hard to infer the following 
result, where the wavelet tree can also reproduce any cell of $C$ (and thus
replace $C$, if we accept its access time).

\begin{solution}[Top-$k$ Heaviest Colors] {\rm \cite{GNP11}} 
\label{thm:topkcolgnp}
The problem can be solved in $O(k\,\lg(D/k))$ time and
$n\lg D+o(n\lg D)$ bits of space. Within this space we can access any
cell of $C$ in time $O(\lg D)$.
\end{solution}

Note that this assumes that we can freely reorder the colors. If this is not
the case we need other $D\lg D$ bits to store the permutation. On the other
hand, by spending linear space ($O(n\lg n)$ bits), we can 
use the improved range quantile algorithms of \citeN{BGJS11}.
In this case, however, an 
optimal-time solution by \citeN{KN11} is preferable.

\begin{solution} [Top-$k$ Heaviest Colors] {\rm \cite{KN11}} 
\label{thm:topkcolkn}
The problem can be solved in real time and
$O(n\lg D)$ bits of space.
\end{solution}

Therefore, we obtain the following results for document retrieval
problems. 

\begin{solution}[Top-$k$ Most Important Documents] {\rm \cite{KN11}} 
\label{thm:topkimportkn}
The problem can be solved in time
$O(\tsearch(m) + k)$ and $|\CSA|+O(n\lg D)$ bits
of space, where $\CSA$ is a CSA indexing $\D$. 
\end{solution}

\begin{solution}[Top-$k$ Most Important Documents] {\rm \cite{GNP11}} 
\label{thm:topkimportgnp}
The problem can be solved in time
$O(\tsearch(m) + k\,\lg(D/k))$ and $|\CSA|+n\lg D+o(n\lg D)$ bits
of space, where $\CSA$ is a CSA indexing $\D$. 
\end{solution}

We defer the results using compressed space to 
Section~\ref{sec:topkcompr} (Solution~\ref{thm:topkimportcomprbn}).

\section{Top-$k$ Document Retrieval}
\label{sec:topk}

We now consider the general case, where the weights of the documents may
depend on $P$ as well.
\citeN{HSV09} introduced a fundamental framework to solve this general
top-$k$ document retrieval problem. All the subsequent work
can be seen as improvements that build on their basic ideas (see
\citeN{HPSW10} for prior work).

Their basic construction enhances the GST of the collection $\D$ so that
the {\em local} suffix tree of each document $T_d$ is {\em embedded} into
the global GST. More precisely, let $u$ be a node of the suffix tree of 
document $T_d$, and let $w$ be its parent in that suffix tree.
Further, let $\occ(u)=\tf(u,d)$ be the number of leaves 
below $u$ in the suffix tree of $T_d$.
There must exist nodes $u'$ and $w'$ in the GST of $\D$ with the same string 
labels, $str(u')=str(u)$ and $str(w')=str(w)$. Then we record a {\em pointer} 
labeled $d$ from $u'$ to $w'$, with weight $\occ(u)$. Thus 
the set of parent pointers for each document $d$ forms a subgraph of
the GST that is isomorphic to the suffix tree of $T_d$. From the fact that
the pointers labeled $d$ correspond to an embedding of the suffix tree of
$T_d$ into the GST, the following crucial lemma follows easily.

\begin{lemma} {\rm \cite{HSV09}} \label{lem:hsv}
Let $v$ be a nonroot node in the GST of $\D$. 
For each document $d$ where $str(v)$ occurs $\tf(v,d)>0$ times, there exists 
exactly one pointer from the subtree of $v$ (including $v$) to a proper 
ancestor of $v$, labeled $d$ and with weight $\tf(v,d)$.
\end{lemma}

To see this, note that 
if $str(v)$ occurs in $T_d$, there must be a node $u$ of the suffix tree
of $T_d$ mapped to $u'$ in GST, which is below $v$ (or is $v$ itself). If we 
follow the successive upward pointers from $u'$, we go over $v$ at
some point,\footnote{Since $T_d \not= \varepsilon$, there are at least two
distinct symbols in $T_d\$$, and thus its root is mapped to the root of the
GST. So we always cross $v$ at some point.}
so there must be at least one such pointer from a subtree of $v$
to a proper ancestor of $v$. On the other hand, there cannot be two such 
pointers leaving from $u''$ and $u'''$, because the LCA of both nodes must
also be in the suffix tree of $T_d$, and hence $u''$ and $u'''$ must point to
this LCA or below it. But since $u''$ and $u'''$ descend from $v$, their LCA 
also descends from $v$, so their pointers point at or below $v$.
Notice, moreover, that the source $u'$ of this unique pointer has a weight
$\tf(u,d)$, which must be $\tf(v,d)$, since all the occurrences of $v$ in
document $d$ are in nodes below $u'$ in the GST.

\begin{figure}[t]
\centerline{\includegraphics[width=0.9\textwidth]{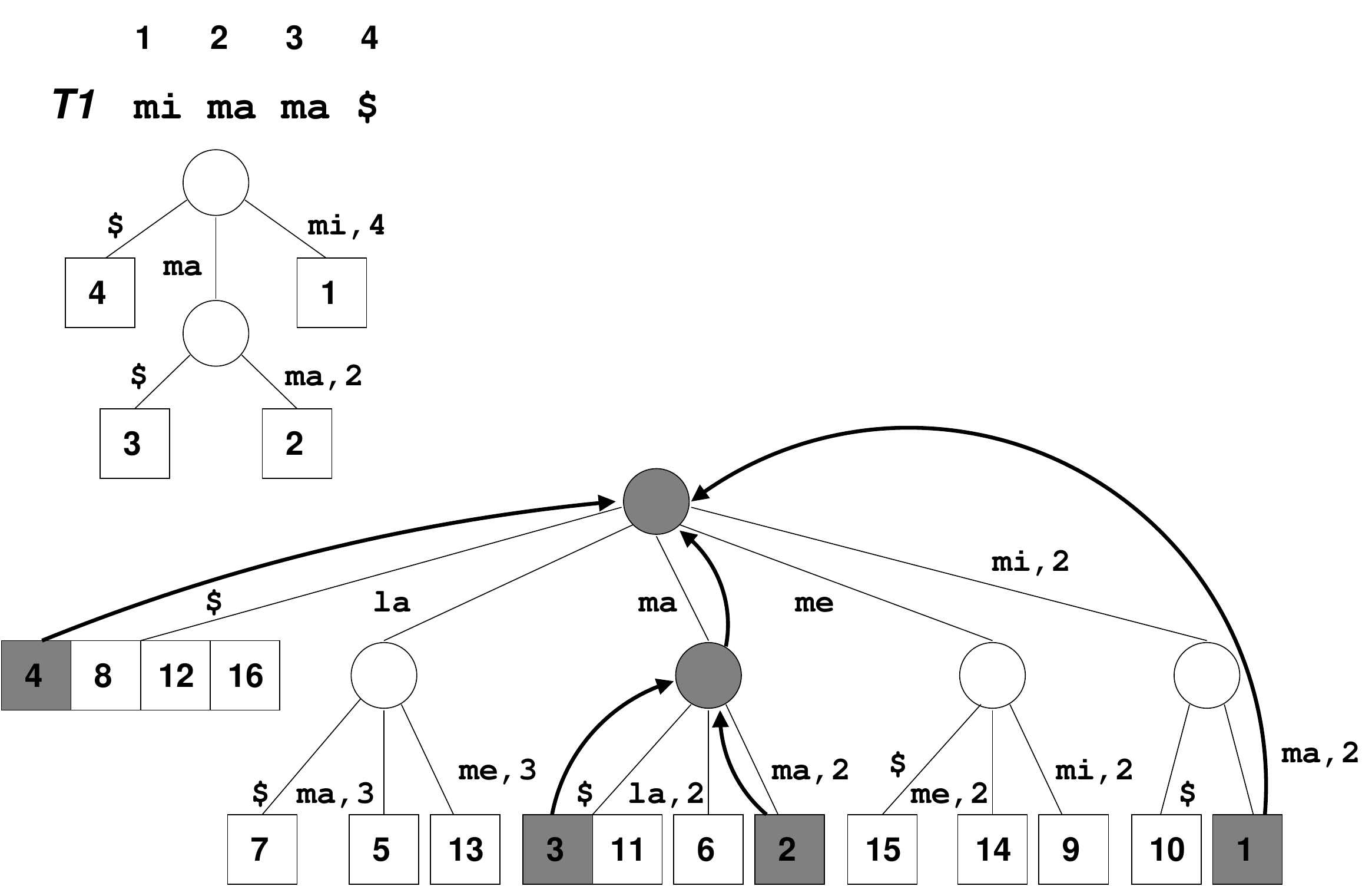}}
\caption{The embedding of the suffix tree of $T_1$ in the global GST. The
nodes are shown grayed and the upward pointers as thick arrows.} 
\label{fig:embed}
\end{figure}

\begin{example}
Fig.~\ref{fig:embed} illustrates how the suffix tree of document $T_1$ is
embedded in the GST, in our running example. The upward pointers describe the
topology of the local suffix tree. Note that for all the patterns that appear
in $T_1$, such as $\textsf{"ma"}$, $\textsf{"mi ma"}$, and $\textsf{"ma ma"}$,
but not $\textsf{"la"}$ nor $\textsf{"me"}$, there is exactly one upward 
pointer leaving from the subtree of the locus and arriving at an ancestor of 
the locus.
\end{example}

\begin{figure}[t]
\centerline{\includegraphics[width=0.9\textwidth]{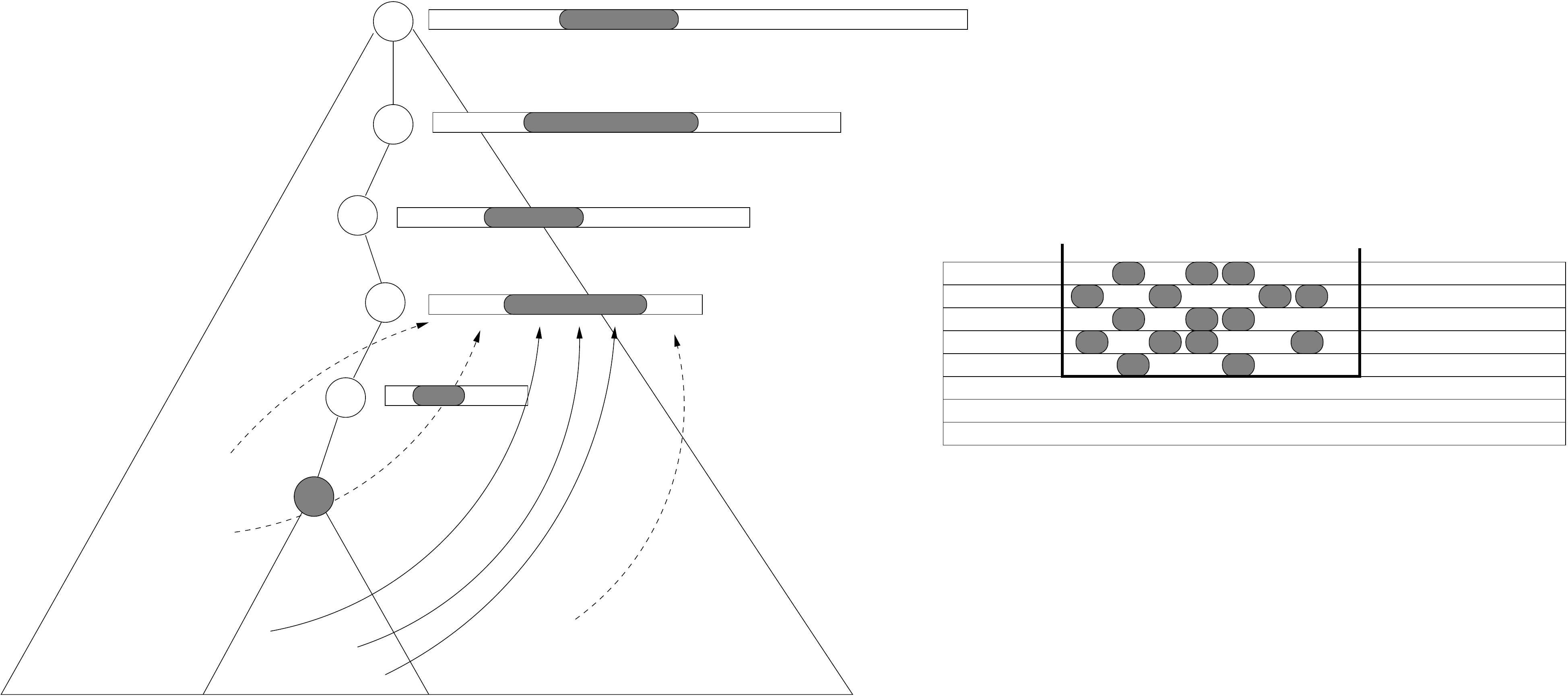}}
\caption{The linear-space schemes for top-$k$ documents. On the left, the 
GST with the locus node 
shadowed. From the arrays of its ancestors (targets of pointers) we choose
the areas (shadowed) of those leaving from the subtree of the locus. We draw
the pointers that go to the grandparent of the locus. Solid arrows leave from
the subtree of the locus and dashed ones from other descendants of the
grandparent node. On the right, the mapping of those arrays into a grid, where
the row is the depth of the target and the column is the preorder of the
source.}
\label{fig:schemetopk}
\end{figure}

\citeN{HSV09} store the pointers at their target nodes, $w'$. Thus, given a 
pattern $P$, we find its locus $v$ in the GST, and then the ancestors 
$v_1, v_2, \ldots$ of $v$ record exactly one pointer labeled $d$ per 
document where $P$ appears, together with the weight $\tf(P,d)$. However, we 
only want those pointers that originate in nodes $u'$ within the subtree of 
$v$. For this sake, the pointers arriving at each node $w'$ are stored in 
preorder of the originating nodes $u'$, and thus those starting from 
descendants $u'$ of $v$ form a contiguous range in the target nodes $w'$. 
Furthermore, we build RMQ data structures (this time choosing maxima, not 
minima) on the $\tf(P,d)$ values associated to the pointers.
Fig.~\ref{fig:schemetopk} (left) illustrates the scheme.

The node $v$ has at most $m$ ancestors. In principle we have to binary search
those $m$ arrays to isolate the ranges of the pointers leaving
from the subtree of $v$. A technical improvement reduces the time to find those
ranges from $O(m\lg D)$ to $O(m)$ time, let $v_i[sp_i,ep_i]$ be those ranges. 
Then, with an RMQ on each such interval we obtain the positions $p_i$ where the
maximum weight in each such array occurs. Those $m$ weights are inserted into
a max-priority queue bounded to size $k$ (i.e., the $(k+1)$th and lower weights
are always discarded). Now we extract the maximum from the queue, which is 
the top-1 answer. We go back to its interval $v_i[sp_i,ep_i]$ and
cut it into two, $v_i[sp_i,p_i-1]$ and $v_i[p_i+1,ep_i]$, compute their RMQ
positions, and reinsert them in the queue. After repeating this process $k$
times, we have obtained the top-$k$ documents. 

\begin{solution}[Top-$k$ Documents] {\rm \cite{HSV09}} \label{thm:topkhsv}
The problem can be solved in $O(m+k\lg k)$ time and
$O(n\lg n)$ bits of space.
\end{solution}

\begin{figure}[t]
\centerline{\includegraphics[width=0.9\textwidth]{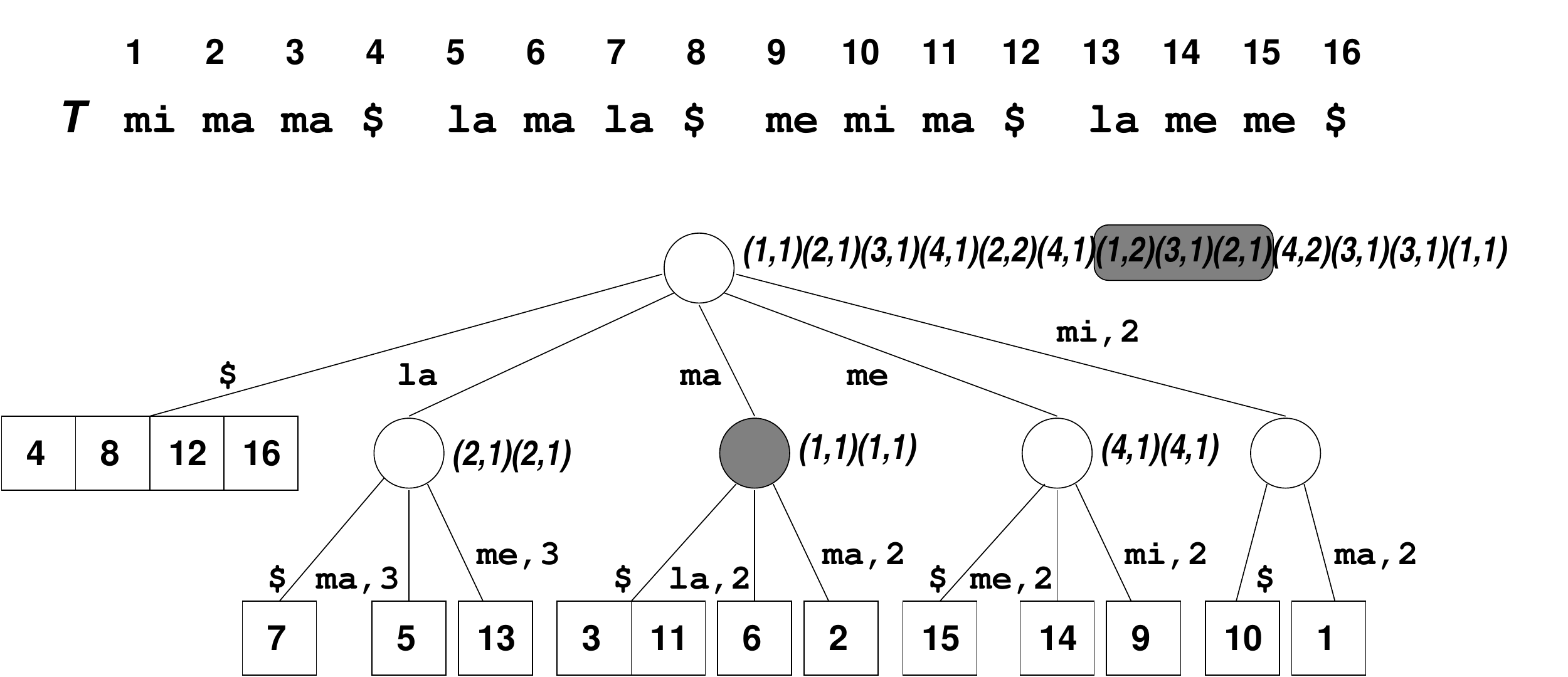}}
\caption{The locus of $P=\textsf{"ma"}$ and the area of the list that
corresponds to it.}
\label{fig:hsv}
\end{figure}

\begin{example}
Fig.~\ref{fig:hsv} shows the arrays of target nodes, in the format 
$(d,\tf(u',d))$, sorted by preorder of the source nodes but omitting the
preorder information for clarity. We shadow the locus node $v$ of 
$P=\textsf{"ma"}$. In this small example the locus has only one proper 
ancestor, the root $v_1$. The range $v_1[sp_1,ep_1] = v_1[7,9]$ in that
array corresponds to the pointers leaving from the subtree of $v$. An RMQ
structure over this array lets us find in constant time the top-1 answer,
$v_1[7]=(d=1,\tf(P,d)=2)$. To get the top-2 answer we split the interval into
$v_1[7,6]$ (empty) and $v_1[8,9]$, and pick the largest of the two.
\end{example}

Note that the weights $\tf(P,d)$ can be replaced by any other measure 
that is a function of the locus of $P$ in the suffix tree of $T_d$. This 
includes some as sophisticated as $\dmin(P,d)$, the minimum distance 
between any two occurrences of $P$ in $d$.

\citeN{NN12} improved the space and time of this solution by using a different
way of storing the pointers. They consider a grid of size $O(n) \times O(n)$ 
so that a pointer from node $u'$ to node $w'$ is stored as a point
$(\mathit{depth}(w'),\mathit{preorder}(u'))$ in this grid, associated to the
document $d$ and with weight $\tf(u',d)$. Then, once the locus $v$ of $P$ is
found in the suffix tree, the problem is reduced to that of finding the $k$
heaviest points in the range $[0,\mathit{depth}(v)-1] \times 
[\mathit{preorder}(v),\mathit{preorder}(v)+\mathit{subtreesize}(v)-1]$ on
the grid (note that there may be several points in a single place of this 
grid, which can be dealt with by creating unique columns for them). 
This is, again, a geometric problem.
We illustrate it in Fig.~\ref{fig:schemetopk} (right).

The key to achieving optimal time is to note that the height of the query range
is at most $m$, and we have already spent $O(m)$ time to find the pattern.
\citeANP{NN12} show that, if we can spend time proportional to the row-size of
the query range, then it is possible to report each top-$k$ point in constant time.
In addition, they manage to slightly reduce the space.

\begin{solution}[Top-$k$ Documents] {\rm \cite{NN12}} \label{thm:topknn}
The problem can be solved in $O(m+k)$ time and
$O(n(\lg D + \lg\sigma))$ bits of space.
\end{solution}

Other weights than $\tf(P,d)$ can also be used in this (and also
\citeANP{HSV09}'s) solution, yet the
space here becomes $O(n(\lg D + \lg\sigma + \lg\lg n))$ bits.

\begin{figure}[t]
\centerline{\includegraphics[width=\textwidth]{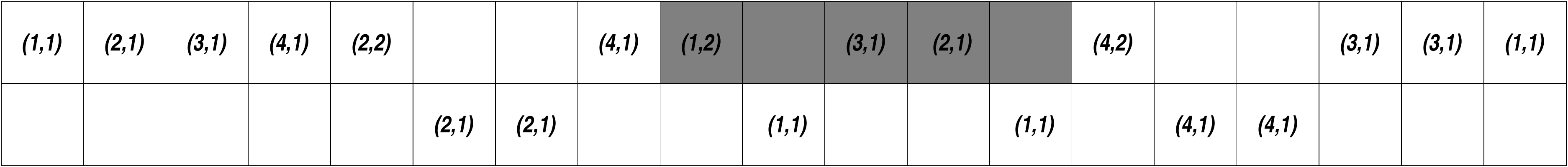}}
\caption{The grid representation of the pointers and the query.} 
\label{fig:nn}
\end{figure}

\begin{example}
Fig.~\ref{fig:nn} illustrates \citeANP{NN12}'s representation of the pointers 
on a grid. In our example the grid is just of height two, and there is no more
than one pointer per cell. The query area is shaded.
\end{example}

We note that Solutions~\ref{thm:topkhsv} and \ref{thm:topknn}
are online, that is, we do not
need to specify $k$ in advance, but can run the algorithms and stop them at
any time, after reporting $k$ documents. Interestingly, the optimal online 
solution
immediately yields optimal solutions to various outstanding challenges left
open by \citeN{Mut02}. First, if we want to list all the documents where a
pattern appears more than $t$ times, we simply run the online algorithm until
it outputs the first document $d$ with $\tf(P,d) < t$. Second, if we want to
list all the documents where two occurrences of the pattern appear within
distance at most $t$, we do the same with the $\dmin$ weighting function.
\citeN{Mut02} had obtained sophisticated solutions for those problems, which
were optimal-time and linear-space only for $t$ fixed at index construction
time. Now we can solve those problems as particular cases, in linear space
and optimal time, defining $t$ at query time, and even in online form. 

\citeANP{Mut02}'s solutions, on the other hand, work for
general color range problems, not only for document retrieval.
This opens the question of whether we can also solve the corresponding top-$k$
problem on colors.

\begin{problem}[Top-$k$ Colors]
Preprocess an 
array $C[1,n]$ of colors in $[1..D]$ so that, 
given a range $[sp,ep]$ and a threshold $k$, we can output $k$ 
colors with highest frequency in $C[sp,ep]$.
\end{problem}

For $k=1$ this problem is called the {\em range mode} problem.
The best exact solution is by \citeN{CDLMW12}, who achieve
linear space but a high time, $O(\sqrt{n/\lg n})$. 
The problem admits, however, good approximations. 
An $(1+\epsilon)$-approximation to the range mode problem is to find an element
whose frequency is at least $1+\epsilon$ times less than that of the range
mode. \citeN{GJLT10} show that one can find such an approximation in 
$O(\lg(1/\epsilon))$ time and $O((n/\epsilon)\lg n)$ bits of space.

\citeN{GKNP13} extends the solution to find approximate top-$k$ colors, where
one guarantees that no ignored color occurs more than $1+\epsilon$ times than 
a reported color. The technique composes the approximate solution for top-1 
color with a wavelet tree on $C$ (Definition~\ref{def:wt}), so that one such
structure is stored for each sequence $C_v$ associated to wavelet tree node
$v$. For the top-1 answer, say $d_1$, it is sufficient to query the root. For 
the top-2 answer, $d_2$, we must exclude $d_1$ from the solution. This is 
done by considering all the siblings of the nodes in the path from the 
root to the leaf $d_1$ of the wavelet tree. The maximum over all those top-1 
queries gives the top-2. For the top-3, we must repeat the process to exclude
also $d_2$ from the set, and so on. We maintain a priority queue with all the 
wavelet tree nodes that are candidates for the next answer. After returning
$k$ answers, the queue has $k\lg D$ candidates (as the root-to-leaf paths 
are of length $\lg D$). By using an appropriate 
priority queue implementation, Solution~\ref{thm:topkcol} is obtained.

\begin{solution}[Top-$k$ Colors] {\rm \cite{GKNP13}} \label{thm:topkcol}
An $(1+\epsilon)$-approximation for the problem takes
$O(k\lg D \lg(1/\epsilon))$ time and
$O((n/\epsilon)\lg D\lg n)$ bits of space.
\end{solution}

\begin{figure}[t]
\centerline{\includegraphics[width=\textwidth]{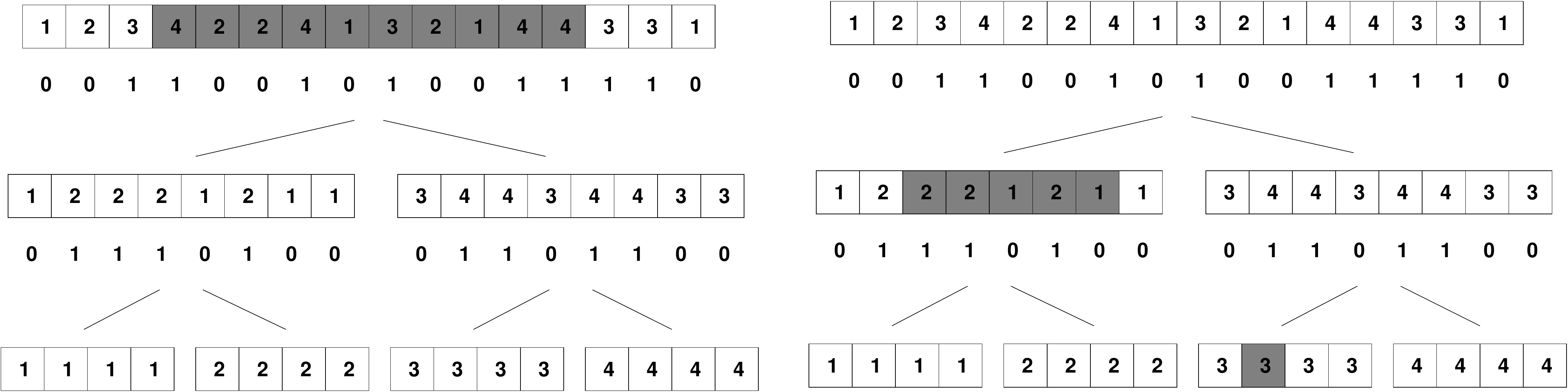}}
\caption{The process of solving top-$k$ colors.}
\label{fig:topcol}
\end{figure}

\begin{example}
Fig.~\ref{fig:topcol} illustrates the process. On the left, we query
$C[4,13]$, finding on the root that $4$ is the most frequent color. To find
the top-2 result, we remove $4$ from the alphabet by partitioning the root
node (which handles symbols $[1..4]$)
into two wavelet tree nodes: that handling $[1..2]$ and that handling $[3]$
(on the right). Now we perform the query on both arrays, finding that $2$ is 
the most frequent color from both nodes. If we wanted the top-3, it would
be decided between wavelet tree leaves handling $[1]$ and $[3]$.
\end{example}

This result is not fully satisfactory for two reasons. First, the space is
super-linear. Second, it is approximate. The second limitation, however,
seems to be intrinsic. \citeN{CDLMW12} show that it is unlikely, even for
$k=1$, to obtain times below $\Theta(n^{\omega/2-1})$ using space below
$\Theta(n^{\omega/2})$, where $\omega$ is the matrix multiplication exponent
(best known value is $\omega=2.376$).
This is a case where document retrieval queries, which 
translate into specific colored range queries (where the possible queries
come from a tree) are much easier than general colored range queries.

On the other hand, there has been much research, and very good results, on 
solving the top-$k$ documents problem in compressed space (for the $\tf$
measure). We describe the most important results next.

\section{Top-$k$ Document Retrieval in Compressed Space}
\label{sec:topkcompr}

In their seminal paper, \citeN{HSV09} also presented the first compressed 
solution to the top-$k$ document retrieval problem under measure $\tf$. The 
idea is to sample some suffix tree nodes and
store the top-$k$ answer for those sampled nodes. The sampling mechanism
guarantees that this answer must be ``corrected'' with just a small number of 
suffix tree leaves in order to solve any query.

Assume for a moment that $k$ is fixed and let
$b = k \lg^{2+\epsilon} n$. We choose the GST leaves whose suffix array
position is a multiple of $b$, and mark the LCA of each pair of consecutive
chosen leaves. This guarantees that the LCA of any two chosen leaves is also
marked.\footnote{We are presenting the marking scheme as simplified by
\citeN{NV12}, where this property is proved.} 
At each of the $n/b$ marked internal nodes
$v$, we store the result $\Top(\D,str(v),k)$. This requires $O(k\lg n)$ bits 
per sampled node, which adds up to $O(n/\lg^{1+\epsilon} n)$ bits.
The subgraph of the GST formed by the marked nodes (preserving ancestorship)
is called $\tau_k$.

Instead of storing the GST, we store the trees $\tau_k$, for
all $k$ values that are powers of 2. All the $\tau_k$ trees add up to
$O(n/\lg^\epsilon n) = o(n)$ bits. Given a top-$k$ query, we solve it in
tree $\tau_{k'}$, where $k'=2^{\lceil\lg k\rceil}=O(k)$ is the power of 2
next to $k$.

Just as the pattern $P$ has a locus node $v$ in the GST, it has a locus $v'$ 
in $\tau_{k'}$, using the same Definition~\ref{def:stree}. It is not hard
to see that, since the nodes of $\tau_{k'}$ are a subset of those of the GST,
$v'$ must descend from $v$ (or be the same).
A way to find $v'$ is to use a CSA and obtain the range $[sp,ep]$ of $P$, then
restrict it to the closest multiples of $b' = k' \lg^{2+\epsilon} n$, 
$[sp'',ep''] \subseteq [sp,ep]$, then take the $(sp''/b)$th and $(ep''/b)$th 
leaves of $\tau_{k'}$, and finally take $v'$ as the LCA of those two leaves. The
following property is crucial.

\begin{lemma} {\rm \cite{HSV09}} \label{lem:tauk}
The node $v' \in \tau_{k'}$ covers a range $[sp',ep']$
such that $[sp'',ep''] \subseteq [sp',ep'] \subseteq [sp,ep]$.
\end{lemma}

It holds that $[sp'',ep''] \subseteq [sp',ep']$ because $v'$ is the LCA of
leaves $sp''$ and $ep''$ in the GST, and it holds that $[sp',ep'] \subseteq 
[sp,ep]$ because $v$ is the LCA of $[sp,ep]$ and it is an ancestor of $v'$.
Moreover, note that $sp'-sp \le b$ and $ep-ep' \le b$.

\begin{figure}[t]
\centerline{\includegraphics[width=0.6\textwidth]{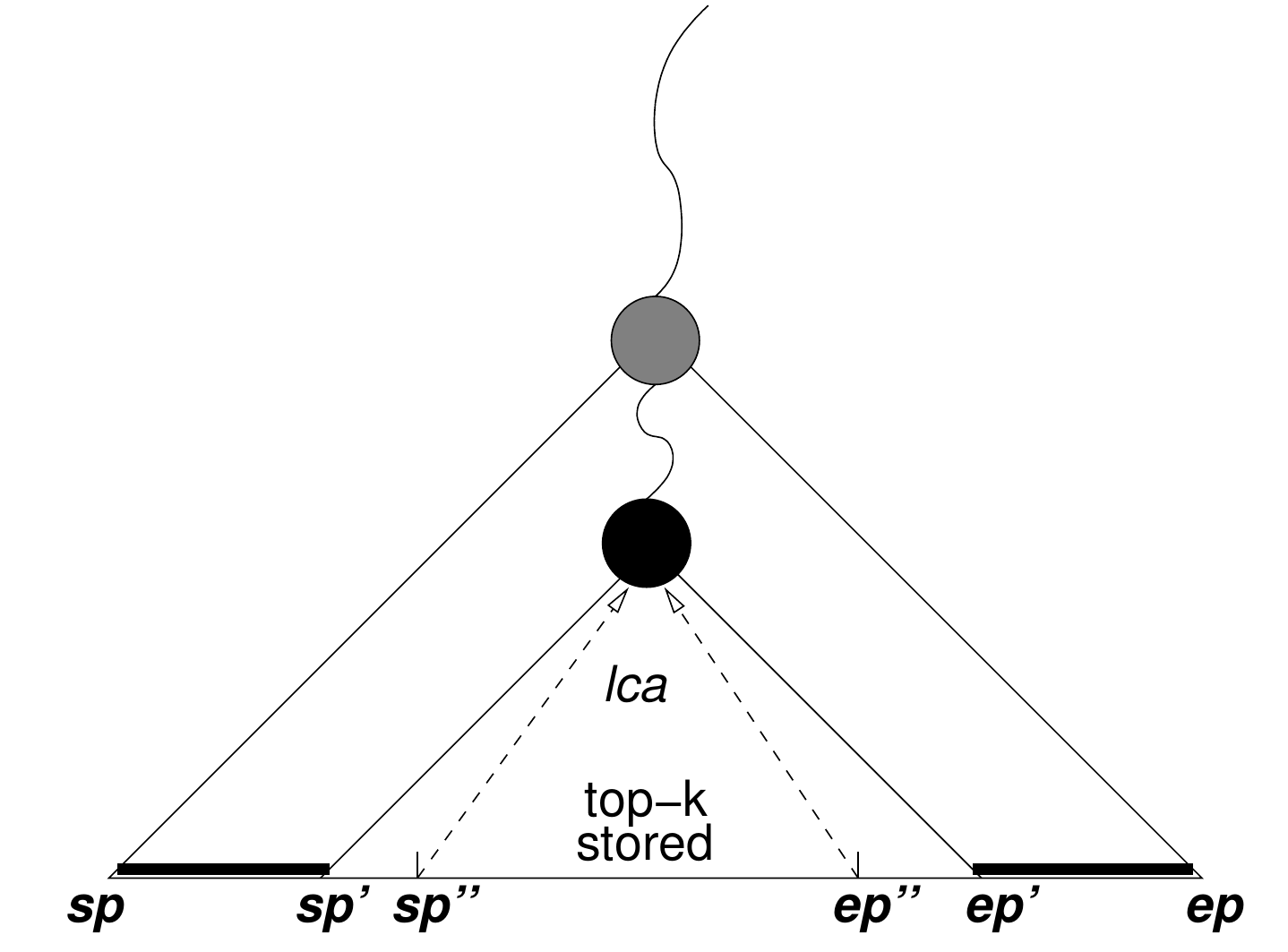}}
\caption{The compressed top-$k$ retrieval scheme.}
\label{fig:comprhsv}
\end{figure}

Node $v'$ has precomputed the top-$k$ answer for $[sp',ep']$ (moreover, the 
top-$k'$ answer). We only need to correct this list with the documents that are 
mentioned in $A[sp,sp'-1]$ and $A[ep'+1,ep]$, and those ranges are shorter
than $b$. It is also possible that $[sp'',ep''] = \emptyset$, but
in this case it holds that $[sp,ep]$ is shorter than $2b$ and thus the answer 
can be computed from scratch by examining those $O(b)$ cells.
Fig.~\ref{fig:comprhsv} illustrates the scheme. The locus $v$ is in gray and
the marked node $v'$ in black. The areas that must be traversed sequentially
are in bold.

In the general case, we have up to $k$ precomputed candidates and must correct 
the answer with $O(b)$ suffix array cells. We traverse those cells one by one
and compute the corresponding document $d$ using $A$ as in
Solution~\ref{thm:dlsada}. Now, each such document $d$ may occur
many more times in $C[sp',ep']$, yet be excluded from the top-$k$ precomputed
list. Therefore, we need a mechanism to compute its frequency $\tf(P,d)$ in 
$C[sp,ep]$.
Note that we cannot use the technique of Solution~\ref{thm:dltfsada}, because 
we do not have access to the first and last occurrence of $d$ in $C[sp,ep]$.

We do have access, however, to either the first (if 
we are scanning $C[sp,sp'-1]$) or the last (if we are scanning $C[ep'+1,ep]$)
position of $d$ in $C[sp,ep]$. Appendix~\ref{app:dltfsada} explains how 
\citeANP{HSV09} compute $\tf(P,d)$ in $O(\taccess\,\lg n)$ time in this case. 
Once we know the frequency, we can consider including $d$ in the top-$k$ 
candidate list (note $d$ might already be in the list, in which case we have 
to update its original frequency, which only considers $C[sp',ep']$).
This yields a total cost of
$O(b\,\taccess \lg n)$ time to solve the query. By using a compressed bitmap
representation (Solution~\ref{thm:comprbitmaps}) for the bitmap $B$ that marks 
the document beginnings, we obtain the following result.

\begin{solution}[Top-$k$ Documents] {\rm \cite{HSV09}} \label{thm:topkcomprhsv}
The problem can be solved in time
$O(\tsearch(m) + k\,\taccess\lg^{3+\epsilon} n)$ and $2|\CSA|+D\lg(n/D)+O(D)
+o(n)$ bits of space, where $\CSA$ is a CSA indexing $\D$, for any constant $\epsilon>0$. 
\end{solution}

Note that the space simplifies to $2|\CSA|+o(n)$ if $D=o(n)$.
There have been several technical improvements over this idea
\cite{GKNP13,BNV13}. The best current results, however, have required
deeper improvements.

One remarkable idea arised when extending the mmphfs of 
Solution~\ref{thm:dltfbn} to top-$k$ retrieval. The idea of \citeN{BNV13} is 
that, if a document $d$ occurs in the left ($[sp,sp'-1]$) and the right 
($[ep'+1,ep]$) tails of the interval, then we know its first and last 
occurrence in $C[sp,ep]$, and thus the mmphfs 
can be used to compute $\tf(P,d)$ fast. The problem are the documents $d$ that
appear only in one of the tails. \citeN{BNV13} prove, however, that there
can be only $k+\sqrt{2bk}$ elements of this kind that can make it to the
top-$k$ list. 

To see this, let $f_{min}$ be the $k$th frequency in the top-$k$
stored set. Then all the other documents have frequency $\le f_{min}$.
The first $k$ documents of the tail can immediately enter the
list, if they now reach frequency $f_{min}+1$. However, the next documents
to enter the top-$k$ list must now reach frequency $f_{min}+2$, and thus we
need to scan at least $2k$ cells of the tail to complete the next batch of
$k$ candidates. Similarly, the next $k$ candidates require scanning at least
$3k$ cells to reach frequency $f_{min}+3$, and so on. 
To incorporate $sk$ elements we need to scan 
$\Omega(s^2 k)$ cells. Since we scan at most $b$ cells, the bound 
$O(\sqrt{bk})$ follows.

With some care, the frequency of all those potential candidates can be stored 
as well (for example, their frequency must be in the narrow range 
$[f_{min}-b+1,f_{min}]$, and instead of storing the document identifiers $d$ 
we can mark one of their occurrences in $[sp,sp'-1]$ or $ep'+1,ep]$, either
of length at most $b$. We can then obtain $d$ using the CSA, using
only $O(\lg\lg n)$ bits to specify $d$ and its frequency.

Recently, \citeN{Tsu13} improved this result further, by noticing that
the idea of limiting the number of candidates can be
extended to the case where the document appears in both tails. This is because
the only interesting ranges are those that correspond to GST nodes, and the
leaves covered by the successive unmarked ancestors of $v'$ (until reaching
the nearest marked ancestor) form $O(b)$ increasing sets of leaves, so the
reasoning of \citeN{BNV13} applies verbatim.

The surprising result is that {\em all} the possible candidates for the
nonmarked nodes, and their frequencies, can be precomputed and stored. There 
is no need at all to use mmphfs (nor local CSAs or document arrays) to solve
a top-$k$ query! This yields the first result for this problem with essentially
optimal space, and moreover very competitive time.

\begin{solution}[Top-$k$ Documents] {\rm \cite{Tsu13}} \label{thm:topkcomprtsur}
The problem can be solved in time
$O(\tsearch(m) + k (\taccess+\lg k+\lg\lg n)\lg k \lg n (\lg\lg n)^4)$ and 
$|\CSA|+D\lg(n/D)+O(D) +o(n)$ bits of space, where $\CSA$ is a CSA 
indexing $\D$. 
\end{solution}

Assuming we use a CSA with $\taccess=O(\lg^{1+\epsilon}n)$, the time simplifies
to $O(\tsearch(m)+k\lg k\lg^{2+\epsilon}n)$. If $D=o(n)$, the space
simplifies to the optimal $|\CSA|+o(n)$. 

\medskip

On the other hand, \citeN{BNV13} show that these ideas can be applied to solve 
the top-$k$ most important problem in compressed space. In this case they sort
the document identifiers by decreasing weight, and each marked node in 
$\tau_k$ stores simply the $k$ smallest document identifiers in the range.
There is no need of the individual CSAs to compute term frequencies. Further,
they speed up the traversal of the blocks of size $O(b)$ by subsampling them and
creating minitrees $\tau'_{k'}$ inside each block. Therefore, instead of 
collecting the candidates from at most one tree and traversing two blocks, we 
must collect the candidates from at most one tree, two minitrees, and two 
subblocks, the latter being sequentially traversed. Those minitrees store the 
top-$k'$ answers for selected nodes, just as the global one, with the 
difference that instead of a document identifier, they store a CSA position 
inside the block where such document appears, as before. This allows them
to encode each document identifier in $O(\lg\lg n)$ bits and thus use 
smaller miniblocks. 

\begin{solution}[Top-$k$ Most Important Documents] {\rm \cite{BNV13}} 
\label{thm:topkimportcomprbn}
The problem can be solved in time
$O(\tsearch(m) + k\,\taccess\lg k \lg^\epsilon n)$ and 
$|\CSA|+D\lg(n/D)+O(D) +o(n)$ bits of space, where $\CSA$ is a CSA 
indexing $\D$, for any constant $\epsilon>0$. 
\end{solution}

With the above assumptions on $\taccess$ and $D$, this simplifies to
$O(\tsearch(m)+k\lg k \lg^{1+\epsilon}n)$ time and $|\CSA|+o(n)$ bits. The 
space is
asymptotically optimal and the time is close to that for document listing
(Solution~\ref{thm:dlsada}).

\medskip

Recently, \citeN{HSTV13} translated this result back
into top-$k$ (most frequent) document retrieval. The solution of \citeN{BNV13}
does not work for this problem because one cannot easily compose two (or, in
this case, three) partial top-$k$ most frequent document answers into the
answer of the union (as a global top-$k$ answer could be not a 
top-$k$ answer in any of the sets). This worked for the top-$k$ most
important document problem, which is easily decomposable. 

However, \citeN{HSTV13} consider the trees and
the minitrees in a slightly different way. There are two kinds of trees, the
original ones, $\tau_k$, and a new set of (also global) trees, $\rho_k$.
This new set of trees uses shorter blocks, of length $c < b$. For each node
$v \in \rho_k$, we consider the highest node $u \in \tau_k$ that descends from
$v$ (there is at most one such highest node, because $\tau_k$ is LCA-closed).
Further, the area covered by $v$ is wider than that of $u$ by at most $c$ 
leaves on each extreme.
For $v$ they store only the top-$k$ answers that are not already mentioned in
the top-$k$ answers of $u$. Those answers must necessarily appear at least
once outside the area of $u$, and thus they can be encoded, similarly to
\citeN{BNV13}, as an offset of $O(\lg \lg n)$ bits. Their frequency is not
stored, but computed using individual CSAs as in 
Solution~\ref{thm:topkcomprhsv}. Then a top-$k$ query requires collecting
results from one $\tau_k$ tree node, from one $\rho_k$ tree node, plus two
traversals over $O(c)$ leaves. Overall, they obtain the same result of
\citeN{BNV13}, but now for the more difficult top-$k$ most frequent document
retrieval.

\begin{solution}[Top-$k$ Documents] {\rm \cite{HSTV13}} 
\label{thm:topkcomprhstv}
The problem can be solved in time
$O(\tsearch(m) + k\,\taccess\lg k \lg^\epsilon n)$ and 
$2|\CSA|+D\lg(n/D)+O(D) +o(n)$ bits of space, where $\CSA$ is a CSA 
indexing $\D$, for any constant $\epsilon>0$. 
\end{solution}

Again, with the above assumptions on $\taccess$ and $D$, this simplifies to
$O(\tsearch(m)+k\lg k \lg^{1+\epsilon}n)$ time and $2|\CSA|+o(n)$ bits.
This is the best time that has been achieved for top-$k$ retrieval when
using this space. 

Very recently, \citeN{NT13} managed to combine the ideas of
Solutions~\ref{thm:topkcomprtsur} and \ref{thm:topkcomprhstv} to obtain
what is currently the fastest solution within optimal space. They define
a useful data structure called the {\em sampled document array}, which
collects every $s$th occurrence of each distinct document in the document
array, and stores it with a $\rank$/$\select$ capable sequence representation
(Solution~\ref{thm:seqs}).
The structure also includes a bitmap of length $n$ that marks the cells of the
document array that are sampled. A compressed representation of this bitmap
gives constant-time $\rank$ and $\select$ within $O((n/s)\lg s)+o(n)$ bits
(Solution~\ref{thm:comprbitmaps}).
By choosing $s=\lg^2 n$, say, the whole structure requires just
$o(n)$ bits, and it can easily compute the frequency of any
document $d$ inside any range $C[sp,ep]$ in time $O(\lg\lg D)$, with a maximum
error of $O(s)$. Therefore, we only need to record $O(\lg s)=O(\lg\lg n)$ bits
to {\em correct} the information given by the sampled document array.

With this tool, they can use Solution~\ref{thm:topkcomprhstv} without the
second $|\CSA|$ bits needed to compute frequencies. Computing frequencies is
necessary for the top-$k$ candidates found in $\tau_k$ and $\rho_k$, and also
for the $O(c)$ documents that are found when traversing the leaves. Now,
\citeN{Tsu13} showed that only $O(\sqrt{ck})$ frequencies need to be stored.
Added to the fact that now they require only $O(\lg\lg n)$ bits to store a
frequency, this allows them to use a smaller block size $c$ for $\rho_k$,
and thus solve queries faster.

\begin{solution}[Top-$k$ Documents] {\rm \cite{NT13}}
\label{thm:topkcomprnt}
The problem can be solved in time
$O(\tsearch(m) + k\,\taccess\lg^2 k \lg^\epsilon n)$ and
$|\CSA|+D\lg(n/D)+O(D) +o(n)$ bits of space, where $\CSA$ is a CSA
indexing $\D$, for any constant $\epsilon>0$.
\end{solution}

This simplifies to $O(\tsearch(m)+k\lg^2 k \lg^{1+\epsilon}n)$ time and
$|\CSA|+o(n)$ bits with the above assumptions.
It is natural to ask whether it is possible to retain this space while
obtaining the time of Solution~\ref{thm:topkcomprhstv},
and even the ideal time $O(\tsearch(m)+k\,\taccess)$.

\medskip

There have been, on the other hand, much faster solutions using the
$n\lg D$ bits of the document array \cite{GKNP13,BNV13}. An interesting solution
from this family \cite{HST12} is of completely different nature: They start 
from the linear-space Solution~\ref{thm:topkhsv} and carefully encode the 
various components. Their result is significantly faster than
any of the other schemes
(we omit an even faster variant that uses $2n\lg D$ bits).

\begin{solution}[Top-$k$ Documents] {\rm \cite{HST12}} \label{thm:topkcomprhst}
The problem can be solved in time
$O(\tsearch(m) + (\lg\lg n)^6 + k(\lg\sigma \lg\lg n)^{1+\epsilon})$ and 
$|\CSA|+n\lg D + o(n\lg D)$ bits of space, where $\CSA$ is a CSA 
indexing $\D$, for any constant $\epsilon>0$.
\end{solution}

The best solution in this line \cite{NT13b}, however, is closer in spirit to
Solution~\ref{thm:topkcomprhstv}. They use the document array and 
$\sqrt{\lg^* n}$ levels of granularity to achieve close to optimal time per
element, $O(\lg^* n)$. A further technical improvement reduces the time to
$O(\lg^* k)$.

\begin{solution}[Top-$k$ Documents] {\rm \cite{NT13b}} \label{thm:topkcompactnt}
The problem can be solved in time
$O(\tsearch(m) + k\lg^* k)$ and 
$|\CSA|+n\lg D + o(n\lg D + n\lg\sigma)$ bits of space, where $\CSA$ is a CSA 
indexing $\D$, for any constant $\epsilon>0$.
\end{solution}

In Section~\ref{sec:dlcompr} we showed how document listing can be carried
out using the optimal $|\CSA|+o(n)$ bits of space, while the solutions 
for document listing with frequencies require significantly more space.
However, for top-$k$ problems we have again
efficient solutions using $|\CSA|+o(n)$ bits of space.
In Appendix~\ref{sec:newdltfcompr} we build on those solutions to
achieve document listing with frequencies within optimal space.

\begin{solution}[Document Listing with Frequencies] \label{thm:dltfopt}
The problem can be solved in time
$O(\tsearch(m) + \docc\,\taccess\lg \docc\lg^\epsilon n)$
and $|\CSA|+D\lg(n/D)+O(D) +o(n)$ bits of space, where $\CSA$ is a CSA
indexing $\D$, for any constant $\epsilon>0$.
\end{solution}

\section{Practical Developments}

Many of the theoretical developments we have described are simple enough to
be implementable, in some cases after some algorithm engineering
tuning. In this section we describe the results obtained by the
implementations we are aware of.

\subsection{Document Listing with Term Frequencies}

There are few doubts that Solution~\ref{thm:dlsada} is a good practical
method for plain document listing (although recent experiments show that
one can do much faster using somewhat more space \cite{FN13}). 
The situation, however, is not so clear
for Solution~\ref{thm:dltfsada}, which gives the term frequencies of the listed
documents. This solution was independently implemented in two occasions
\cite{CNPT10,NPV11} and found in both cases to use much more space than 
expected. This is probably because the individual indexes $\CSA_d$ pose
a constant-space overhead that is significant for relatively small
documents (see a discussion of implementation issues of CSAs in a previous
survey \cite{FGNV09}). While this solution is expected to be slower than using
an explicit document array, the fact that it also uses more space might
be an artifact that can be solved with a more careful implementation that
represents all the $\CSA_d$ arrays as a single structure.

\citeN{VM07} presented the first experimental results showing that, as
expected, the document array was a fast but space-consuming structure for
this problem (Solution~\ref{thm:dltfvm}). 
\citeN{CNPT10} implemented the quantile-based listing algorithm
(Solution~\ref{thm:quantile}), showing its superiority in time and
space over Solutions~\ref{thm:dltfsada} and \ref{thm:dltfvm}, yet space was
still an issue. They also
considered a baseline solution storing inverted indexes for
$q$-grams, which was also inferior. 

\citeN{NPV11} introduced a technique to compress the wavelet tree of the
document array. The idea is based on an observation by \citeN{GN07}, who
compressed suffix arrays by differentially encoding and then 
grammar-compressing them. The reason why this work has deep roots
(see a thorough discussion in \citeN{NM07}), but it can be summarized as:
repetitive texts induce long areas in the suffix array that are identical to
other areas, yet with the values shifted by 1. In a document array, most of
those areas become just plain repetitions, as the suffix shifted by 1 usually
falls in the same document \cite{GKNP13}. 
However, just grammar-compressing the document 
array is not enough, because one needs $\rank$ operations on the sequence in
order to run the document listing algorithm on it. \citeN{NPV11} 
grammar-compressed the bitmaps $B_v$ stored at the wavelet tree nodes, when
this was convenient (they showed that the repetitions at the root level faded
away at deeper levels). They also replaced the quantile-based 
Solution~\ref{thm:quantile} by the always faster and simpler
Solution~\ref{thm:dltfgnp}. As a result, they obtained document listing
solutions that were about twice as slow and required about half the space,
when the text collection was compressible.

Finally, \citeN{BNV13} implemented Solution~\ref{thm:dltfbn} using mmphfs.
Although it loses to the compressed wavelet trees on compressible texts, it
offers a more stable performance, always using less space than uncompressed 
wavelet trees. Their time performance, however, is significantly worse.

\begin{figure}[t]
\includegraphics[width=\textwidth]{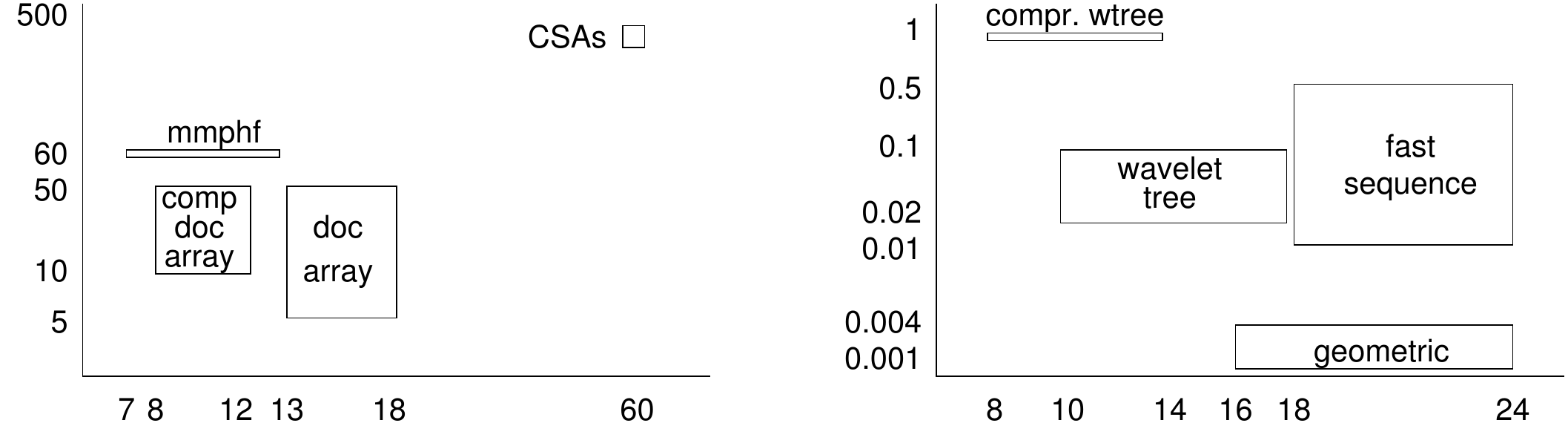}
\caption{Schema of experimental results for document listing (left) and
top-$k$ document retrieval (right). 
The $x$-axis is the space in bits per character (bpc) 
and does not include the 4--8 bpc of the global CSA. The time, on the $y$-axis,
is on the left the milliseconds (msec) to solve the whole document listing 
query, and on the right the msec per each of the $k$ results of the top-$k$
query.}
\label{fig:plots}
\end{figure}

Fig.~\ref{fig:plots} (left) gives some rough numbers on a typical desktop 
computer (a 2GHz Intel Xeon with 16GB RAM and 2MB cache, using C/C++ 
with full optimization and no multicore parallelism),
averaging over various types of text collections.%
\footnote{These experimental results are newer than those in \citeN{NV12}.}
Here ``doc array'' stands for the uncompressed document array,
``comp doc array'' for the compressed document array,
``mmphf'' for the solution based on mmphfs,
and ``CSAs'' for the implementation using individual CSAs.
It is likely that an implementation of Solution~\ref{thm:dltfopt} will require
less space than all these solutions, yet probably it will be significantly
slower than document-array-based solutions.

\subsection{Top-$k$ Documents}

The most interesting ideas of \citeN{CNPT10} were two heuristics for top-$k$
document retrieval that used just the plain document array. The best was
a prioritized document listing that took advantage that, as we backtrack in
the wavelet tree, we can know the sums of the $\tf(P,d)$ values
over all the documents $d$ represented in each wavelet tree node (this is 
just the size of the interval $[sp,ep]$ on the node). In the traversal they
gave priority to the nodes with higher $ep-sp+1$. They kept
in a priority queue the wavelet tree nodes about to be traversed (initially
just the root). Then they extracted the node with largest $ep-sp+1$ value.
If it was a leaf, they reported the corresponding document. Otherwise they
inserted both children in the queue. Then the first $k$ leaves extracted are
the answer.
%
%
Later, \citeN{CPS12} showed that the idea works well at a much
larger scale and even competes, up to some degree, with inverted-index top-$k$
based solutions, on natural language text collections. See also the preceding
work by \citeN{PTSHVC11}.

The first implementation of the succinct top-$k$ framework of \citeN{HSV09} 
was by \citeN{NV12}. They focused on
Solution~\ref{thm:topkcomprhsv}, yet using a document array instead of 
the individual CSAs (which, as explained, did not work well in practice). 
They made several optimizations,
in particular replacing the brute-force scanning of the blocks by an adaptation
of the prioritized document listing of \citeN{CNPT10}.
This traversal is stopped as soon as it
delivers documents with frequencies below the $k$th candidate already known.

Another optimization was to factor out the redundancies that arise because the
same node may store the top-$k$ answers in some $\tau_k$ and then the
top-$2k$ answers in $\tau_{2k}$. They store all the top-$k$ precomputed 
solutions in $\tau_1$ (which contains the other trees) for the maximum $k$ 
where each node stores answers. The other trees store only their topology
plus pointers to $\tau_1$. 

The experiments compare all these solutions, including several variants to
store the $\tau_k$ trees and representations of the document array (including
one based on Solution~\ref{thm:seqs}, where only the basic brute-force block
traversal algorithm can be implemented). The results show that the data
structures of \citeN{HSV09} pose very little extra space and considerably 
improve upon the time performance of the basic heuristics \cite{CNPT10}. They
also show that the improvements make a significant difference with brute-force
scanning, even when this is implemented over the faster sequence representation
of Solution~\ref{thm:seqs}. The different variants of wavelet trees, compressed
or not, dominate the time/space map.

Fig.~\ref{fig:plots} (right) gives some rough numbers.
Here ``wavelet trees'' refers to plain wavelet trees and
``compr.\ wtree'' to compressed wavelet trees (on compressible texts).
The basic heuristics without the structures 
$\tau_k$ use almost the same space, but require more time (in some cases very
little, in others up to 10 times more).
The brute-force solution on the fast sequence implementation is called
``fast sequence''.

Recently, some attempts at directly implementing the linear space solutions
have been made. \citeN{KN13} implemented Solution~\ref{thm:topknn}, replacing
the complex optimal-time top-$k$ algorithm on the grid by a simpler one 
\cite{NNR13}. The fact that the suffix tree height is $O(\lg n)$ with 
high probability (whp) on rather general assumptions on the text \cite{Szp93}, 
yields time complexity $O(m+(k+\lg\lg n) \lg\lg n)$ (whp).
They also improve the space by
removing from the grid the points with term frequency 1 (this idea was also 
mentioned by \citeN{HST12}). If the grid returns less than
$k$ answers, a normal document listing on the document array is used
to obtain further documents to complete the answer. Their performance is 
shown in Fig.~\ref{fig:plots} (right) under the name ``geometric''.
The space is not small, but the structure is an order of magnitude faster
than the others.

Another probably practical solution, yet to be implemented
\cite{HST12}, departs from the linear-space solution of \citeN{HSV09}.
The results, summarized in Solution~\ref{thm:topkcomprhst}, will probably 
translate directly into a practical result, comparable to that of \citeN{KN13}.

It is illustrative to take a look at the direct implementations of other
proposals to see how drastic the above space savings are. For example,
\citeN{HPSW10} implemented a predecessor of the linear-space index of
\citeN{HSV09}. While they obtain times in the range 0.01$k$--0.02$k$ 
msec, their index takes about 4,000 bpc (i.e., 250 
times the space of the text!). Still, we can probably reduce current spaces
further, as suggested by Solutions~\ref{thm:topkcomprtsur} and
\ref{thm:topkcomprnt},
but possibly at the price of a significantly higher time, since a CSA is much
slower than a wavelet tree.

\section{Extensions}

We have focused on a few fundamental 
document retrieval problems. However, there are more complex and challenging
ones. Arguably, the most important extension are queries
formed by more than one pattern string. The document listing problem then
becomes listing the documents where some of the patterns appear (union
queries), or all of the patterns appear (intersection queries) or, 
generalizing, where at least $t$ of the patterns appear (thresholded queries). 
The corresponding top-$k$ problems aim at finding the $k$ highest weighted
documents among those that qualify, where the weight considers
all the patterns appearing in the document.

Those problems are well known in the inverted index literature,
where most of the research focuses on intersections (see \citeN{BLOLS09} for
a recent survey). There is a difficulty measure (i.e., a lower
bound) that applies to the corresponding problem of intersecting (inverted)
lists: the number of jumps from one list to another when reading the documents 
from all the lists in increasing order. \citeN{GKNP13} consider the intersection
problem on color arrays (and hence on document arrays) represented as wavelet 
trees, and show that they can achieve a complexity close to that lower bound.
With respect to top-$k$ algorithms on inverted indexes, those for intersections
usually carry out a boolean intersection first and then filter the highest
frequencies, so they can be extended to document retrieval using \citeANP{GKNP13}'s
solution. The algorithms for top-$k$ unions generally process the documents 
in decreasing frequency order. It is not difficult to adapt the top-$k$
document retrieval techniques we have covered to retrieve the results online,
that is, without knowing $k$ a priori. This lets them simulate the sequential
access to an inverted list where the documents are listed in decreasing
frequency order, and thus any algorithm on inverted lists can be simulated on
top of these iterators.

A more theoretical line of research aims at time complexities that are
independent of the inverted lists. \citeN{FKMS03} show that the intersection
problem for
two patterns of length $O(m)$ can be solved using $O(n^{3/2}\lg n)$ words of 
space and $O(m+\sqrt{n}+\docc)$ time. This space is impractical in most cases. 
It was improved to $O(n\lg n)$ words by \citeN{CP10}, although the time raised
to $O(m+\sqrt{n\,\docc}\lg^{2.5} n)$. Then it was further improved 
\cite{HSTV10} to linear space and $O(m+\sqrt{n\,\docc}\lg^2 n)$ time, by
extending the succinct framework (Solution~\ref{thm:topkhsv}) so that all
pairs of marked nodes are preprocessed (thus the nodes must be much larger now,
hence the complexity). \citeN{HSTV10} also generalize the solution
to handle top-$k$ retrieval (where $\docc$ becomes $k$ in the time) and
to handling up to $t$ patterns, for $t$ fixed at indexing time. Here the
time worsens to $O(mt+n^{1-1/t}\docc^{1/t}\polylog(n))$, which quickly tends
to linear time.
\citeN{FGKLMSV12} showed that this problem is indeed much more difficult than
those involving one pattern: any solution using $O(m+\docc+\polylog\,n)$ time
on a pointer machine requires $\Omega(n(\lg n / \lg\lg n)^3)$ bits of space.
It is likely that the problem is actually much harder than what the lower bound
suggests.

An interesting variant of this problem is to list the documents where a pattern
does {\em not} appear. \citeN{Mut02} shows that this can be solved
in real time and linear space, yet
his solution does not seem amenable to the space-reduction techniques that
have been developed for the simple document listing problem. However, this
query is most interesting when combined with other patterns that must appear
in the documents.
This was addressed by \citeN{FGKLMSV12} and improved to linear space by 
\citeN{HSTV12}, who achieve $O(m+\sqrt{n}\lg\lg n+\sqrt{n\,\docc}\lg^{2.5} n)$
time to handle one ``positive'' and one ``negative'' pattern. This problem 
seems as difficult as handling two ``positive'' patterns, but no lower bound
is known.

Another extension that has been considered is to add further restrictions to
the documents to be retrieved, for example adding to each document a numeric
parameter (such as date, version number, etc.) and let queries specify a range
of acceptable values for this parameter. \citeN{KN11} and \citeN{NN12} consider
this kind of extensions, and show they can be solved with a slightly higher
cost than the version without parameters. For example, parameterized top-$k$
queries can be solved in time $O(m+\log^{1+\epsilon} n + k\log^\epsilon n)$
and linear space, for any constant $\epsilon>0$ \cite{NN12}.

Another dimension the current research is just starting to explore are the 
computation models. Most of the current work refers to static indexes and
deterministic
worst-case sequential algorithms in main memory. For example, a recent
work \cite{SSTV13} addresses the top-$k$ problem under the external memory 
model. They obtain linear
space and the almost optimal I/O time $O(m/B + \lg_B n + \lg^{(h)} n + k/B)$ 
for any constant $h$ ($\lg^{(h)}n$ means taking logarithm $h$ times), where
$B$ is the disk block size. By using very slightly superlinear space,
$O(n\lg^*n)$ words, they reach the optimal I/O time
$O(m/B+\lg_B n + k/B)$. They also show how to solve top-$k$ queries in main
memory in $O(k)$ time once the locus is known, which enables other more 
complex queries. As another example, in a recent update to their
work, \citeN{NN13topk} further improve their top-$k$ solution to the RAM-optimal
time $O(m/\lg_\sigma n + k)$, and also consider the dynamic scenario where
documents can be inserted and removed from the collection. They obtain query
time $O(m(\lg\lg n)^2/\lg_\sigma n + \lg n + k\lg\lg k)$, and 
$O(\lg^{1+\epsilon}n)$ time per character inserted or deleted for any constant
$\epsilon>0$, which are not far from optimal as well.

\section{Conclusions}

While document retrieval on several Western natural languages can be handled
with simple inverted indexes, extending this task to other languages and 
scenarios requires solving document retrieval on general sequence collections.
This has proven to be much more algorithmically challenging and has stimulated
a fair amount research in the last decade. Many of those problems can be 
reduced to ``range color'' problems, which have many additional applications 
in data mining.

Table~\ref{tab:colors} gives the time/space complexities achieved for the 
different range color problems we have considered in this survey. While the
solutions to the variants of color listing are rather satisfactory, those 
for color counting and top-$k$ heaviest colors have good time complexities, but
their space is linear. Worse, there exist only approximate efficient solutions
to top-$k$ colors (and this seems to be intrinsic), which in addition require superlinear space.

\begin{table}[t]
\begin{center}
\begin{tabular}{lrcc}
Problem & Sol. & Time & Space \\[1ex]
\hline\\[-2ex]
Color Listing & \ref{thm:clsada}~~& Real & Data + $O(n)$ \\[1ex]
\hline\\[-2ex]
Color Listing with Frequencies & \ref{thm:cltfbn}~~& Optimal & Data + $O(n\lg\lg D)$ \\[1ex]
\hline\\[-2ex]
Color Counting & \ref{thm:colcount}~~& $\lg n / \lg\lg n$ & $n\lg n + o(n\lg n)$ \\[1ex]
	       & \ref{thm:colcount2}~~& $\lg(ep-sp+1)$ & $n\lg D + o(n\lg D)+O(n)$ \\[1ex]
\hline\\[-2ex]
Top-$k$ Heaviest Colors & \ref{thm:topkcolkn}~~& Real & $O(n\lg D)$ \\[1ex]
			& \ref{thm:topkcolgnp}~~& $k\lg(D/k)$ & $n\lg D + o(n\lg D)$ $^*$ \\[1ex]
\hline\\[-2ex]
Top-$k$ Colors ($(1{+}\epsilon)$-approximation) & \ref{thm:topkcol}~~& $k\lg D \lg(1/\epsilon)$ & $O((n/\epsilon)\lg D\lg n)$
\end{tabular}
\end{center}

\vspace*{-2mm}
\caption{Best time/space complexities achieved for range color problems.
The asterisk means that the structure contains sufficient information to
reproduce any array cell in $O(\lg D)$ time.}
\label{tab:colors}
\end{table}

Table~\ref{tab:docs} gives simplified time/space complexities for the document
retrieval problems we have considered. Each category is ordered by decreasing
space (considering, when they are not directly comparable, the most common case).
There are optimal-time solutions to all the problems, yet requiring linear 
space (except, notably, computing document frequency). On the other hand, 
there are asymptotically space-optimal solutions in all cases, adding just 
$o(n)$ bits on top of a CSA (except, again, document frequency, which adds
$O(n)$). Some of those solutions have been assembled or precised along this 
survey.  The space-optimal solutions require, however, polylogarithmic
time. There are various tradeoffs in between using more space and less time,
but it is unknown whether they are optimal.

On the practical side, all the successful implemented solutions use the
document array represented with wavelet trees. For document listing with
frequencies, these solutions require about 3 times the size of the
collection, and replace it. The most space-efficient solutions may reach, on
compressible collections, as little as 12 bpc, and take a few tens of
msec to run a query.
The more space-demanding solutions (and all the solutions on incompressible
collections) take as much as 26 bpc and
solve the queries in a few msec. For top-$k$ queries the most
compressed solutions use about $k$ msec,
whereas the most space-demanding ones take $k$--$4k$ microseconds.

\begin{table}[t]
\begin{center}
\begin{tabular}{lrcc}
Problem & Sol. & Time ($+\tsearch(m)$) & Space ($+|\CSA|$) \\[1ex]
\hline\\[-2ex]
Document Listing & \ref{thm:dlsadalarge}~~& $1$ & $n\lg D$ \\[1ex]
                 & \ref{thm:dlhsv}~~& $\lg^{1+\epsilon}n$ & $o(n)$ \\[1ex]
\hline\\[-2ex]
Document Listing with Frequencies & \ref{thm:dltfbn-large}~~& $1$ & $n\lg D$ \\[1ex]
				  & \ref{thm:dltfsada}~~& $\lg^{1+\epsilon} n$ & $|\CSA|+O(n)$ \\[1ex]
				  & \ref{thm:dltfbn}~~& $\lg^{1+\epsilon} n$ & $O(n\lg\lg\lg D)$ \\[1ex]
				  & \ref{thm:dltfopt}~~& $\lg\docc\lg^{1+\epsilon} n$ & $o(n)$ \\[1ex]
\hline\\[-2ex]
Document Frequency & \ref{thm:docfreq}~~& $1$ & $O(n)$ \\[1ex]
\hline\\[-2ex]
Top-$k$ Most Important Documents & \ref{thm:topkimportkn}~~& $1$ & $O(n\lg D)$ \\[1ex]
			& \ref{thm:topkimportgnp}~~& $\lg(D/k)$ & $n\lg D$ \\[1ex]
			& \ref{thm:topkimportcomprbn}~~& $\lg k\lg^{1+\epsilon}n$ & $o(n)$ \\[1ex]
\hline\\[-2ex]
Top-$k$ Documents & \ref{thm:topknn}~~& $1$ & $O(n\lg D+n\lg\sigma)$ \\[1ex]
		  & \ref{thm:topkcompactnt}~~& $\lg^* k$ & $n\lg
D+o(n\lg\sigma)$ \\[1ex]
		  & \ref{thm:topkcomprhstv}~~& $\lg k\lg^{1+\epsilon} n$ & $|\CSA|+o(n)$ \\[1ex]
		  & \ref{thm:topkcomprnt}~~& $\lg^2 k\lg^{1+\epsilon} n$ & $o(n)$ 
\end{tabular}
\end{center}

\vspace*{-2mm}
\caption{Best time/space complexities achieved for document retrieval problems.
We assume to simplify that $\taccess = \lg^{1+\epsilon} n$ for some constant
$\epsilon>0$, and that $D=o(n)$. Times are in addition to
$\tsearch(m)$ (this is just $m$ for Solution~\ref{thm:topknn}) and per
element output. We write space $n\lg D$ for $n\lg D + o(n\lg D)$.}
\label{tab:docs}
\end{table}

In general, the current solutions give satisfactory time performance, but their
space requirements are still too high. This is in sharp contrast with the 
pattern matching problem, which the CSAs solve efficiently in optimal 
space (i.e., the entropy of the collection) and in addition replace the
collection \cite{NM07,FGNV09}. We
believe that the recent optimal-space solutions should be the
basis of the next generation of practical document retrieval indexes. Still,
their theoretical time complexities suggest that a good deal of algorithm
engineering will be necessary to render their times acceptable.

To conclude, an interesting question is whether these more general data 
structures will ever be able to compete with inverted indexes in the very same 
niche the latter have been designed for. While we do not expect inverted indexes
to be overthrown in typical natural language queries, we do believe the 
suffix-array-based solutions for general sequences will prove more efficient 
in handling more sophisticated queries, where the simple bag-of-words model is 
insufficient. 
This includes features that are becoming standard in search engines, such as 
phrase patterns \cite{PTSHVC11,FBNCPR11,CPS12}, 
approximate pattern matching \cite{NBYST01,CB09}, 
and autocompletion queries \cite{BMW08,HO13},
to name a few. The given references show either that suffix-array-based
solutions are superior to inverted indexes, or that the latter have to be
significantly extended to cope with those types of queries.


%
%

\bibliographystyle{acmtrans}
\bibliography{paper}

\appendix

\section{Range Minimum Queries and Lowest Common Ancestors} \label{app:rmq}

The RMQ problem has a fascinating history, intimately related to the LCA
problem. \citeN{HT84} showed that the LCA problem could be solved in 
constant time after a linear-time preprocessing, but
their algorithm was quite complicated. \citeN{SV88} managed to somehow
simplify the algorithm while retaining its optimality. \citeN{BV93} showed
that, if one traverses the tree in preorder and writes down the depths of the
touched nodes in an array, then the LCA problem on the tree becomes an RMQ
problem on the array of depths (one also needs a pointer from each tree node
to its first occurrence in the array). This RMQ problem is particular because
consecutive array entries differ by $\pm 1$, and this was exploited by
\citeANP{BV93} to solve it. \citeN{GBT84} showed that, in turn, a general RMQ 
problem could be converted into an LCA problem, by considering the {\em 
Cartesian tree} \cite{Vui80} of the array (Definition~\ref{def:cartesian}). 
The observation \citeANP{GBT84}\ made is that $\rmq_L(sp,ep)$ corresponds 
to the LCA of the nodes corresponding to array positions $i$ and $j$ of the 
Cartesian tree of $L$.

Finally, \citeN{BF00} simplified the
methods up to a point that the solution was considered to be practical:
To solve the general RMQ problem, one builds the Cartesian tree and solves the
LCA problem on it instead. To solve that LCA problem, one traverses
the tree and converts it into a restricted RMQ ($\pm 1$) problem.
To solve that problem, one precomputes solutions to all intervals whose length
is a power of 2, $\rmq_L(i,i+2^\ell-1)$, so that any interval $L[sp,ep]$ is 
covered by two such (overlapping) intervals, $[sp,sp+2^\ell-1]$ and 
$[ep-2^\ell+1,ep]$, for $\ell = \lfloor \lg(ep-sp+1) \rfloor$. To avoid using 
$O(n\lg n)$ words of space with all those precomputed intervals, one 
precomputes only the intervals starting at multiples of $\frac{1}{2}\lg n$. 
Queries then cover a number of whole intervals plus two partial ones at the 
tails. For the part covered by whole intervals we have two candidates to be
the answer, as explained. For the tails, since this is a restricted ($\pm 1$) 
RMQ problem, there are only $2^{\frac{1}{2}\lg n}=\sqrt{n}$ possible distinct 
intervals, and thus
a precomputed table of sublinear size can answer RMQ queries on 
all the possible ranges within any possible interval. 
We then compare the (up to) 4 
candidate cells in $L$ and return the minimum.
This explains Solution~\ref{thm:rmqlinear}.

A new twist to the problem was given by \citeN{FH11}, who aimed at solving
the RMQ problem {\em without accessing the array $L$}. They showed that it
is sufficient to store $2n+o(n)$ bits in order to answer RMQ queries in
constant time. The idea is to start with the Cartesian tree and convert it
into a general tree by the well-known isomorphism: We create a special root
for the general tree, and the nodes in the leftmost branch of the binary tree
are the children of the root. Their right subtrees are recursively converted
into general trees. It can then be proved that $\rmq_L(i,j)$ is 
obtained by first computing the LCA of the $i$th and $j$th nodes (in preorder)
of this general tree, and then taking the child of this node in the path to
node $j$. Various compact tree representations can carry out those 
operations in $O(1)$ time and using $2n+o(n)$ bits (e.g., that of \citeN{SN10}).

Note we must solve the LCA problem on those compact tree representations. It 
is particularly convenient that the representation resorts to balanced 
parentheses: one traverses the tree in preorder, opens a parenthesis when
arriving at a node and closes it when leaving the node. Since the tree depth
increases by 1 when opening a parenthesis and decreases by 1 when closing it,
the LCA problem on the tree becomes naturally a restricted ($\pm 1$) RMQ
operation on the very same $2n$ bits of the parenthesis representation. Thus 
we can solve it by adding $o(n)$ bits on top of the $2n$ bits of the parentheses
representation (i.e., the array of $\pm 1$ values), as explained.
Note that we have not accessed $L$ at all, and
this explains Solution~\ref{thm:rmq}.

Furthermore, by noting that each Cartesian tree yields different RMQ answers
for some query, and that there are about $4^n/n^{3/2}$ different binary trees
on $n$ nodes, one can see that $\lg (4^n/n^{3/2}) = 2n-O(\lg n)$ bits are
necessary to distinguish between all the possible arrays $L$. 
Thus Solution~\ref{thm:rmq} is asymptotically optimal.

\section{Proof of Correctness of Sadakane's Algorithm}
\label{app:sadaproof}

While \citeN{Mut02} gives a detailed proof 
of the correctness of his algorithm, the variant of \citeN{Sad07} is not
clearly proved correct. As this is a nonobvious issue, here we 
prove \citeANP{Sad07}'s algorithm correct. More precisely, we prove it is
correct when it visits the left subtree first. 

Both \citeANP{Mut02}'s and \citeANP{Sad07}'s algorithms are 
described in detail Section~\ref{sec:doclist}.
It is useful to realize that both algorithms reconstruct
the top part of the Cartesian tree of $L[sp,ep]$. \citeANP{Mut02}'s is known
to reconstruct precisely the nodes $p$ where $L[p] < sp$ (by the definition
of $L$, these are the leftmost occurrences of each document), and we prove now
that \citeANP{Sad07}'s algorithm does the same. Let us call $CT$ the part of
the Cartesian tree reconstructed by the algorithms.

We assume \citeANP{Mut02}'s algorithm also visits the left subtree first
(although in its case this does not make a difference), and prove by induction
that both algorithms process the same intervals $[i,j]$, in the same
order, and perform the same action. Both start with $[sp,ep]$,
compute the position $p = \rmq_L(sp,ep)$, and split the interval at $p$, 
because there must be some value smaller than $sp$ in $[sp,ep]$, and thus
$L[p] < sp$ (\citeANP{Mut02}'s algorithm), and because bitmap $V$ is all 
zeroed in the beginning (\citeANP{Sad07}'s algorithm).

Now consider a general interval $[i,j]$, where by inductive hypothesis both
algorithms have performed exactly the same steps until now. Both algorithms
will compute $p=\rmq_L(i,j)$. Now there are two cases:
\begin{itemize}
\item $L[p]<sp$. Then \citeANP{Mut02}'s algorithm will report document
$C[p]$, which means it is the first time it sees this document. By the 
inductive hypothesis, \citeANP{Sad07}'s algorithm has visited the same 
documents until now, thus it is also the first time it sees document $C[p]$.
Hence it holds $V[C[p]] = 0$, and the algorithm also reports $C[p]$. Then both
split the interval at $p$ and continue processing it.
\item $L[p] \ge sp$. This means that the first occurrence of $C[p]$ in
$[sp,ep]$ is to the left of $p$. Moreover, it is to the left of $i$ (otherwise
the $\rmq$ operation would have given that position instead of $p$). Since 
\citeANP{Mut02}'s algorithm finds the leftmost occurrence of each document in 
the interval, and we assume it reconstructs $CT$ in left-to-right preorder,
any node to the left of $i$ must have already been reconstructed, because it
will not be visited later. Then, \citeANP{Mut02}'s algorithm has already
output $C[p]$ and, by the inductive hypothesis, \citeANP{Sad07}'s algorithm 
has already output $C[p]$ as well. Thus $V[C[p]]=1$, and then both algorithms 
terminate the recursion in this node.
\end{itemize}

This completes the proof.
The following example shows that indeed an error can occur if we process the
recursive calls in the wrong order.

\begin{example}
In Fig.~\ref{fig:muthu2}, consider color listing in the
interval $C[5,11]$, where the four colors appear. Run \citeANP{Sad07}'s 
algorithm going left first and verify that it behaves exactly as 
\citeANP{Mut02}'s algorithm. Now consider processing the right interval 
first. We start with $p=\rmq_L(5,11)=8$, report $C[8]=1$ and set $V[1]
\leftarrow 1$. Now we go right and process $C[9,11]$. Here we find 
$p=\rmq_L(9,11)=9$, report $C[9]=3$ and set $V[3]\leftarrow 1$. Then we
process $C[10,11]$, find $p=\rmq_L(10,11)=10$, and since $C[10]=2$ and
$V[2]=0$, we report color $2$ and set $V[2]\leftarrow 1$ (note that
\citeANP{Mut02}'s algorithm
would not have reported color $2$ here because $L[p] \ge sp$).
Finally it processes $C[11,11]$, where color $C[11]=1$ is not reported
because $V[1]$ is already $1$. Now we finally go to the left child of the
initial recursive call, interval $C[5,7]$. Here we compute $p=\rmq_L(5,7)=5$,
and since $C[5]=2$ and $V[2]=1$, the recursion terminates without having
reported color $C[7]=4$.
\end{example}

\section{Constant-Time Array Initialization in Linear-Bit Space}
\label{app:constinit}

The classical solution to this problem uses linear space, that is, $O(D\lg D)$
extra bits on top of the array (see Ex.\ 2.12, page 71 in \citeN{AHU74},
or a complete description by \citeN{Meh84}, Sec.\ III.8.1). 
We show here how to implement the solution using only $O(D)$ 
bits. To be precise, we are interested in implementing the following data 
structure. 

\begin{definition}[Initializable Array]
An {\em initializable array} is a data structure $V[1,D]$ that supports the
operations $init(V,D,v)$, $read(V,i)$ and $write(V,i,v)$. The first operation
initializes $V[i]\leftarrow v$ for all $1 \le i \le D$; the second obtains $V[i]$
and the third sets $V[i] \leftarrow v$. 
\end{definition}

The classical technique is as follows. We use a second
array $U[1,D]$ and a stack $S[1,D]$, both storing indices in $[1..D]$. An
additional variable $0 \le t \le D$ tells the current size of $S$, and
variable $I$ stores the initialization value.

\begin{figure}[t]
\includegraphics[width=0.49\textwidth]{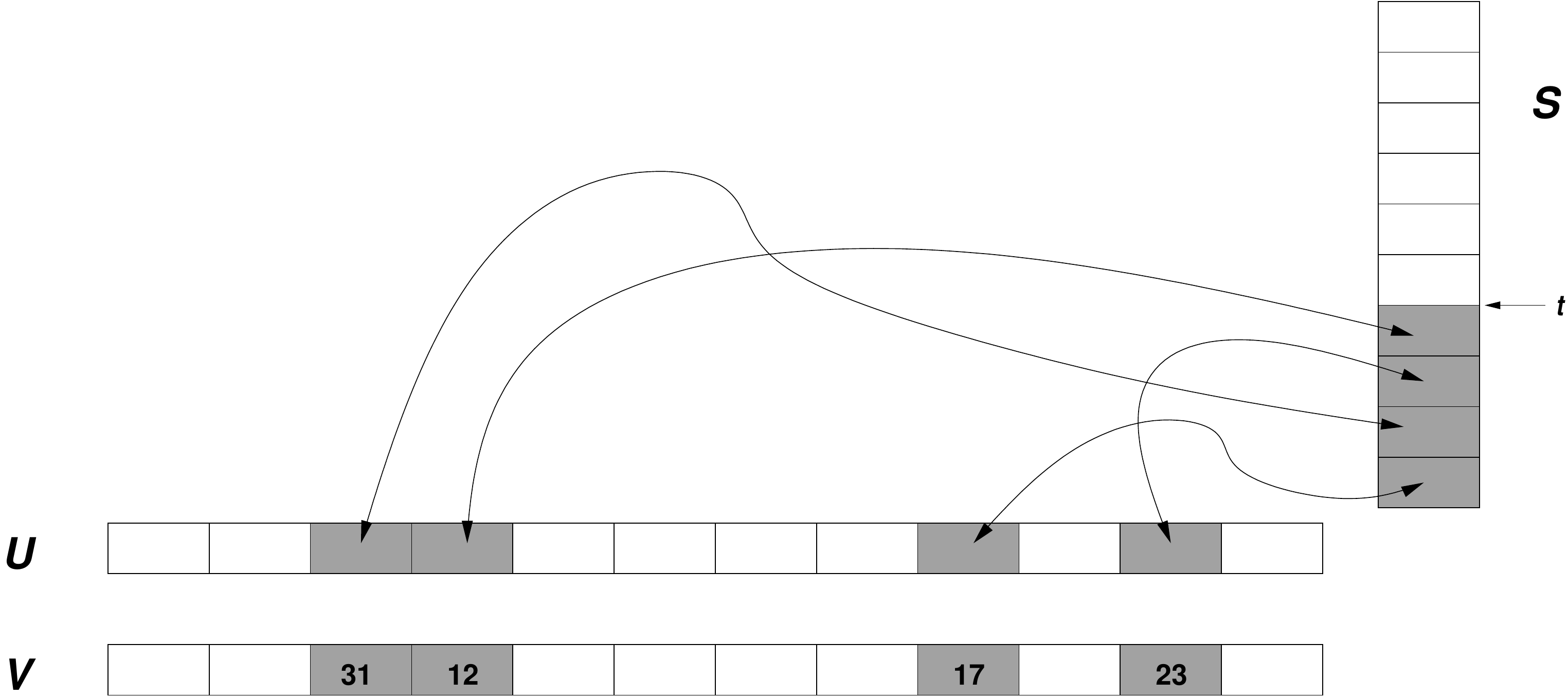}
\includegraphics[width=0.49\textwidth]{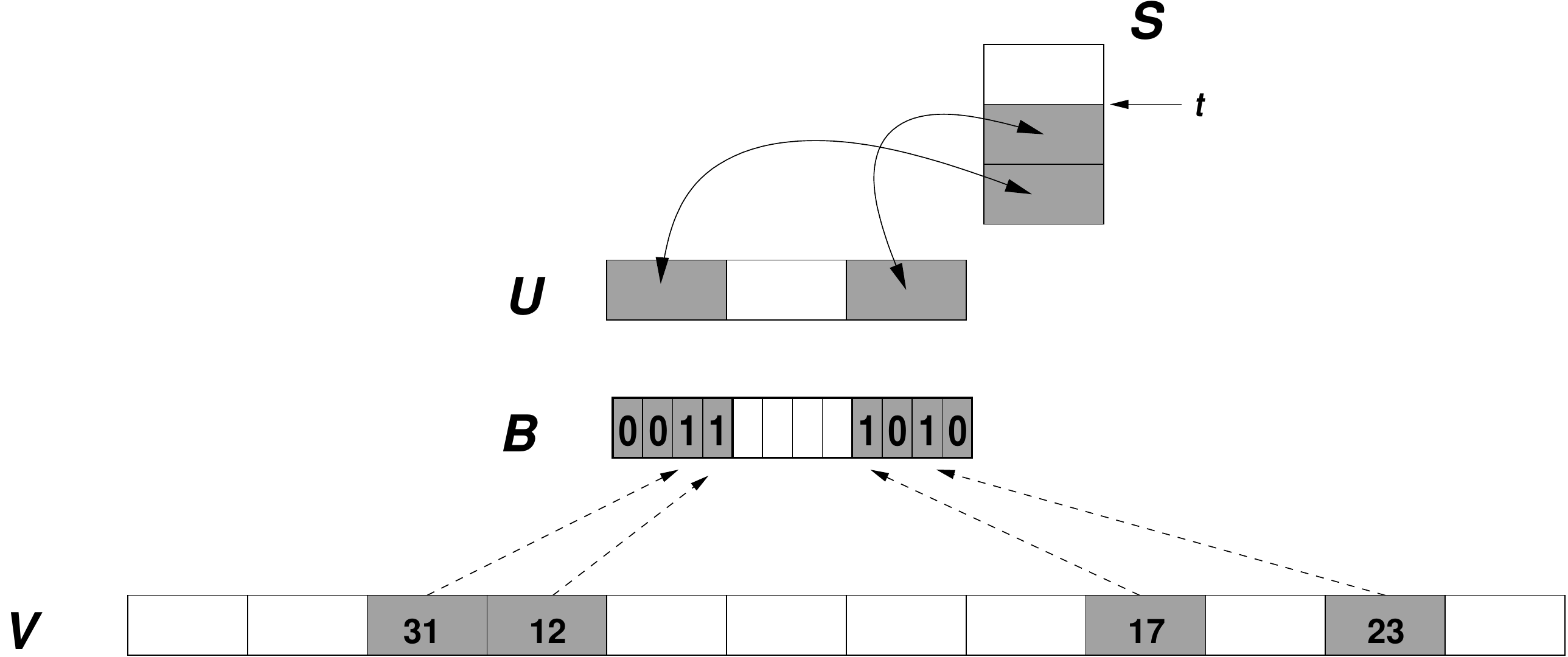}
\caption{On the left, the classical scheme, tripling the space. On the right,
our compact solution using $3n$ extra bits.}
\label{fig:constinit}
\end{figure}

Initialization of the whole structure, $init(V,D,v)$, consists of setting
$t \leftarrow 0$ and $I \leftarrow v$.
The invariant maintained by the structures is as follows:
$$ V[i] ~\textrm{is initialized} ~~\Longleftrightarrow~~
   \left(1 \le U[i] \le t ~\land~ S[U[i]] = i\right),
$$
which is immediately correct once we set $t=0$. The idea is to distinguish
initialized entries $V[i]$ because $U[i]=j$ and there is a back pointer
$S[j] = i$. Let us take an uninitialized entry $V[i]$. Value $U[i]$ is not
initialized either. If $U[i]<1$ or $U[i]>t$, we know for sure that $V[i]$ is not
initialized. Yet it could be that $1 \le U[i] \le t$. But then it is not
possible that $S[U[i]] = i$ because entry $U[i]$ in $S$ has been used
to initialize another entry $V[i']$ and then $S[U[i]] = i' \not= i$.
Operation $read(V,i)$ is as follows. If $V[i]$ is initialized, then it returns
$V[i]$, otherwise it returns $I$. Operation $write(V,i,v)$ is as follows. If
$V[i]$ is not initialized, it first sets $t \leftarrow t+1$, $U[i]\leftarrow t$,
and $S[t] \leftarrow i$. After the possible initialization, it sets
$V[i] \leftarrow v$.
It is interesting that one can even uninitialize $V[i]$,
by setting $S[U[i]] \leftarrow S[t]$, $U[S[t]] \leftarrow U[i]$, $t \leftarrow
t-1$.
Fig.~\ref{fig:constinit} (left) illustrates the basic technique.

\medskip

To reduce space
we use, instead of array $U$ and stack $S$, a bit vector $B[1,D]$ so that
$B[i] = 1$ iff $V[i]$ has
been initialized. This way we require only $D$ bits in addition to $V$. Of
course the problem translates into initializing $B[i] \leftarrow 0$ for all
$i$. We now take advantage of RAM model of computation, with word size 
$w \ge \lg D$.

As $B$ is stored as a contiguous sequence of bits, let us interpret this
sequence as an array $B'[1,D']$ of $D' = \lceil D/w \rceil$ entries, each
entry holding a computer word of $w$ bits of $B$.
We can apply now the classical solution to $B'$, so
that $B'$ can be initialized in constant time (at value $B'[i] = 0$). The extra
space on top of $B'$ is $2D'\lg D' \le 2 D$ bits. Together with $B'$, the
space overhead of the solution is $3D$ bits. Now, in order to determine whether
$V[i]$ is initialized, we just check $B[i]$: We compute $q = \lfloor i/w\rfloor$
and $r = i~\textrm{mod}~w$ and check the bit number $r+1$ of $B'[q+1]$. If
$B'[q+1]$ is not yet initialized, we know $B[i] = 0$ and thus $V[i]$ is not
yet initialized. To initialize $V[i]$ we set the $(r+1)$-th bit of $B'[q+1]$,
previously initializing $B'[q+1] \leftarrow 0$ if needed.
Fig.~\ref{fig:constinit} (right) illustrates our solution.

We note that this idea can be carried out further one more level to achieve
$D+o(D)$ extra bits of space, instead of $3D$.

\section{Proof of Correctness of Hon et al.'s Algorithm}
\label{app:hsvproof}

We prove that the document listing using $o(n)$ additional bits of
\citeN{HSV09} is correct if we first process the left subinterval, then the
block where the minimum lies, and then the right subinterval. Moreover, we
show that we obtain at least one new document to list per block scanned.%
\footnote{This is a joint result with Djamal Belazzougui and will be published
elsewhere. We include it here for completeness.} The need to process the 
recursion in this order is not clear at all in the original paper \cite{HSV09}.

We show by induction on the size of the current subinterval $L[i,j]$ that,
if we start the procedure with the elements that already appear in $C[sp,i-1]$
marked, then we find and mark the leftmost occurrence of each distinct symbol
not yet marked, spotting at least one new element per block scanned.

This is trivial for the empty interval. Now consider
the minimum position in $L'[k']$, which contains the leftmost occurrence of some
element $C[k]$, for some $k$ within the block of $L'[k']$. If $C[k]$ is already
marked, it means it appears in $C[sp,i-1]$, and thus it holds
$L[k] \ge sp$, and the same holds for all the values in $L[i,j]$. Thus if all the
elements in the block of $L'[k']$ are marked, we can safely stop the procedure.

Otherwise,
before doing any marking, we recursively process the interval to the left of
block $k'$, which by inductive hypothesis marks the unique elements in that
interval. Now we process the current block, finding at least
the new occurrence of element $C[k]$ (which cannot appear to the left of $k'$).
Once we mark the new elements of the current block, we process the interval
to the right of $k'$, where the inductive hypothesis again holds.

The following is an example where an incorrect result is obtained if one 
processes the block before the left branch of the recursion.

\begin{example}
Consider the color array 
$C=\langle 2,3,3,3,2,2,2,2,2,1,1,1\rangle$,
its predecessor array
$L=\langle 0,0,2,3,1,5,6,7,8,0,10,11\rangle$,
and grouping factor $b=2$. Thus, we have
$L'=\langle 0,2,1,6,0,10 \rangle$.
If the query is for the interval $[3,12]$, it is mapped to
$[2,6]$ in $L'$. The minimum is in $L'[5]=0$, which makes
the algorithm mark colors $C[9]=2$ and $C[10]=1$, setting
$V[1] \leftarrow 1$ and $V[2] \leftarrow 1$. Now we
go left and consider subinterval $L'[2,4]$, with the minimum
in $L'[3]=1$. The corresponding colors are $C[5]=2$ and $C[6]=2$.
But since $V[2]=1$, both cells are already reported and the
algorithm finishes in this branch of the recursion, missing
color $C[3]=C[4]=3$.
\end{example}

\section{Compact Dynamic Dictionaries in Constant Worst-Case Time}
\label{app:dict}

We describe a dynamic structure that stores $r$ distinct keys over a universe
$[1..D]$, with $t$-bit satellite data, using $O(rt)+o(r\lg D+D)$ bits.
The structure
supports searches and updates in constant worst-case time. This is folklore but
a clear explanation is not easy to find, to the best of our knowledge.

We split the universe of keys $[1..D]$ into buckets of size $b=\lg^2 D$.
To each bucket we associate a pointer to a data structure
managing the keys that fall in that bucket (this can be initialized in constant
time, see Appendix~\ref{app:constinit}). Given a key we find its bucket in
constant time with a simple division.

The structure that handles each bucket is a B-tree with arity
$a=\sqrt{\lg D}$. Since the local universe
of the bucket is of size $b$, the height of the B-tree is
$h=O(\lg_a b)=O(1)$. Furthermore, the keys inside the
bucket require only $O(\lg\lg D)$ bits, therefore all the keys in an
internal B-tree node fit in $O(a\lg\lg D) = o(\lg D)$ bits, and thus we
can find the subtree containing a key in $O(1)$ time using a precomputed
table of $O(2^{o(\lg D)}b\lg b) = o(D)$ bits.
Thus the B-tree searches require constant time $O(h)$.

Each B-tree is in its own memory area, which can contain up to $O(b)$ nodes,
so the tree pointers require only $O(\lg\lg D)$ bits. Hence the $a$
pointers to the children can also be updated in constant time in the RAM model.
Insertions and deletions of keys can, therefore, be carried out in constant
worst-case time. Since each key may induce the creation of $O(h)$
nodes and also stores its satellite data, the total number of bits stored in 
B-tree nodes is $rt + O(ra\lg\lg D) = r(t+o(\lg D))$.

The $O(D/b)$ memory areas for the B-trees,
adding up to $r(t+o(\lg D))$ bits, are stored in ``extendible arrays''
\cite{RR03}. These pose an overhead of $O(rt)+o(r\lg D)+O((D/b)\lg D)$
bits and handle the accesses and grow/shrink operations on the memory areas
in constant worst-case time. Since $(D/b)\lg D = o(D)$, we have
the promised total space of $O(rt)+o(r\lg D + D)$ bits, and constant
worst-case time for all the operations. In addition, we can maintain a linked
list of the nonempty buckets to allow for $O(r)$-time traversal of the set.

Note that it is easy to carry out predecessor searches in the B-trees. By
adding the structure of \citeN{vEBKZ77} on the buckets, we 
retain the same asymptotic space and support predecessor searches and updates
in time $O(\lg\lg D)$.

\section{Computing a Term Frequency in Compressed Space} 
\label{app:dltfsada}

\citeN{Sad07} maintains, in addition to the global CSA, a CSA for
the local suffix array $A_d$ of each document $T_d$. Since the pointers of 
$A_d$ are interspersed in $A$ in the same order, we can do the following to map 
a global entry $A[i]$ to the corresponding local entry $A_d[j]$ in its 
document $d$. First, compute the global text position $p = A[i]$. Then find
the corresponding document $d = 1+\rank_1(B,p-1)$ and its starting position
in $\T$, $s = 1+\select(B,d-1)$. Finally, use the inverse suffix array to map
the local offset $p-s+1$ into the local suffix array of $d$, 
$j = A_d^{-1}[p-s+1]$. This is done in time $\taccess$ with the CSAs
of $A$ and $A_d$. 

\begin{figure}[t]
\centerline{\includegraphics[width=0.8\textwidth]{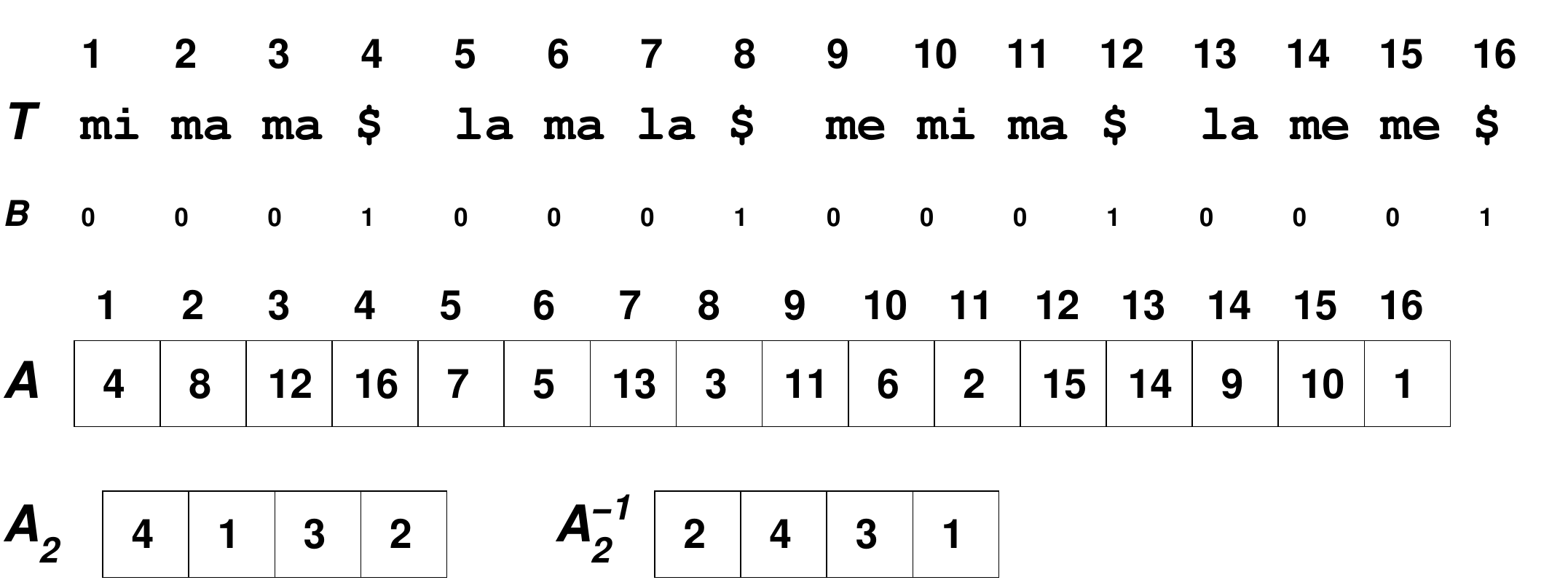}}
\caption{The structures for mapping global to local CSA positions.}
\label{fig:mapcsa}
\end{figure}

\begin{example}
Let us map $A[10]$ to its local suffix array. Fig.~\ref{fig:mapcsa} illustrates
the necessary components. We compute $p=A[10]=6$ using the global CSA. Now
we compute the corresponding document, $d=1+\rank_1(B,6-1)=2$, its
starting position $s = 1+\select(B,2-1)=5$, and the local offset $p-s+1=2$.
Now we obtain $A_2^{-1}[2] = 4$ with the local CSA of document $2$.
Certainly, $A_2[4]=2$ points to the local suffix $\textsf{"ma la \$"}$, which
is the same global suffix $A[10]=6$.
\end{example}

Now imagine we know the first and last occurrence positions of a document $d$ in
$C[sp,ep]$, say $i$ and $i'$, as explained in Section~\ref{sec:tf}. 
We use the above procedure to map them to
positions $j$ and $j'$ in $A_d$. Those are clearly the first and last 
suffixes of $A_d$ that start with $P$, and therefore $\tf(P,d)=j'-j+1$.
Since for all reasonable CSAs it holds $\sum_d |\CSA_d| \le |\CSA|$, we have 
Solution~\ref{thm:dltfsada}.

\citeN{HSV09} show how the same $\tf(P,d)$ can be computed, in time
$O(\taccess\,\lg n)$, if we know only the leftmost position $i$ or the
rightmost position $i'$ of the document in the range. Say it is the position 
$i$. We map this position to $A_d[j]$, and carry out an exponential search 
for the largest position $j' \ge j$ such that $A_d[j']$ is mapped back to 
some $A[i']$ with $i' \le ep$. When we find such $j'$, we have
$\tf(P,d)=j'-j+1$ (we also discover $i'$ in the process). This mapping from
$A_d[j']$ to $A[i']$ is the inverse of the one we described from $A[i]$ to
$A_d[j]$, and it is computed analogously.

\section{Document Listing using Wavelet Trees} \label{app:wtree}

We start with an example of the wavelet tree structure and then show how we
use it for document listing.

\begin{figure}[t]
\centerline{\includegraphics[width=0.6\textwidth]{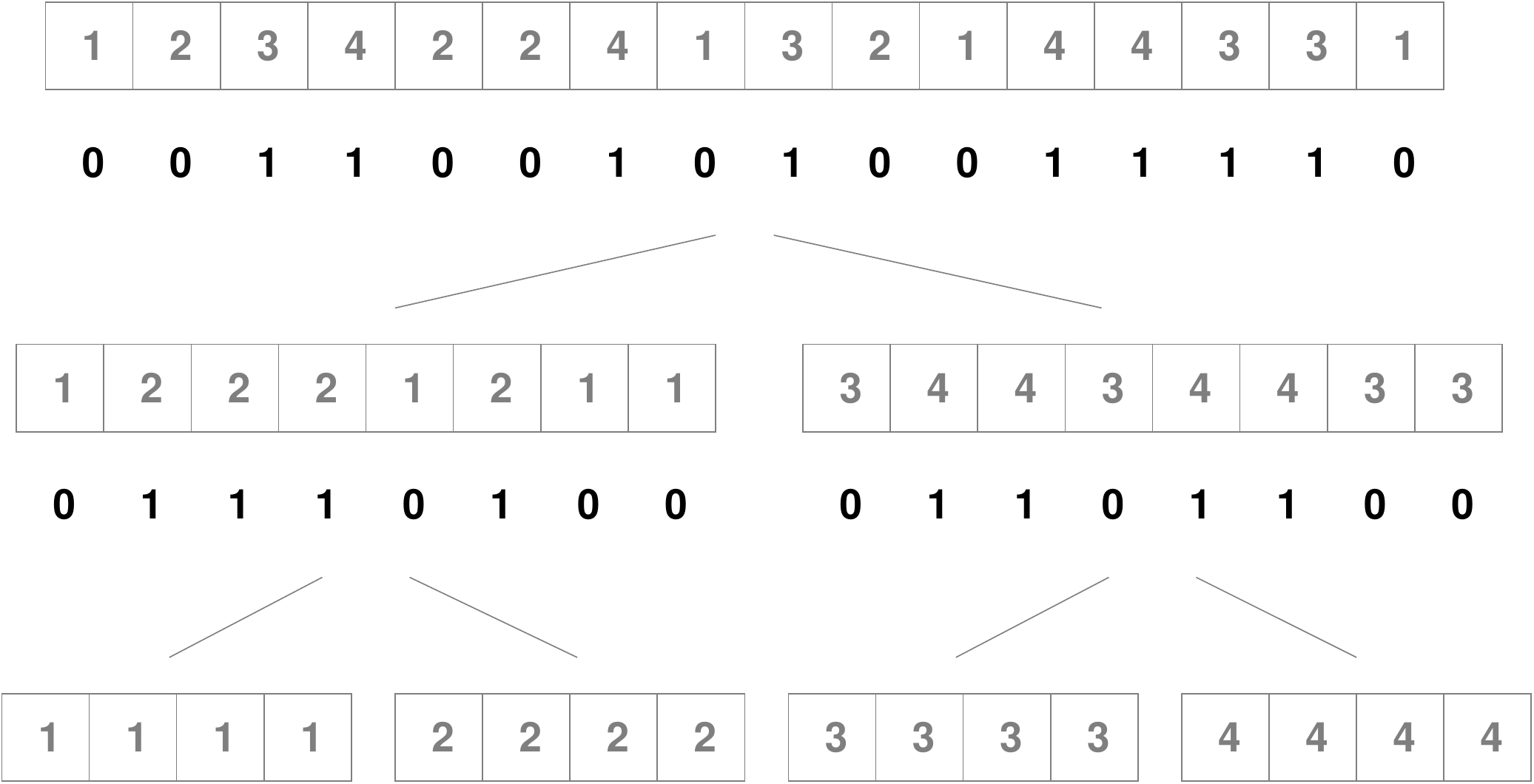}}
\caption{The wavelet tree of array $C$ in our running example. Grayed data
is not represented.}
\label{fig:wavelet}
\end{figure}

\begin{example}
Fig.~\ref{fig:wavelet} displays the wavelet tree for our array $C$. Since we
have $D=4$ symbols, the wavelet tree has 2 bitmap levels (plus the leaves,
which are virtual and not stored). Each node shows its $C_v$ sequence
(which is not stored) in gray and its $B_v$ bitmap (which is stored)
in black. Each level stores $n$ bits. To recover $C[9]$ we start at the root
node $v$ and read $B_v[9]=1$. Thus we go to the right child $u$ of $v$, where
the original position $9$ becomes $\rank_1(B_v,9)=4$ in $C_u$. Now $B_u[4]=0$,
so we go to the left child of $u$ and reach the leaf of symbol $3$, thus
$C[9]=3$.
\end{example}

\citeN{GPT09} showed that the wavelet tree of $C$ allows for a completely
different document listing algorithm, based on the following basic problem
(also known as range selection queries).

\begin{problem}[Range Quantile] \label{prob:quantile}
Preprocess an array $C[1,n]$ of
integers so that, given a range $[sp,ep]$ and an index $q$, we can
return the $q$th smallest element in $C[sp,ep]$.
\end{problem}

\citeANP{GPT09}\ solved this problem over a wavelet tree representation
of $C$ as follows. We start at the root $v$ with interval $[sp,ep]$. We count
the number of 0s in $B_v[sp,ep]$, with $z = \rank_0(B_v,ep)-\rank_0(B_v,sp-1)$.
Now, if $q \le z$, then the answer belongs to the integers handled by the 
left child of the root, thus we remap the interval to 
$sp' = \rank_0(B_v,sp-1)+1$ and $ep' = \rank_0(B_v,ep)$ and continue
recursively on the left child of the root. Otherwise, we remap the interval to
$sp' = \rank_1(B_v,sp-1)+1$ and $ep' = \rank_1(B_v,ep)$ and continue
recursively on the right child of the root, this time looking for $q-z$ 
instead of $q$. When we arrive at a leaf, this is the $q$th smallest value
in $C[sp,ep]$. The process takes $O(\lg D)$ time. Observe that the final
$ep-sp+1$ value is the frequency of the $q$th element in the array.

\begin{example}
We obtain the median of $C[5,11]$, that is, $q=4$. At the root node $v$ we 
count the number of $0$s in $B_v[5,11]$ using 
$\rank_0(B_v,11)-\rank_0(B_v,5-1)=5$. Since $5\ge 4=q$, the $q$th
element of $C[5,11]$ is on the left child $u$ of $v$. Thus we descend to $u$,
mapping the interval $[5,11]$ to $[\rank_0(B_v,5-1)+1,\rank_0(B_v,11)]=[3,7]$.
Now we count the number of $0$s in $B_u[3,7]$ using
$\rank_0(B_u,7)-\rank_0(B_u,3-1)=2$. Since $2<4=q$, the $q$th element of
$C_u[3,7]$ is on the right child $w$ of $u$. Now we remap the interval $[3,7]$
to $[\rank_1(B_u,3-1)+1,\rank_1(B_u,7)]=[2,4]$, and since we move to the right,
we set $q \leftarrow q-2 = 2$. Now we arrive at $w$ and, since it corresponds to
the leaf of symbol $2$, we conclude that the median of $C[5,11]$ is $2$. 
Moreover, it occurs $4-2+1=3$ times in $C[5,11]$.
\end{example}

It is possible to solve the range quantile problem in
time $O(\lg n / \lg\lg n)$ using $O(n\lg n)$ bits (assuming $D=O(n)$)
\cite{BGJS11}, and this is optimal within $O(n\,\polylog(n))$ space \cite{JL11}.
In this survey we are more interested in solutions using sublinear extra
space, thus we state the wavelet tree result. It is an open problem to
obtain optimal time within sublinear extra space.

\begin{solution}[Range Quantile] {\rm \cite{GPT09}} \label{thm:quantile}
The problem can be solved in $O(\lg D)$ time on a representation
of $C$ that uses $n\lg D + o(n\lg D)$ bits of space.
\end{solution}

\citeANP{GPT09}\ use range quantile queries to solve document listing as
follows. They first ask for the $q=1$st quantile of $C[sp,ep]$, thus obtaining
the smallest document in the range, $d$, and its frequency, $\tf(P,d)$. Then
they report $d$ and set $q \leftarrow q + \tf(P,d)$. Now they ask for the 
$q$th element in $C[sp,ep]$, which gives the second smallest document in the 
range with its frequency, and so on, until all the documents
are reported. Note they do not resort to \citeANP{Mut02}'s technique, nor 
they need RMQs. Moreover, they return the documents in increasing order.

While their complexity is not competitive with our previous solutions, a
simplification of their method improves it \cite{GNP11}. Instead of quantile
queries, just traverse all the wavelet tree paths from the root, towards left
and right children, stopping either when the interval $[sp,ep]$ becomes empty,
or when we arrive at a leaf handling an integer $d$, where we report document
$d$ with $\tf(P,d)=ep-sp+1$. 

\begin{example}
Let us list the different values in $C[12,15]$ using range quantile queries.
We ask for the $q=1$st element and get that it is $3$, which appears $2$ times.
Thus we set $q \leftarrow q+2=3$ and ask for the $q=3$rd element, getting $4$,
which occurs $2$ times. Thus we set $q \leftarrow q+2=5$, which is larger than
our interval, so we finish.

Let us now proceed by a recursive traversal. We map the interval $C[12,15]$
to the left child of the root, obtaining $[8,7]$, which is empty, so there are
no elements to report in this subtree. On the right child of the root, the
interval is mapped to $[5,8]$. Now we map to its left child, obtaining
$[3,4]$ at the leaf of symbol $3$, so we report $3$ with frequency $4-3+1=2$.
Now we go to the right child, mapping $[5,8]$ to $[3,4]$. Now we 
arrive at the leaf of symbol $4$, which we also report with frequency $2$.
\end{example}

It is shown that the time per item output of the recursive traversal 
improves as more documents are listed. This results in Solution~\ref{thm:dltfgnp}.

\section{Computing Document Frequency in Compressed Space}
\label{sec:dfsada}

The idea is as follows \cite{Hui92}. For a node $v$ of the suffix tree, 
let $\tf(v,d) = \tf(str(v),d)$ for short.
Thus, if $v$ is the locus of pattern $P$, it holds $\tf(P,d) = \tf(v,d)$. Now 
let us define $u(v) = \sum_{d, \tf(v,d)>0} \tf(v,d)-1$. It is not hard to see 
that $\occ(P,\T) = \sum_{d, \tf(P,d)>0} \tf(P,d) = \df(P) + u(v)$.
Thus we can easily compute $\df(P) = \occ(P,\T) - u(v)$ in a suffix tree where
$u(v)$ is computed for all the nodes.

\citeN{Sad07} shows how to store this information compactly as follows.
Value $u(v)$ is also the number of times the LCA of two consecutive leaves
of $T_d$ descend from $v$ in the GST, for any $d$. Thus, if we define $h(w)$ 
as the number of times the LCA of two such consecutive leaves 
is exactly $w$, it holds $u(v) = \sum_{w~\mathrm{descends~from}~v} h(w)$.
By storing the $h(w)$ values in preorder, we can compute $u(v)$ by adding up
all the $h(w)$ values in a contiguous range. Furthermore, one can see that
each pair of consecutive leaves of the same document adds 1 to some $h(w)$
value, and thus $\sum_w h(w) = n-D<n$. This allows us to store all the $h(w)$
values in unary, that is, concatenating $1^{h(w)}0$ to a bitmap $H[1,2n]$,
so that if $p$ is the preorder of node $v$ and $s$ is the number of nodes in
its subtree, we have $u(v) = \select_0(H,p+s-1)-\select_0(H,p-1)-s$. The
suffix tree topology can also be represented with $O(n)$ bits so that the
required operations on the suffix tree, including finding the node $v$ that
covers the suffix array interval $[sp,ep]$, can be carried out in constant time
\cite{SN10}.
On top of any CSA, this yields Solution~\ref{thm:docfreq}.

\begin{figure}[t]
\centerline{\includegraphics[width=0.9\textwidth]{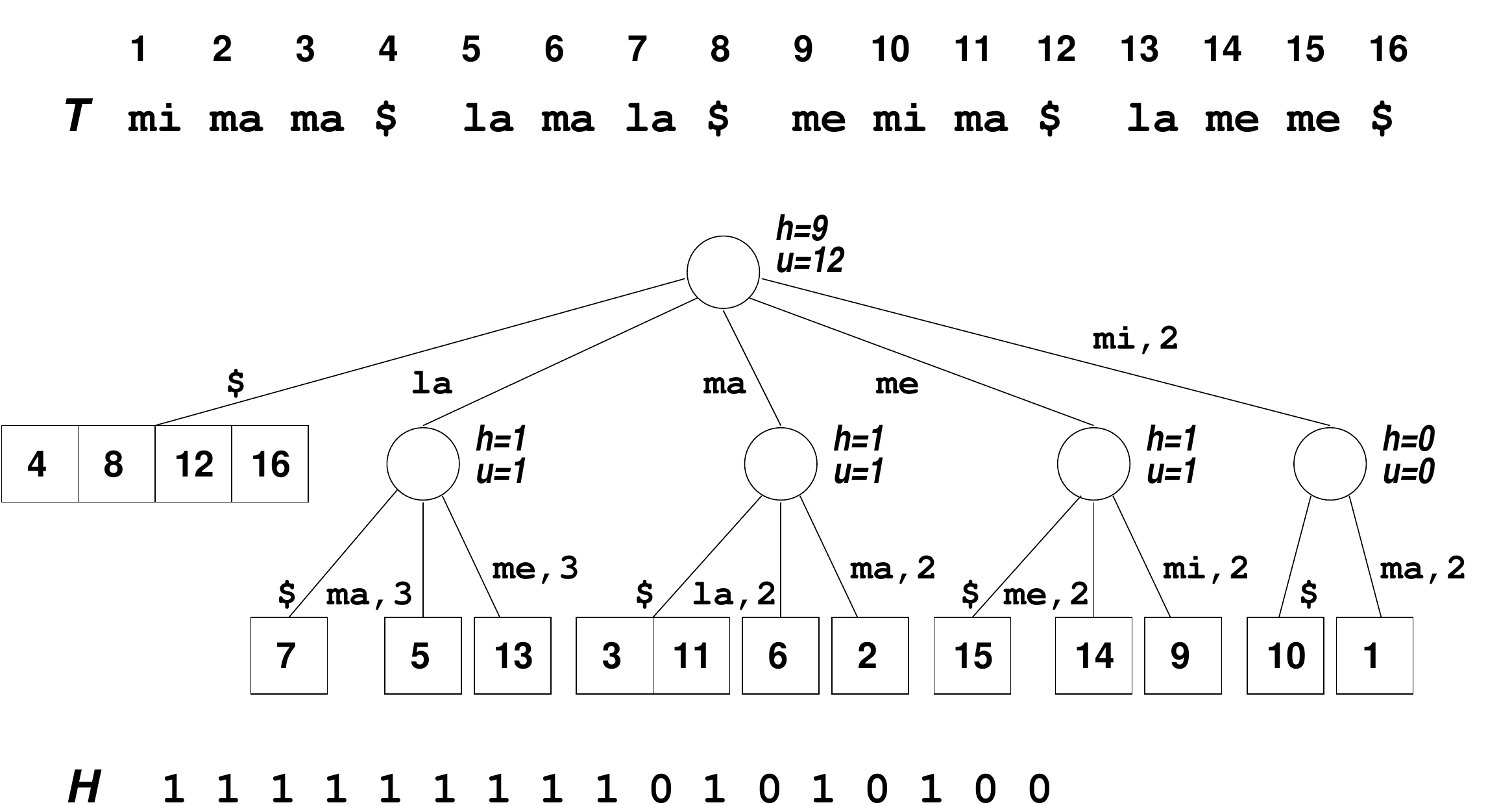}}
\caption{The data structure to compute document frequencies (only $H$ and the
suffix tree topology are represented) on our running example GST.}
\label{fig:dfsad}
\end{figure}

\begin{example}
Fig.~\ref{fig:dfsad} illustrates the data structure on our example GST. We show
$u$ and $h$ besides each internal node, and the bitmap representation $H$ on
the bottom. For example, $\textsf{"ma"}$, with locus $v$, occurs in 
$\df(\textsf{"ma"})=\occ(\textsf{"ma"},\T)-u(v)=4-1=3$ distinct documents, 
where $\occ(\textsf{"ma"},\T)$ is the number of leaves below $v$.
\end{example}

\section{Optimal-Space Solution to Document Listing with Frequencies}
\label{sec:newdltfcompr}

In this section we build on the solutions of Section~\ref{sec:topkcompr} to
achieve document listing with frequencies within optimal space, obtaining
Solution~\ref{thm:dltfopt}.

Very roughly, the idea is to use Solution~\ref{thm:topkcomprhsv}, and more
precisely the refinement of Solution~\ref{thm:topkcomprnt}, and ask for
the top-$k$ documents in the range, for successive powers of 2 for $k$ until
all the documents in the range are returned with their frequencies.

Let $c_k$ be the block size for tree $\rho_k$,
for $k=1,2,4,\ldots, D$. After determining $[sp,ep]$ in $O(\tsearch(m))$ time,
we obtain the corresponding locus of $P$ in the successive trees $\rho_k$, each
in constant time using an LCA operation, as explained, on the maximum
$c_k$-aligned range $[sp'',ep'']$ contained in $[sp,ep]$. We continue until the
locus for some $\rho_k$ stores less than $k$ candidates, which means that there
are less than $k$ distinct documents in the range, or when reaching
$k=D$. For that $k$, the top-$k$ candidate list includes all the distinct
documents in $[sp',ep']$, with their frequencies. Moreover, since the
locus in tree $\rho_{k/2}$ still had $k/2$ distinct elements in its $[sp',ep']$
range, it holds $k/2 \le \docc$, and thus $k = O(\docc)$.

To complete this precomputed set (which is actually computed on the fly with
the help of $\tau_k$ and the sampled document array), we must traverse the
$O(c_k)$ elements of $A$ that are in $[sp,ep]\setminus [sp',ep']$ and add them
to the result. In those leaves we will find $(i)$ elements that already appear
in $[sp',ep']$, and thus their frequency has been obtained from the top-$k$
candidate list, and $(ii)$ new elements that do not appear in $[sp',ep']$.
This means that we do not need to store the frequencies of the $O(\sqrt{ck})$
candidates of \citeN{Tsu13}, but just the $O(k\lg\lg n)$ bits to correct the
frequency information of the top-$k$ list. Documents found in
$[sp,ep]\setminus [sp',ep']$
and not in the top-$k$ list are known to have frequency zero in $[sp',ep']$.

To manage the set of candidates, insert the new
elements that appear, increase the counters of the elements visited, and
finally collect them all for listing, we store the document identifiers in a
dictionary, which offers constant worst-case time for the operations
(see Appendix~\ref{app:dict}).

The time is related to the block size $c_k = k\,\ell_k$.
Since tree $\rho_k$ has $O(n/c_k)$ nodes, the sum
of the space for the lists stored in it is
$O((n/c_k)k\lg\lg n) = O((n/\ell_k)\lg\lg n)$. Choosing
$\ell_k = \lg k \lg^\epsilon n$, the space for $\rho_k$ is
$O(n\lg\lg n/(\lg k \lg^\epsilon n))$ bits. Added over all the trees
$\rho_k$ for $k=2^i$, this gives total space
$O(n\lg\lg n/\lg^\epsilon n) \sum_{i=0}^{\lg D} 1/i =
O(n\lg\lg n\lg\lg D / \lg^\epsilon n) = o(n)$ bits.

\end{document}